%% file: paper.tex
\documentclass[conference]{IEEEtran}
\IEEEoverridecommandlockouts

\usepackage{cite}
\usepackage{amsmath,amssymb,amsfonts}
\usepackage{algorithmic}
\usepackage{graphicx}
\usepackage{textcomp}
\usepackage{listings}
\usepackage{fancyvrb}
\usepackage{fvextra}
\usepackage[ruled, linesnumbered, noend]{algorithm2e}
\usepackage{caption}
\usepackage{subcaption}
\usepackage{threeparttable}
\usepackage{multirow}
\usepackage{hyperref}
\usepackage{longtable, booktabs}

\def\BibTeX{{\rm B\kern-.05em{\sc i\kern-.025em b}\kern-.08em
    T\kern-.1667em\lower.7ex\hbox{E}\kern-.125emX}}

\usepackage{color}
\usepackage[table]{xcolor}
\definecolor{deepblue}{rgb}{0,0,0.5}
\definecolor{deepred}{rgb}{0.6,0,0}
\definecolor{deepgreen}{rgb}{0,0.5,0}
\definecolor{cadmiumorange}{rgb}{0.93, 0.53, 0.18}
\definecolor{lightblue}{rgb}{0.68, 0.85, 0.90}

\newcommand{\Grid}{\emph{Grid}}
\newcommand{\Function}{\emph{Function}}
\newcommand{\TimeFunction}{\emph{Time\-Function}}
\newcommand{\SparseFunction}{\emph{Sparse\-Function}}

\newcommand{\Eq}{\emph{Eq}\xspace}
\newcommand{\Operator}{\emph{Operator}\xspace}

\newcommand{\data}{\emph{data}\xspace}
\newcommand{\halo}{\emph{halo}\xspace}
\newcommand{\padding}{\emph{pad\-ding}\xspace}

\newcommand{\HaloSpot}{\emph{Halo\-Spot}}
\newcommand{\HaloSpots}{\emph{Halo\-Spots}}

\newcommand{\HaloTouch}{\emph{Halo\-Touch}\xspace}

\hypersetup{breaklinks=true,
            pdfauthor={},
            pdftitle={Automated MPI-X code generation for scalable finite-difference solvers},
            colorlinks=true,
            citecolor=black,
            urlcolor=blue,
            linkcolor=black,
            pdfborder={0 0 0}}

\newcommand\pythonstyle{\lstset{
    language=Python,
    commentstyle=\ttm,
    escapechar=|,
    basicstyle=\ttfamily\footnotesize,
    aboveskip=5pt,
    belowskip=5pt,
    otherkeywords=
    {self, Operator, Eq, Constant}, %
    keywords=[1]{
        TimeFunction, Grid, Function
    },
    keywordstyle=[1]\color{cadmiumorange},
    keywords=[2]{
        HaloUpdateCall, HaloUpdateList, HaloWaitList, HaloSpot,
        HaloTouch
    },
    keywordstyle=[2]\color{deepgreen},
    keywords=[3]{
        Expression
    },
    keywordstyle=[3]\color{red},
    keywords=[4]{
        Iteration
    },
    keywordstyle=[4]\color{blue},
    commentstyle=\color{deepgreen}\ttfamily,
    emph={Pdb, stdout},          %
    emphstyle=\ttfamily\color{deepred},    %
    morestring=[s]\\\%,
    stringstyle=\color{deepgreen},
    numbers=left,
    numbersep=2pt,
    frame=tb,                         %
    showstringspaces=false,            %
    float=htpb!,
    breaklines=true,
    captionpos=b
}}

\newcommand\Pseudostyle{\lstset{
    language=C,
    commentstyle=\itshape,
    basicstyle=\footnotesize,
    aboveskip=5pt,
    belowskip=5pt,
    otherkeywords={self},             %
    keywordstyle=\color{deepblue},
    emph={min, max},          %
    emphstyle=\color{deepred},    %
    stringstyle=\color{deepgreen},
    numbers=left,
    numbersep=2pt,
    frame=tb,                         %
    showstringspaces=false            %
    float=htpb!,
    captionpos=b,
    escapeinside={\%*}{*)}
}}

\newcommand\Cstyle{\lstset{
    language=C,
    commentstyle=\itshape,
    basicstyle=\ttfamily\scriptsize,
    aboveskip=5pt,
    belowskip=5pt,
    otherkeywords={self},             %
    keywordstyle=[2]\color{deepgreen},
    keywords=[2]{
        HaloUpdateCall, HaloUpdateList, HaloWaitList, HaloSpot,
        HaloTouch, Halo,
    },
    keywordstyle=[3]\ttfamily\color{red},
    keywords=[3]{
        Eq},
    emph={MIN, MAX, pragma, omp, parallel, schedule},          %
    emphstyle=\ttfamily\color{deepred},    %
    stringstyle=\color{deepgreen},
    numbers=left,
    numbersep=2pt,
    frame=tb,                         %
    showstringspaces=false            %
    float=htpb!,
    captionpos=b,
    breaklines=true,
    escapeinside={\%*}{*)}
}}

\lstnewenvironment{python}[1][]
{
    \pythonstyle
    \lstset{#1}
}{}

\lstnewenvironment{mypseudo}[1][]
{
    \Pseudostyle
    \lstset{#1}
}{}

\lstnewenvironment{clang}[1][]
{
    \Cstyle
    \lstset{#1}
}{}

\def\BibTeX{{\rm B\kern-.05em{\sc i\kern-.025em b}\kern-.08em
    T\kern-.1667em\lower.7ex\hbox{E}\kern-.125emX}}

\begin{document}

\title{Automated MPI-X code generation for scalable finite-difference solvers
}

\makeatletter
\newcommand{\linebreakand}{%
  \end{@IEEEauthorhalign}
  \hfill\mbox{}\par
  \mbox{}\hfill\begin{@IEEEauthorhalign}
}
\makeatother

\author{\IEEEauthorblockN{George Bisbas}
\IEEEauthorblockA{
\textit{Imperial College London}\\
London, United Kingdom \\
g.bisbas18@imperial.ac.uk}
\and
\IEEEauthorblockN{Rhodri Nelson}
\IEEEauthorblockA{
\textit{Imperial College London}\\
London, United Kingdom \\
rhodri.nelson@imperial.ac.uk}
\and
\IEEEauthorblockN{Mathias Louboutin}
\IEEEauthorblockA{
\textit{Devito Codes} \\
Atlanta, USA \\
mathias@devitocodes.com}
\linebreakand
\IEEEauthorblockN{Fabio Luporini}
\IEEEauthorblockA{
\textit{Devito Codes} \\
Lucca, Italy \\
fabio@devitocodes.com}
\and
\IEEEauthorblockN{Paul H.J. Kelly}
\IEEEauthorblockA{
\textit{Imperial College London}\\
London, United Kingdom \\
p.kelly@imperial.ac.uk}
\and
\IEEEauthorblockN{Gerard Gorman}
\IEEEauthorblockA{\textit{Imperial College London}\\
London, United Kingdom \\
g.gorman@imperial.ac.uk}
}

\maketitle

\author{\IEEEauthorblockN{Anonymous Authors}}

\thispagestyle{plain}
\pagestyle{plain}

\begin{abstract}
Partial differential equations (PDEs) are crucial in modeling
diverse phenomena across scientific disciplines, including seismic and medical imaging,
computational fluid dynamics, image processing, and neural networks.
Solving these PDEs at scale is an intricate and time-intensive process that demands careful tuning.
This paper introduces automated code-generation techniques specifically tailored
for distributed memory parallelism (DMP) to execute explicit finite-difference (FD) stencils at scale,
a fundamental challenge in numerous scientific applications.
These techniques are implemented and integrated into the Devito DSL and compiler framework,
a well-established solution for automating the generation of FD solvers based on a
high-level symbolic math input.
Users benefit from modeling simulations for real-world applications at a high-level symbolic abstraction
and effortlessly harnessing HPC-ready distributed-memory parallelism without altering
their source code. This results in drastic reductions both in execution time and developer effort.
A comprehensive performance evaluation of Devito's DMP via MPI demonstrates highly competitive strong and weak scaling on CPU and GPU clusters,
proving its effectiveness and capability to meet the demands of large-scale scientific simulations.
\end{abstract}


\IEEEpeerreviewmaketitle

\section{Introduction}\label{sec:introduction}

Driven by the continued effort to build supercomputers capable of solving real-world problems, the need for scalable and portable software development has become increasingly fundamental to efficiently using those resources.
In order to solve real-world problems on such large systems, distributed memory parallelism (DMP) is mandatory.
However, building scalable solutions that preserve numerical accuracy while harnessing computational performance can be tedious and error-prone,
even for high-performance computing (HPC) specialists.
To tackle these layers of complexity, abstractions are the natural solutions and have led to recent advances in computational fluid dynamics \cite{firedrake2017, powell2023}, electromagnetics \cite{warren2016,warren2019}, iterative solvers \cite{petsc-user-ref, unat2017} and machine learning \cite{moses2023high, grady2023}.

Motivated by that consideration, we introduce automated code generation for DMP extension to Devito \cite{devitoTOMS2020, devito-api}.
Devito is a symbolic stencil DSL and just-in-time (JIT) compiler that allows the high-level definition of complicated physical simulations such as wave-based inversion \cite{witteJUDI2019} or ultrasound medical imaging \cite{cueto2022ultrasound, stride2022}.
Thanks to its abstracted compiler, Devito generates HPC code for CPUs and GPUs, allowing for accelerated development.
By conserving the levels of abstractions in Devito and introducing DMP within the Devito compiler itself,
we show that the automatically JIT-generated and compiled code efficiently scales to modern supercomputers for industry-scale applications such as full-waveform inversion (FWI), high-frequency reverse-time migration (RTM) or real-time medical ultrasound imaging.

With these motivations in mind, this paper makes the following contributions:
\begin{itemize}
\item A novel end-to-end software stack that automates and abstracts away Message-Passing
Interface (MPI) code generation from a high-level symbolic specification within the Devito compiler.
\item Seamless integration of DMP via MPI, coupled with OpenMP and OpenACC, SIMD vectorization,
cache-blocking, flop-reducing arithmetic, and various other performance optimizations (refer to \autoref{fig:devito-outline}).
\item Support for distributed NumPy arrays enabling domain decomposition without requiring changes to user code.
The data is physically distributed, but from the user's perspective, it remains a logically centralized entity.
Users can interact with data using familiar indexing schemes (e.g., slicing) without concern about the underlying layout.
\item Enabling automatic code generation for three alternative DMP communication strategies (\emph{basic, diagonal, and full}) to facilitate high performance for kernels having differing computation and communication requirements, such as operational intensity (OI) and memory footprint.
\item DMP support for operations beyond stencils, such as local/sparse operations that can be limited to a single domain.
Such operations are imperative for real-world applications that simulate point operations, such as the source of wave propagation.
\item A comprehensive cross-comparison of strong scaling DMP strategies for four wave propagator stencil kernels with varying memory and computation needs,
conducted on CPU and GPU clusters, scaling up to 128 GPU devices on Tursa \cite{tursa}, and 128 nodes (16384 CPU cores) on Archer2 \cite{archer2}.
All contributions, code, and benchmarks are open source and available online at \cite{devito_zenodo_4.8.10}.

\end{itemize}

This paper is organized as follows: we first introduce the Devito DSL and compiler framework through illustrative examples and highlight the key features of the DSL and Devito's internal representation (IR) (\autoref{sec:devito-api-codegen}).
Second, we detail the core contribution of this work, the methodology, and the implementation
of the DMP within the Devito compiler (\autoref{sec:dmp}).
Finally, we present the performance evaluation (\autoref{sec:evaluation}) and discuss related work (\autoref{sec:related-work}).

\section{The Devito DSL and compiler framework}\label{sec:devito-api-codegen}

Devito~\cite{devitoTOMS2020,devito-api} is Python-embedded and operates at the intersection of DSLs, compiler technology,
computational science and HPC, providing a streamlined path for the rapid development of highly optimized PDE solvers.
This approach bridges the gap between interdisciplinary scientists, enabling mathematicians, (geo-)physicists,
computer scientists and others to focus on their respective areas of expertise, maximizing collaboration and overall efficiency.

Devito leverages a symbolic mathematical language based on SymPy \cite{SymPy}, facilitating the concise expression of FD stencil operations.
While the primary focus of Devito revolves around the construction of PDE solvers, its functionality extends to various operations, including tensor linear algebra,
convolutions, boundary conditions, and interpolations. 
The building of an explicit FD solver requires several key objects,
including \Grid, \Function, \TimeFunction, \Eq, and \Operator,
which are comprehensively documented, among others, in the Devito API Reference \cite{devito_documentation}.
\autoref{lst:dsl} presents an example of modeling a heat diffusion operator.
This \Operator employs the Laplacian of {\tt u} with respect to {\tt x,y},
using a 2$^{nd}$ order accurate discretization in space (SDO: spatial discretization order)
on a structured grid with a domain size of $4$x$4$.

\begin{python}[label=lst:dsl, caption={A diffusion operator uses the symbolic notation {\tt u.laplace},
 which is the second derivative of {\tt u} with respect to all spatial dimensions.}]
from devito import *
# Some variable declarations
nx, ny = 4, 4
nt = 1
nu = .5
dx, dy = 2. / (nx - 1), 2. / (ny - 1)
sigma = .25
dt = sigma * dx * dy / nu
# Define the structured grid and its size
grid = Grid(shape=(nx, ny), extent=(2., 2.))|\label{lst:trivial-grid}|
# Define a discrete function encapsulating space-
# and time-varying data, and initialize its data
u = TimeFunction(name="u", grid=grid, space_order=2)
u.data[1:-1,1:-1] = 1|\label{lst:trivial-slice}|
# Define the equations to be solved
eq = Eq(u.dt, u.laplace)
stencil = solve(eq, u.forward)
eq_stencil = Eq(u.forward, stencil)
# Generate C-code using the Devito compiler
op = Operator([eq_stencil])|\label{lst:trivial-op}|
# JIT-compile and run
op.apply(time_M=1, dt=dt)|\label{lst:trivial-end}|
\end{python}

The Devito compiler employs a multi-step compilation process (\autoref{fig:devito-outline}) to translate the symbolic representation into efficient C code.
To enhance modularity and maintainability, optimization passes are decomposed into separate tasks performed at different levels of abstraction called Intermediate Representations (IRs).
This section focuses on two primary IRs pivotal in most of the compiler's optimizations.

\begin{figure}[htpb!]
    \centering
    \includegraphics[width=0.44\textwidth]{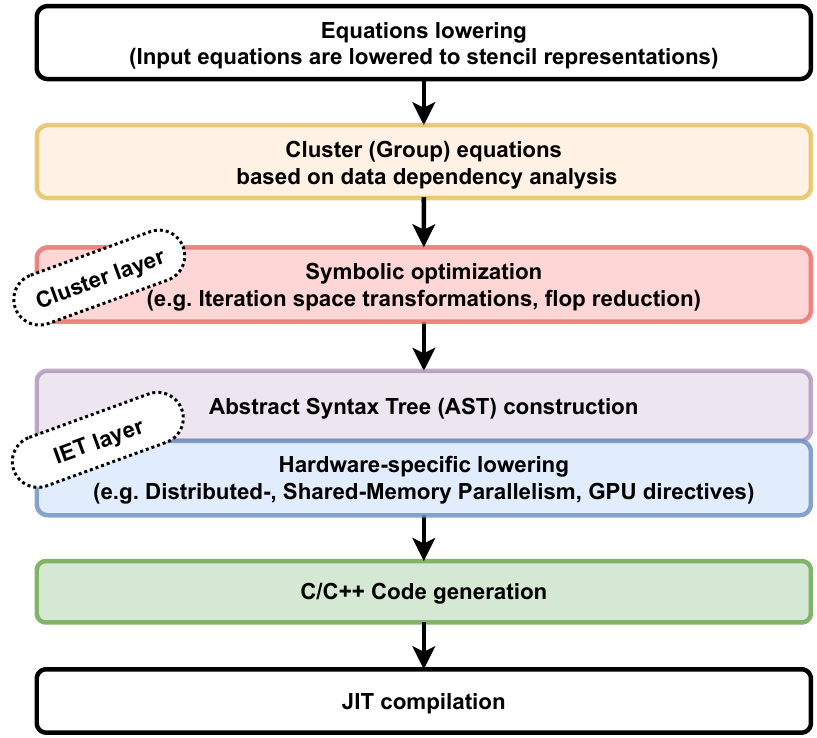}
    \caption{A high-level overview of the Devito compilation framework: from high-level symbolic maths to HPC-ready optimized code.}
    \label{fig:devito-outline}
\end{figure}

First, the Cluster-level IR (see \autoref{fig:devito-outline}),
named after clustering (grouping) symbolic mathematical expressions based on their computational properties.
The compiler performs advanced data dependence analysis at this level, facilitating the transformations in most of the following passes.
This layer reduces the arithmetic intensity of stencil kernels via loop invariant code motion, blocking for data locality, factorization, 
extracting increments to eliminate cross-iteration redundancy (CIRE) and common sub-expression elimination (CSE).

The second IR is the Iteration/Expression Tree (IET) IR,
which establishes control flow and assembles an immutable Abstract Syntax Tree (AST)
consisting mainly of iterations (loops) and expressions.
The optimization focus shifts to the loops rather than the mathematical expressions, as observed in the Cluster-level IR.
Here, optimizations are tailored to the target hardware, e.g., SIMD vectorization, OpenMP shared memory parallelism for CPUs, and OpenACC for GPUs and their tuning.
This level incorporates the synthesis of DMP, the core contribution of this work, thoroughly presented in \autoref{sec:dmp}.
Several computation and communication patterns of MPI code generation are supported, and all can be applied seamlessly with zero alterations to the user code.

\section{Automated Distributed-memory Parallelism}\label{sec:dmp}

This section details the methodology to implement automated DMP.
While our work is implemented within the Devito framework,
the compiler concepts and ideas are general enough to apply to other projects.
Devito employs MPI to implement DMP.
The Python package {\em mpi4py} \cite{mpi4py-cite} is utilized for Python-level message-passing.
At the C-level, explicit MPI-based halo exchanges are incorporated in the generated C code.
DMP is designed to facilitate a seamless transition for users, allowing them to run at scale with zero changes to their code.
Users are expected to execute the MPI launch command, specifying the desired number of ranks and other possible arguments:
\emph{mpirun, mpiexec, srun}, e.g.: \emph{mpirun -np $<$num\_processes$>$ [options] $<$executable$>$ [arguments]}.
While some pre- and post-processing may be rank-specific (e.g., plotting on a given MPI rank only),
we anticipate that future Devito releases will further streamline these processes by providing enhanced support APIs.
Next, the features implemented to automatically generate C-code leveraging DMP are detailed.
The Devito DSL example in~\autoref{lst:dsl} will be referenced to exemplify these features.
Readers can get a more comprehensive tutorial experience in available Jupyter notebooks \cite{devito48-mpi}.

\paragraph{Domain decomposition}\label{par:domain_decomposition}
Devito leverages the MPI Cartesian topology abstraction to logically partition a grid to the available MPI ranks.
Domain decomposition occurs during the creation of the \Grid~object (refer to~\autoref{lst:dsl}, line~\ref{lst:trivial-grid}).
While a default strategy is employed, users have the flexibility to customize domain decomposition by passing the ``topology'' argument to a
\Grid, as in Grid(\dots, topology=(\dots)).
Examples of 3D custom decompositions are illustrated in \autoref{fig:custom_decomposition}.

\begin{figure}[!htbp]
\centering
\begin{subfigure}[b]{0.15\textwidth}
\includegraphics[width=\textwidth]{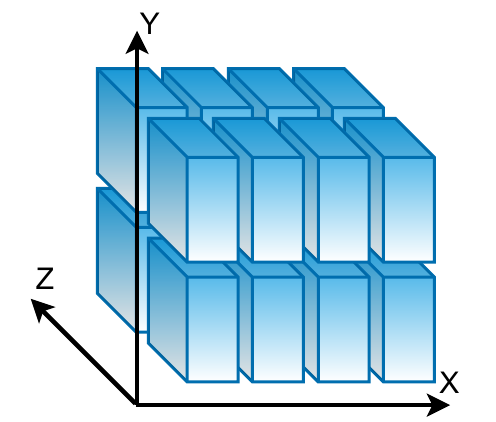}
\caption{topology=(4,2,2)}
\label{fig:custom_4_2_2}
\end{subfigure}
\begin{subfigure}[b]{0.15\textwidth}
\includegraphics[width=\textwidth]{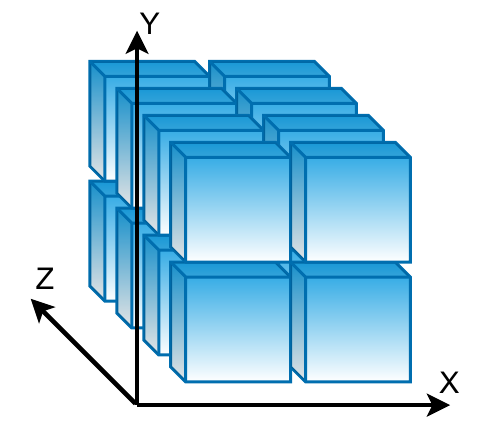}
\caption{topology=(2,2,4)}
\label{fig:custom_2_2_4}
\end{subfigure}
\begin{subfigure}[b]{0.15\textwidth}
\includegraphics[width=\textwidth]{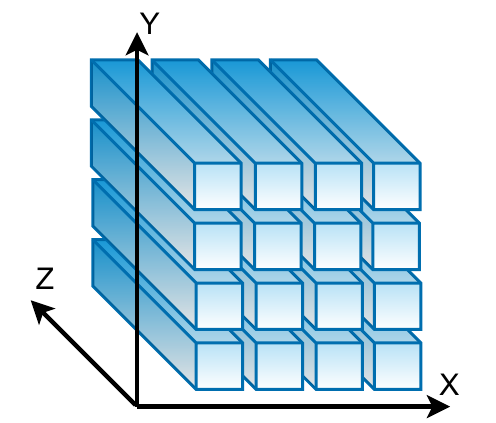}
\caption{topology=(4,4,1)}
\label{fig:custom_4_4_1}
\end{subfigure}
\caption{Users can tailor the domain decomposition using various configurations,
 such as (4,2,2), (2,2,4), and (4,4,1), each suitable for 16 MPI ranks.}
\label{fig:custom_decomposition}
\end{figure}

\paragraph{Data access (read/write)}\label{par:data_access}
Devito relies on NumPy arrays for data access.
To seamlessly integrate with MPI, NumPy arrays have been extended via subclasses,
enabling data distribution to processes according to the domain decomposition.
The data is physically distributed, but from the user's perspective, the array is, logically, a centralized entity.
Data decomposition is abstracted away from the user, providing a 
transparent interaction with data as if using standard NumPy arrays.
Users can employ various indexing schemes, including basic and slicing operations.
Underneath, robust global-to-local index conversion routines are employed,
ensuring that read and write accesses are directed to the relevant subset of ranks.
The code snippet in~\autoref{lst:dsl} is an illustrative example.
For instance, a slicing operation is carried out in line \ref{lst:trivial-slice}.
For a decomposition across 4 ranks, the resulting data is shown in~\autoref{lst:dense-data-0}.

\begin{python}[label=lst:dense-data-0, numbers=none,
    caption={Immediately after executing line~\ref{lst:trivial-slice} of Listing~\ref{lst:dsl}.
    Each MPI rank has converted the global, user-provided slice into a local slice, then used to perform the \emph{write}.
    The 0s are unaffected entries. ({\tt u.data} is allocated and initialized to 0 the first time it is accessed).}]
[stdout:0]      [stdout:1] 
[[0.00 0.00]    [[0.00 0.00]
 [0.00 1.00]]    [1.00 0.00]]

[stdout:2]      [stdout:3] 
[[0.00 1.00]    [[1.00 0.00]
 [0.00 0.00]]    [0.00 0.00]]
\end{python}

After constructing (\autoref{lst:dsl}:line~\ref{lst:trivial-op}) and running (\autoref{lst:dsl}:line~\ref{lst:trivial-end}) an \Operator,
the data is updated according to the standard 3d7pt Jacobi Laplacian stencil kernel.
\autoref{lst:dense-data-1} shows the resulting, rank-local view.

\begin{python}[label=lst:dense-data-1, numbers=none,
 caption={Output of {\tt u.data} after a single application of the \Operator given in Listing~\ref{lst:dsl}, line~\ref{lst:trivial-end},
 to the distributed data illustrated in Listing~\ref{lst:dense-data-0}.
 Entries are updated according to the standard 3d7pt Jacobi Laplacian stencil kernel.
 }]
[stdout:0]      [stdout:1] 
[[0.50 -0.25]   [[-0.25 0.50]
 [-0.25 0.50]]   [0.50 -0.25]]

[stdout:2]      [stdout:3] 
[[-0.25 0.50]   [[0.50 -0.25]
 [0.50 -0.25]]   [-0.25 0.50]]
\end{python}

\paragraph{Sparse data}\label{par:sparse_data}

\SparseFunction s represent a sparse set of points that may not align inherently with the points of the computational FD grid.
This abstraction facilitates various operations such as injection and interpolation to structured grid point positions \cite{bisbas2021}.
While this DSL object is not shown in the example provided, it plays a vital role throughout the evaluated benchmarks.
A sparse point could lie within a grid quadrant and, therefore, has suitable coordinates relevant to the physical domain of the problem.
Before executing an \Operator with DMP enabled, the coordinates of a specific point, $P$, are assigned to their respective owner MPI process.
This assignment corresponds to the rank-local data, determined by the domain decomposition strategy.
To illustrate this distribution process, consider \autoref{fig:sparse_global_local}.
In this representation, the sparse point $C$ is situated within an area shared by all four ranks.
Points $B$ and $D$ are shared by two ranks, and point $A$ is exclusively owned by rank 0.
This strategic distribution ensures that each MPI process works with the data relevant to its computational domain.

\begin{figure}[!htb]
        \centering
        \includegraphics[width=0.42\textwidth]{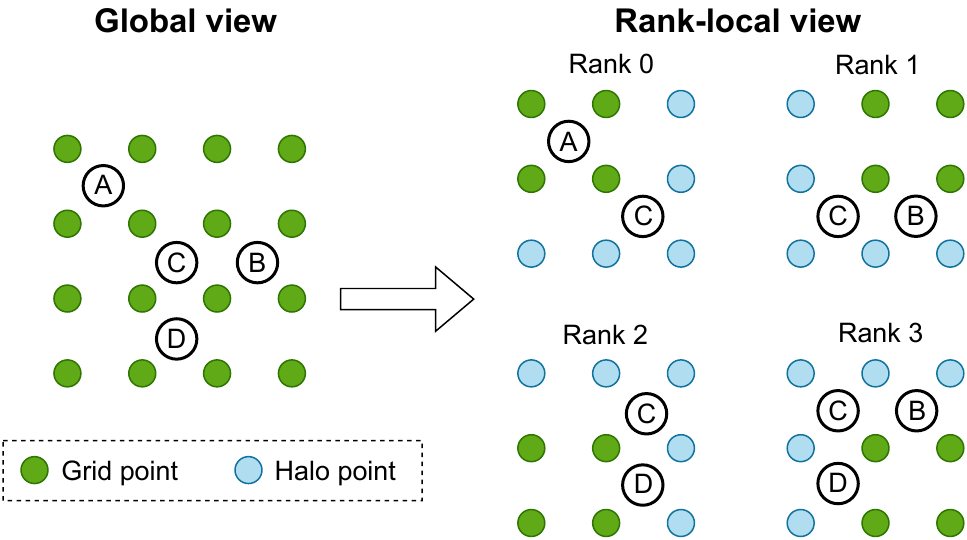}
        \caption{The compiler analyzes data dependencies to schedule the ownership of non-aligned sparse points.
 Points at shared boundaries are scheduled to the respective involved ranks.}
        \label{fig:sparse_global_local}
\end{figure}

\paragraph{Access alignment}\label{par:data_alignment}
Devito's \Function{s} and \TimeFunction{s} are structured into three distinct data regions:
\data, \halo, \padding. The \data region aligns with the underlying \Grid. For instance, in Listing~\ref{lst:dsl}, the \TimeFunction
{\tt\ u} has a domain region with a shape of $4\times4$.
This region represents the area where an \Operator is allowed to write to.

Surrounding the \data region are ghost cell regions--first \halo and then \padding.
The primary purpose of \padding is to optimize performance, offering advantages such as improved data alignment.
In contrast, \halo regions serve two essential functions: (i) in a non-distributed run, \halo regions are read-only for an \Operator when iterating in
the proximity of the domain boundary; (ii) in a distributed run, \halo regions
contain the necessary data that is exchanged between neighboring processes, a prerequisite to updating local grid points.
Given the presence of \halo and \padding, array accesses necessitate a shift by a specific offset.
Therefore, ensuring access alignment is critical to index data accesses accurately.

Hence, the compiler's first task is to ensure that all user-provided equations
align with the \data region, regardless of the presence of the ghost regions.
For example, in Listing~\ref{lst:dsl}, at line~\ref{lst:trivial-op},
the generated stencil includes accesses to {\tt u[t,x,y]}.
Assuming {\tt u} has an SDO of 2, it has, by default, a \halo of size 2 along dimensions x and y.
Assume there is no \padding, then the compiler, during the {\it Equations lowering} stage (refer to~\autoref{fig:devito-outline}),
transforms this access into {\tt u[t,x+2,y+2]}.
Consequently, with a grid shape of $4\times4$ and when iterating within the user's intended space $x, y: [0, 3]\times[0, 3]$, aligned array accesses within the \data region are ensured.

\paragraph{Data regions}\label{par:data_regions}
To facilitate the compiler's reasoning about rank-local data regions,
we use aliases for the \data regions to be computed or/and exchanged.
These aliases, schematically illustrated in \autoref{fig:dataregions}, include:
HALO for the ghost cell region, CORE for the points in the core,
that do not require reading from HALO, OWNED, for the points that require reading from the HALO,
DOMAIN, comprising of the CORE and OWNED, the actual computational domain to write, and finally
FULL, including CORE, OWNED and HALO.

\begin{figure}[!htb]
        \centering
        \includegraphics[width=0.35\textwidth]{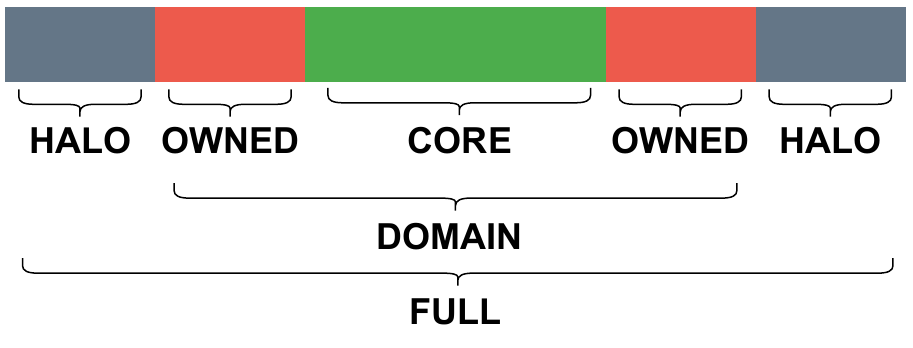}
        \caption{Aliases for data regions facilitate reasoning about and modeling halo exchanges}
        \label{fig:dataregions}
\end{figure}   

\paragraph{Detecting halo exchanges}\label{par:halo-analysis}

The generation of data communications via halo exchanges follows an analysis and synthesis approach.
During the early compilation stages, specifically at the Cluster level,
the compiler analyses a Cluster's properties and accesses (reads/writes) to identify potential areas where data exchanges are required.
At this level, expressions still need to be optimized, and the data dependence analysis needed to detect halo exchanges is more straightforward than it could be at later stages of the compiler.
A halo exchange here is represented as yet another Cluster, having its own iteration space and expressions.
The iteration space hints to the compiler for the position where halo exchanges should occur before propagating to the next IR level.
The Schedule tree in Listing~\ref{lst:comms} is an example of the first abbreviated form of an IR to structure the information by placing a halo in a loop structure.

\begin{clang}[label=lst:comms,
    caption={The compiler has detected a domain decomposition of the spatial dimensions that should occur for every timestep before the stencil computation}]
|-- [Eq,Eq,...]
|-- time++
    |-- <Halo>
        |-- x++
            |-- y++
                |-- [Eq,Eq]
\end{clang}

The objects in the representation in \autoref{lst:comms} contain metadata employed at the IET level.
The IET is comprised of nodes embodying iterations/loops and mathematical expressions.
During the IET construction, we utilize metadata propagated from the objects representing halo exchanges.
These are subsequently lowered to the so-called \HaloSpots, as illustrated in \autoref{lst:comm_iet-0}, a simplified version of an IET.
Each \HaloSpot~conveys crucial information about the required exchange operation (e.g., \emph{send, recv, wait}, etc.) and data structures indicating which \Function{s} necessitate a halo exchange.
Furthermore, it specifies the points to be transferred in each dimension and where they should be inserted upon reception (e.g., at $t+1$, $t$, $t-1$, or other time-buffers).

\begin{python}[label=lst:comm_iet-0,
                caption={The IET structure contains a \HaloSpot.
                Its metadata will be used to lower it to the required exchange and finally transform it to callables.}]
<Callable Kernel>  
  <Expression r0 = 1/dt>
  <Expression r1 = 1/(h_x*h_x)>
  <Expression r2 = 1/(h_y*h_y)>
 <[affine,sequential] Iteration time...>
  <HaloSpot(u)>
  <[affine,parallel] Iteration x...>
   <[affine,parallel,vector-dim] Iteration y...>
    <Expression r3 = -2.0*u[t0,x + 2,y + 2]>
    <Expression u[t1, x + 2, y + 2] =
            dt*(r0*u[t0, x + 2, y + 2] + ...)>
\end{python}

\paragraph{Building and optimizing halo exchanges}

The optimization of exchanges occurs in a subsequent compiler pass at the IET layer, where we manipulate \HaloSpots.
This pass may drop, merge, or reposition \HaloSpots~to enhance the performance of halo exchanges.
For instance, we may drop a \HaloSpot~especially if a prior \HaloSpot~
has already performed the same update, and the data is not yet ``dirty''.
Additionally, we gather information regarding whether and how a \HaloSpot~ could be anticipated in the control flow.
This info might involve hoisting a \HaloSpot~to an earlier position in the control flow or give rise to various computation and communication patterns, elaborated in the following paragraphs.

We replace \HaloSpots~with calls to corresponding sets or standalone MPI message exchanges according to the selected computation communication pattern.
Furthermore, this is the pass we assess the potential for overlapping computation and communication.
The combination of hoisting and overlapping analysis is particularly potent, indicating whether a computation/communication overlap technique can be applied.
For an abbreviated example of the optimized IET corresponding to a standard (\emph{basic}) MPI decomposition pattern, refer to \autoref{lst:comm_iet}.

\begin{python}[label=lst:comm_iet, caption={HaloSpots have been lowered to more specialized nodes containing info on exchange operations}]
<Callable Kernel>
 <Expression r0 = 1/dt>
 <Expression r1 = 1/(h_x*h_x)>
 <Expression r2 = 1/(h_y*h_y)>
 <[affine,sequential] Iteration time...>
  <HaloUpdateList (0, 1, 0)>
   <HaloUpdateCall>
   <[affine,parallel] Iteration x...>
    <[affine,parallel,vector-dim] Iteration y...>
     <Expression r3 = -2.0*u[t0,x + 2,y + 2]>
     <Expression u[t1, x + 2, y + 2] =
            dt*(r0*u[t0, x + 2, y + 2] + ...)>
  <HaloWaitList (0, 0, 0)>
\end{python}

\paragraph{Computation/Communication patterns}\label{par:patterns}

Devito incorporates various computation and communication patterns, each
exhibiting superiority under specific conditions for a kernel, such as operational intensity,
memory footprint, the number of utilized ranks, and the characteristics of the cluster's interconnect.
In this work, we assess three patterns that consistently demonstrate enhanced performance compared to others, at least within a subset of the benchmark evaluations.
These patterns, namely \emph{basic}, \emph{diagonal}, and \emph{full}, have proven to be effective in improving the efficiency and scalability of computations.
These three scenarios are schematically represented in \autoref{fig:mpi-modes}, and a summary of their characteristics is in \autoref*{tab:summary_patterns}.
Our evaluation in \autoref{sec:evaluation} provides valuable insights into their applicability and performance characteristics across various computational scenarios.

\begin{table*}[t]
\centering
\caption{Summary of communication/computation patterns.}
\begin{tabular}{ p{1.3cm}||p{1.3cm}|p{2.8cm}|p{2.1cm}|p{1.8cm}|p{2.3cm}  }
\hline
\hline
MPI mode & Target & Communication & Message batches & \#messages (3D) & Buffer allocation \\
\hline
\hline
Basic   & CPU, GPU & Sync, No comp overlap  & Multi-step     &  6  & runtime (C/C$++$)  \\ \hline
Diagonal & CPU  & Sync, No comp overlap  & Single-step       & 26  & pre-alloc (Python) \\ \hline
Full    & CPU & ASync, comp overlap & Single-step   & 26  & pre-alloc (Python)  \\
\hline
\end{tabular}
\label{tab:summary_patterns}
\end{table*}

\begin{figure*}[!htbp]
\centering
\begin{subfigure}[b]{0.31\textwidth}
\includegraphics[width=\textwidth]{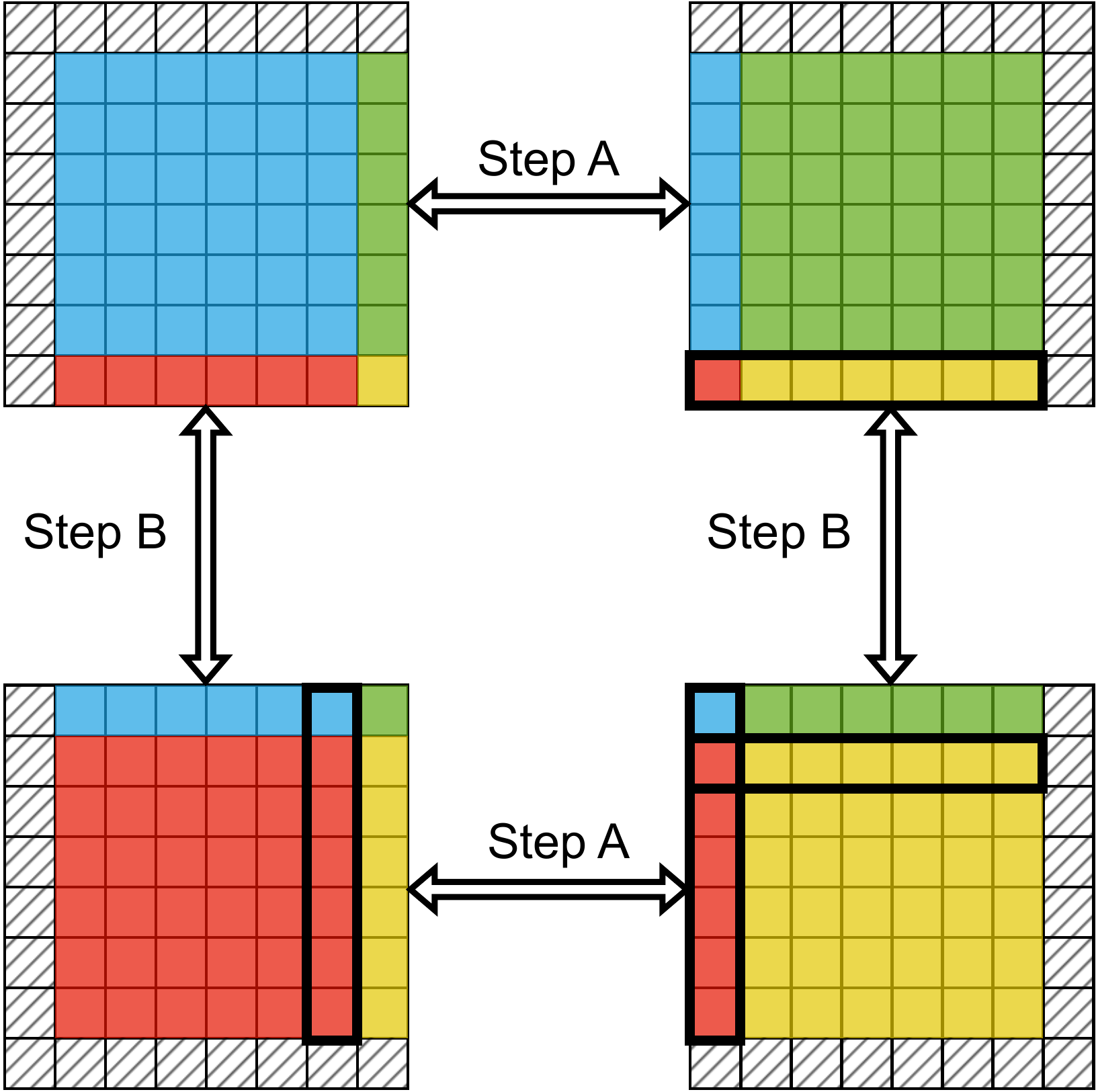}
\caption{Basic mode}
\label{fig:mpi-basic}
\end{subfigure}
\hfill
\begin{subfigure}[b]{0.31\textwidth}
\includegraphics[width=\textwidth]{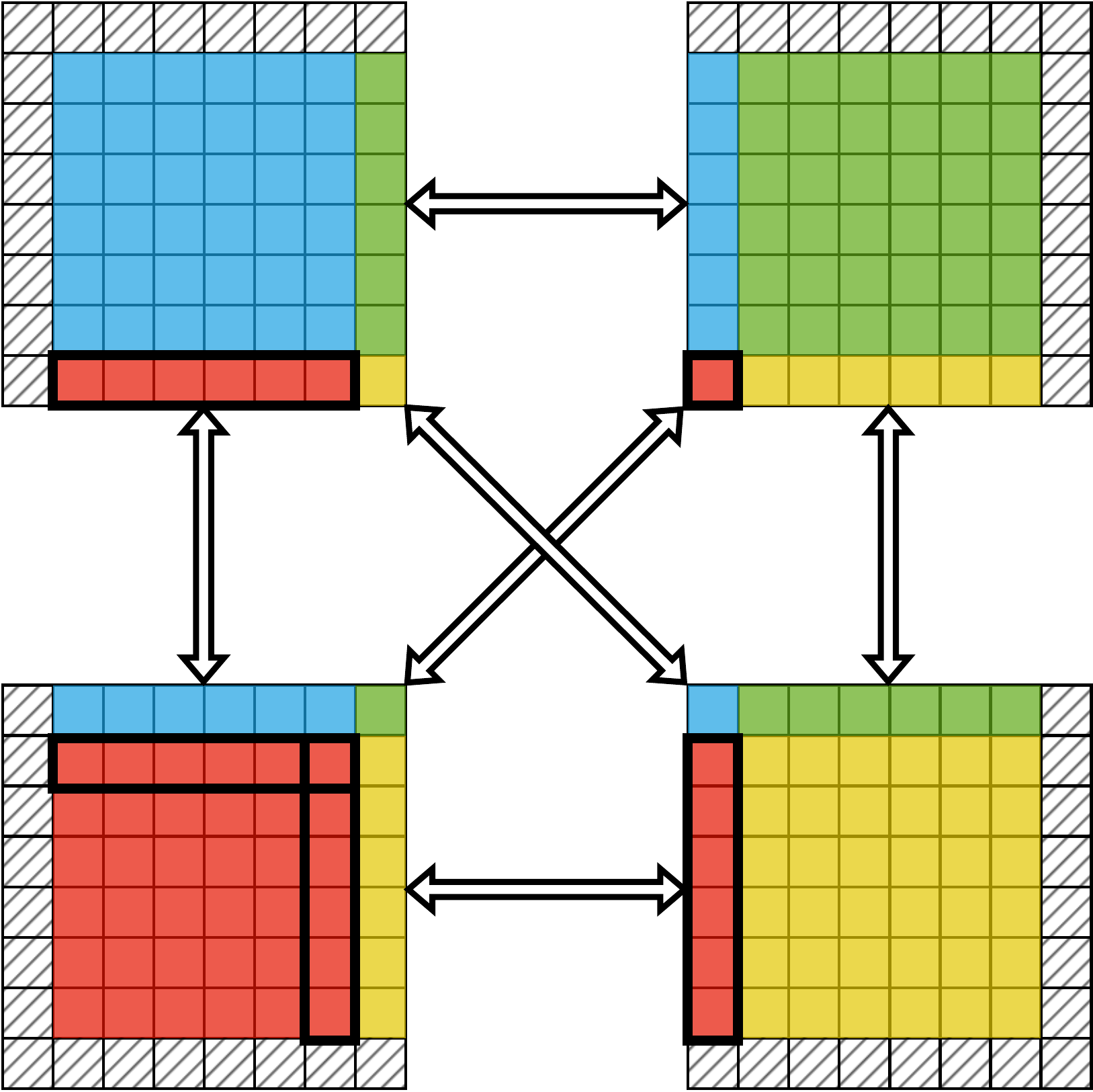}
\caption{Diagonal mode}
\label{fig:mpi-diag2}
\end{subfigure}
\hfill
\begin{subfigure}[b]{0.31\textwidth}
\includegraphics[width=\textwidth]{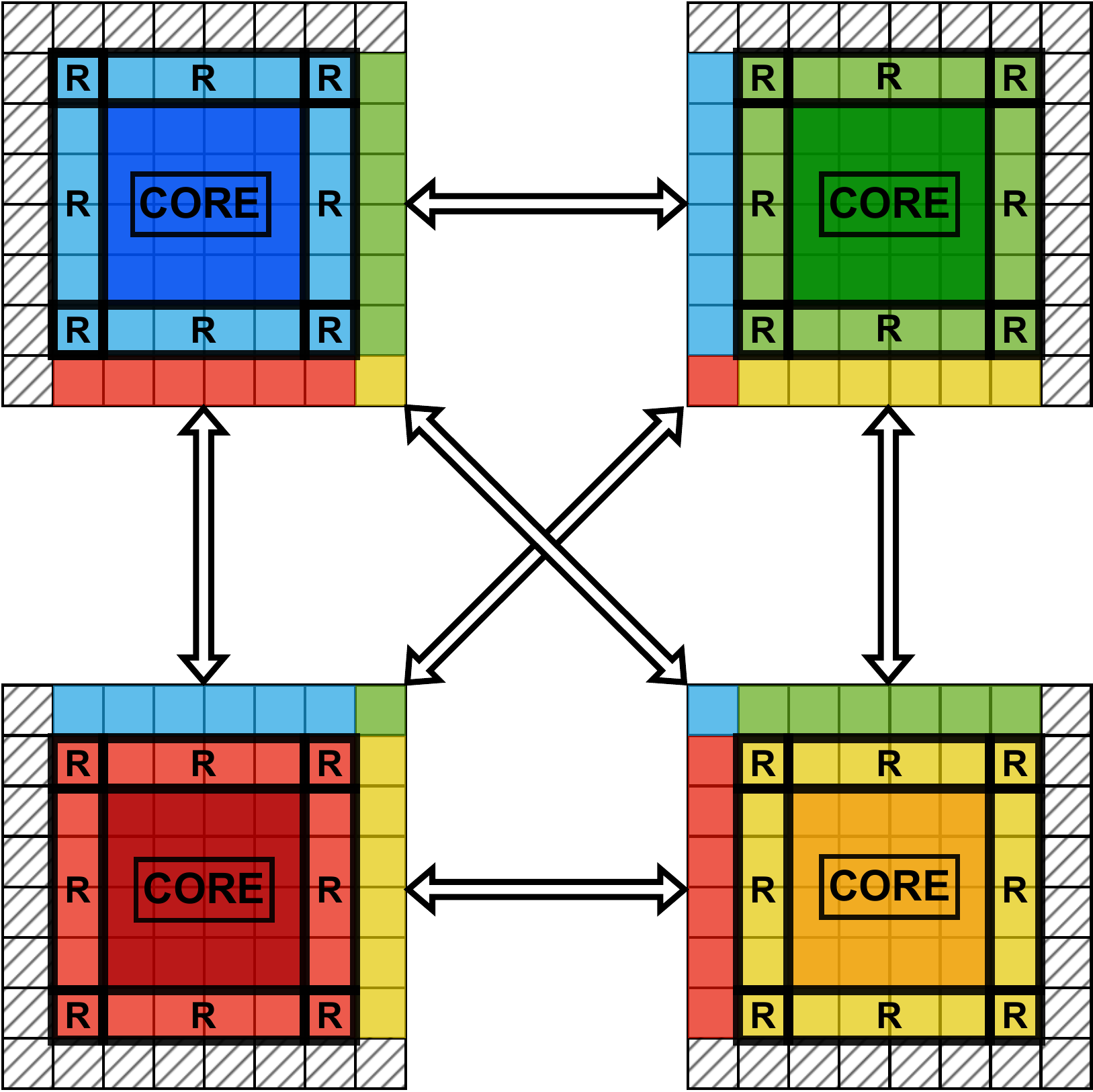}
\caption{Full mode}
\label{fig:mpi-full}
\end{subfigure}
\caption{Different colors indicate data owned and exchanged by different ranks.
        Matching colors on different ranks shows the data updated from neighbors.
        \emph{Basic} mode communicates exchanges in a multi-step synchronous manner.
        \emph{Diagonal} performs additional diagonal communications.
        \emph{Full} mode performs communication/computation overlap.
        The domain is split into CORE and REMAINDER (R) areas.
        REMAINDER areas are communicated asynchronously with the CORE computation.
        }
\label{fig:mpi-modes}
\end{figure*}

\paragraph*{Basic mode}
The \emph{basic} pattern is the simplest among the methods presented in this section and targets CPU and GPU.
This mode, illustrated in \autoref{fig:mpi-basic}, involves blocking point-to-point (P2P) data exchanges perpendicular to the 2D and 3D planes of the Cartesian topology between MPI ranks.
For example, each rank issues $4$ in 2D and $6$ communications in 3D.
While this mode benefits from fewer communications, it may encounter synchronization bottlenecks during grid updates before computing the next timestep.
This method allocates the memory needed to exchange halos in C-land before every communication, only adding negligible overhead.

\paragraph*{Diagonal mode}
Compared to the \emph{basic}, this pattern also performs diagonal data exchanges, facilitating the communication of the corner points in our domains in a single step.
This results in more communications, with $8$ in 2D and $26$ in 3D.
Although it involves more communications, they are issued in a single step, and the messages are smaller compared to \emph{basic}.
Compared to \emph{basic}, this mode slightly benefits from preallocated buffers in python-land, eliminating the need to allocate data in C-land before every communication.
The latter is why this version is not supported on GPUs since the mechanism of pre-allocating buffers on device memory still needs to be supported.

\begin{python}[breaklines=True, label=lst:basic,
        caption={\emph{basic}, one of the simplest to implement patterns, issues a multi-step communication set,
                 while \emph{diagonal} issues a single-step set, having additional diagonal exchanges.}]
# Synchronous non-blocking send/receive to update the domain
# (Multi-step for Basic)/(Single-step for Diagonal)
halo_update()  
# MPI_Wait for halos
halo_wait()
# Compute stencil on local domain
compute()  
\end{python}

\paragraph*{Communication/computation overlap (full)}
This pattern, referred to as \emph{full} in this paper, leverages communication/computation overlap.
The local-per-rank domain is logically decomposed into an inner (CORE) and an outer (OWNED/remainder) area.
In a 3D example, the remainder areas take the form of faces and vector-like areas along the decomposed dimensions.
The number of communications is the same as in the diagonal mode.
This mode benefits from overlapping two steps: halo updating and the stencil computations in the CORE area.
After this step, stencil updates are computed in the ``remainder'' areas.
In the ideal case, assuming that communication is perfectly hidden, the execution time should converge to the time needed to compute the CORE plus the time needed to compute the remainder.
An important drawback of this mode is the slower GPts/s achieved at the remainder areas.
The elements in the remainder are not contiguous; therefore, we have less efficient memory access patterns (strides) along vectorizable dimensions.
These areas have lower cache utilization and vectorization efficiency.
To prod the asynchronous progress engine, we sacrifice an OpenMP thread from the pool that progresses with computation.
This thread occasionally calls {\tt MPI$\_$Test}, ensuring halo exchanges progress effectively during computation.
By default, the compiler places a call to {\tt MPI$\_$Test} before executing a new loop tiling block.
However, the ideal frequency of calls may vary according to many parameters.
Data packing and unpacking, before and after halo exchanges, are OpenMP-threaded.
Again, this is a CPU-only mode, as the preallocation of buffers on device-memory is not yet supported for GPUs.

\begin{python}[label=lst:full, caption={\emph{full} mode overlaps the CORE computation with communicating the REMAINDER (R) areas.}]
# Asynchronous communication
halo_update()
# Compute CORE region
compute_core()
# Wait for halos
halo_wait()
# Compute the REMAINDER (R) regions
remainder()
\end{python}

\section{Performance evaluation}\label{sec:evaluation}

This section presents the systems used for performance evaluation (refer to \autoref{sec:hardware}), 
the wave propagator stencil kernels used for evaluating the efficiency of the generated code (\autoref{sec:benchmarks}),
the problem setup (\autoref{sec:problem-setup}) for these models and finally, the strong (\autoref{sec:strong-scaling}) and weak (\autoref{sec:weak-scaling}) scaling performance evaluation.

\subsection{Hardware}\label{sec:hardware}

\subsubsection{CPU cluster}\label{sec:cpu-system}
Each Archer2 compute node is equipped with a dual-socket AMD Zen2 (Rome) EPYC 7742 64-core 2.25GHz processor,
which has 128 cores and is a highly parallel scalable processor designed for multi-threaded workloads.
Every compute node has 8 NUMA regions and 16 cores per NUMA region, featuring
32kB of L1 cache per core, 512kB of L2 cache per core and 16MB L3 cache for every 4 cores.
Archer2's interconnect employs HPE Slingshot with 200Gb/s signaling with dragonfly topology.
Nodes are grouped into sets of 128, each hosting electrical links between the Network Interface Card (NIC) and switch,
with 16 switches per group and 2 NICs per node.
All-to-all connections among switches in a group are facilitated using electrical links,
while those between groups utilize optical links.
The compiler used was the Cray Clang version 11.0.4 and Cray's MPICH.
Each rank was assigned to a single NUMA region out of the 8 available on a node and every OpenMP worker on a physical thread within this
designated NUMA region.
This results in 8 MPI ranks per node and 16 OpenMP workers per rank, yielding 128 OpenMP threads per node.
We evaluate weak and strong scaling up to 128 nodes (16,384 cores).

\subsubsection{GPU cluster}\label{sec:gpu-system}

Each Tursa compute node has 2x AMD 7413 EPYC 24c processor and 4x NVIDIA Ampere A100-80 GPUs with NVLink intranode GPU interconnect and 4x200 Gbps NVIDIA Infiniband interfaces.
These GPUs have a peak FP32 performance of 19.5 TFLOPS and 80GB HBM2e.
We evaluate weak and strong scaling up to 32 nodes, a total of 128 Ampere A100-80s.
The compiler used was nvc++ 23.5-0.

\subsection{Wave propagator stencil kernels}\label{sec:benchmarks}

To evaluate the DMP schemes, we selected four models
extensively employed in geophysics applications, covering a variety of computation and communication demands.
We introduce the stencils arising from these models, with all corresponding equations
available at the Appendix \ref{sec:app:waves}.
All models for these wave propagators are open-source and available online at \cite{devito_zenodo_4.8.10}.

\subsubsection{Isotropic acoustic}\label{sec:acoustic}

The isotropic acoustic wave equation is fundamental for wave propagation in an isotropic acoustic medium.
This single-scalar partial differential equation (PDE) includes a Laplacian operator,
resulting in a Jacobian ``star'' stencil pattern.
This stencil, commonly employed as a benchmark in literature, is depicted in \autoref{fig:isostencil}.

The equations can be expressed concisely using the Devito symbolic API as demonstrated in \autoref{lst:wave-equation}.
This stencil kernel is typically characterized as memory-bound due to the standard Laplacian operation involved in the equation and has a low operational
intensity \cite{Louboutin2017a, Williams2009}.
This is a second-order in-time model, and the working set for this model consists of 5 fields.

\begin{python}[label=lst:wave-equation, caption=Symbolic definition of the wave-equation]
...
eq = m * u.dt2 - u.laplace
stencil_eq = Eq(u.forward, solve(eq, u.forward))
\end{python}

\subsubsection{Anisotropic acoustic}\label{sec:tti}

The anisotropic acoustic model, also known as TTI (Tilted Transversely Isotropic), finds extensive use
in industrial applications such as subsurface imaging, including Full Waveform Inversion (FWI) and Reverse Time Migration (RTM)
\cite{zhang2011stable, louboutin2018effects, duveneck, Alkhalifah2000AnAW, Bube2016self}.
This model is designed to capture the complexities of layered geological strata.
It is second-order accurate in time, but in contrast to the more straightforward isotropic acoustic model,
the discretized TTI equation involves a rotated anisotropic Laplacian kernel,
significantly increasing the number of floating point operations \cite{Louboutin2017a}.
The rotated Laplacian results in a stencil pattern where memory reads span three 2-dimensional planes,
as illustrated in \autoref{fig:ttistencil}.
This model needs 12 fields while standing out as the most arithmetically intensive, with the highest computation-to-communication ratio.

\begin{figure}[!htbp]
	\centering
	\begin{subfigure}[b]{0.20\textwidth}
            \centering
            \includegraphics[width=\textwidth]{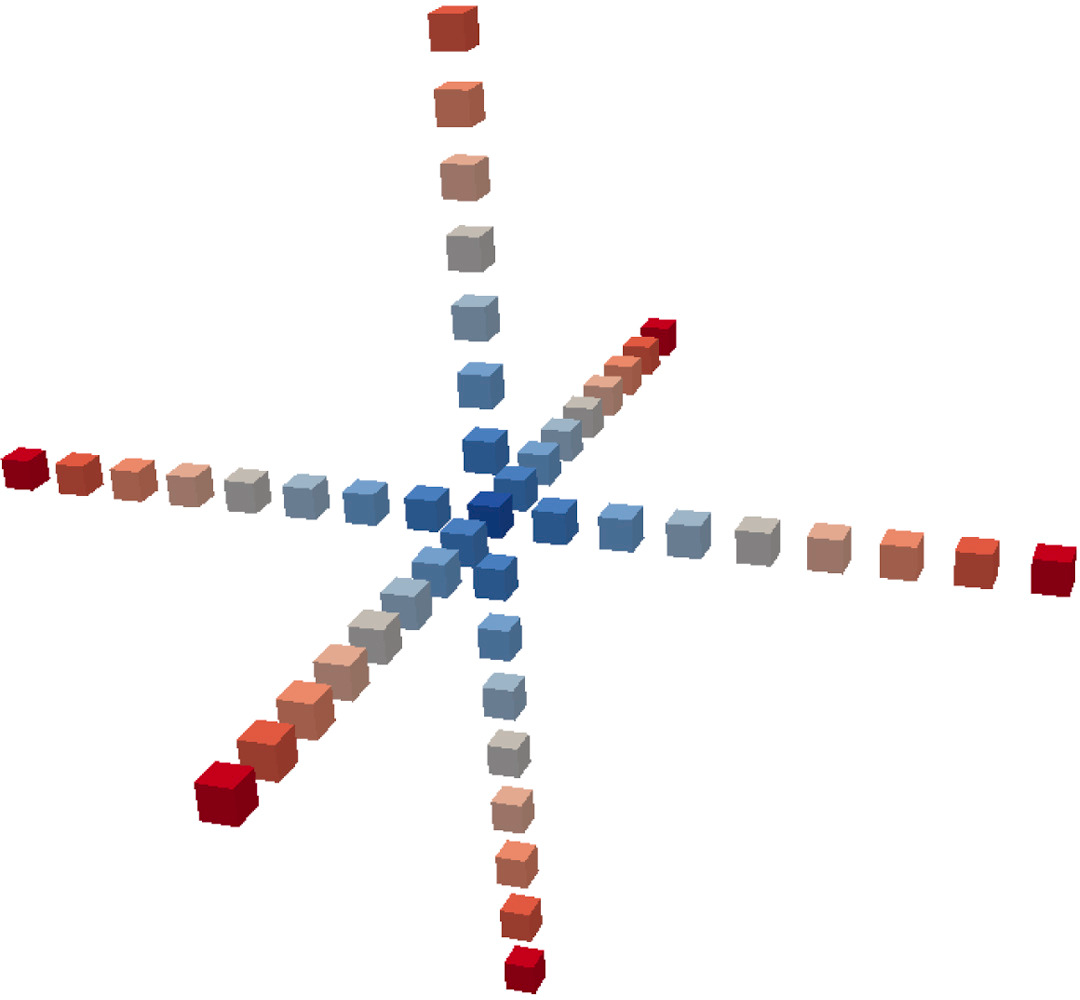}
            \caption{The iso-acoustic stencil}
            \label{fig:isostencil}
            \end{subfigure}
	\begin{subfigure}[b]{0.20\textwidth}
            \centering
            \includegraphics[width=\textwidth]{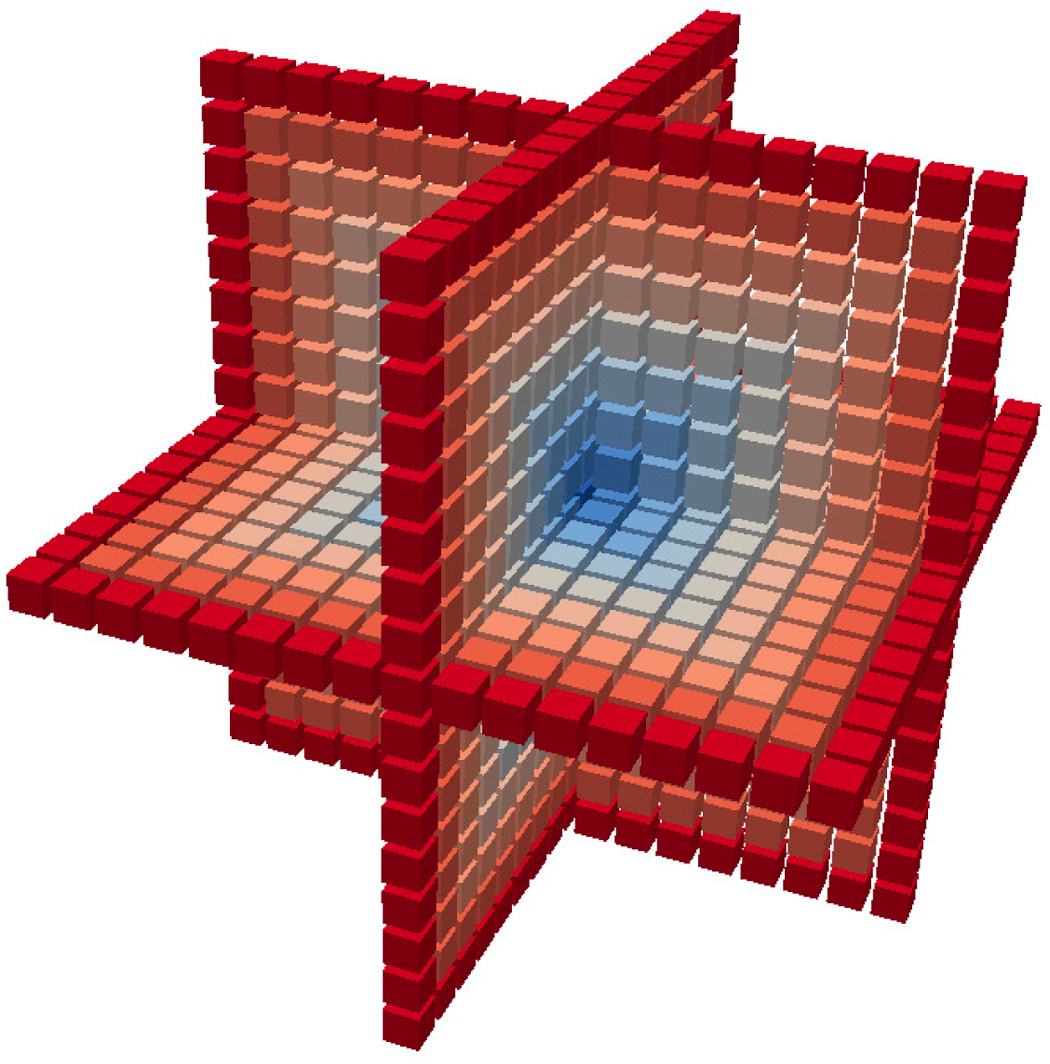}
            \caption{The TTI stencil}
            \label{fig:ttistencil}      
    \end{subfigure}
    \caption{A 16th-order (769pt) TTI stencil, 256 accesses (16*16) per plane,
        has way higher operational intensity than the isotropic acoustic star stencil (49pt).
        Figures adapted from \cite{Louboutin2017a}.       
        }
    \label{fig:stencils}
\end{figure}

\subsubsection{Isotropic elastic}\label{sec:elastic}

The isotropic elastic model encapsulates the complete physics of elastic waves,
including compressional and shear waves \cite{virieux1986p}.
Unlike the previous acoustic approximations, this equation has two notable characteristics.
Firstly, it uses a first-order discretization in time, allowing us to extend our work to a smaller range of local data dependencies over time.
Compared to second-order in time systems requiring three buffers to store dependencies, first-order systems only require one buffer.
Secondly, this equation constitutes a coupled system consisting of a vectorial and a tensorial PDE, resulting in drastically increased data
movement.
While the isotropic acoustic equation has one or two state parameters, the isotropic elastic equation involves nine parameters in the wavefield.
Therefore, we need to apply a star stencil (refer to \autoref{fig:isostencil}) nine times per timestep.
The resulting stencil is highly memory-bound.
The isotropic elastic model has a bigger working set compared to the other models (22 fields) and a high number of floating point operations even at its arithmetically optimized version.

\subsubsection{Visco-elastic}\label{sec:visco-elastic}

The visco-elastic wave propagation model, modeled upon the work introduced in~\cite{Robertson:Viscoelastic94},
has additional tensor fields and requiring a total of 15 stencils to update the necessary fields.
This model employs a first-order accurate discretization in time and is computationally very interesting,
featuring a higher memory footprint (36 fields) and peak operation intensity.
It employs a staggered grid with significantly increased communication costs compared to other models.
Moreover, it is utilized for high-fidelity modeling, closely aligning with physical models of interest to the scientific community.

\subsection{Problem Setup}\label{sec:problem-setup}

All models were evaluated for SDO 8, a commonly used simulation setup.
We used the biggest possible models fitting to the single-node available memory.
The computational domain grid for the isotropic acoustic: $1024^3$ for CPU, $1158^3$ for GPU, for elastic: $1024^3$ for CPU, $832^3$ for GPU,
for TTI: $1024^3$ for CPU, $896^3$ for GPU, and for the viscoelastic: $768^3$ for CPU, $704^3$ for GPU, with a 40-point deep absorbing boundary condition (ABC) layer, resulting in domains 80 points bigger per side alongside the read-only halo area, which depends on the SDO. Models were simulated for $512ms$.
For the isotropic and anisotropic acoustic (TTI) models, the simulation time results in $290$ timesteps, while for the elastic, $363$, and $251$ for the viscoelastic.
Source injection was modeled using a Ricker wavelet, a commonly used seismic source wavelet in geophysics and seismic exploration \cite{Gholamy2014}.

We evaluate the forward modeling achieved throughput in GPts/s for all the benchmarks.
Single-node performance was optimized and tuned through a combination of flop-reduction transformations and loop blocking autotuning (refer to \autoref{sec:devito-api-codegen}),
resulting in efficient implementations exhibiting highly competent throughput and optimized computation-to-communication ratio, as established on roofline plots in related work
\cite{devitoTOMS2020, bisbas2021} ensuring we compare against a highly competitive baseline.

\autoref{fig:roofline} shows the achieved GFlops/s for the single-node execution of kernels in a CPU/GPU integrated roofline.
For the CPU points (red), we used the Empirical Roofline toolkit \cite{ert} to get the roofline bounds and Devito's performance report
for GFlops/s and OI. GFlops/s is in accordance with well-established profiling tools like Likwid \cite{likwid, likwid2}.
The reported OI for the CPU runs was precomputed at compile-time from Devito by examining the code's abstract syntax tree (AST) to identify operations and memory accesses and compute the ratio of computation to the amount of memory traffic.
At the time of the experiments' execution, hardware counters were unavailable to obtain the operational intensity through some profiling tool at the underlying platform.
We used metrics from the NVIDIA Nsight Systems profiler for the GPU-reported OI and reported the kernel launch with the higher runtime.
Again, flop-reduced optimized kernels may exhibit a worse roofline percentage peak but achieve better throughput (GPts/s) and time to solution.
Similarly, when it comes to strong scaling, it is worth noting that unoptimized kernels would be dominated by computation time versus communication and exhibit near-perfect strong scaling efficiency but worse time to completion.
Thus, throughput (GPts/s) is strongly recommended as the primary metric to evaluate FD-stencil kernels.
The software version used for benchmarks is open source and available online at [\href{https://zenodo.org/records/14605097}{zenodo.org/records/14605097}].

\begin{figure}[!htbp]
	\centering
        \includegraphics[width=0.48\textwidth]{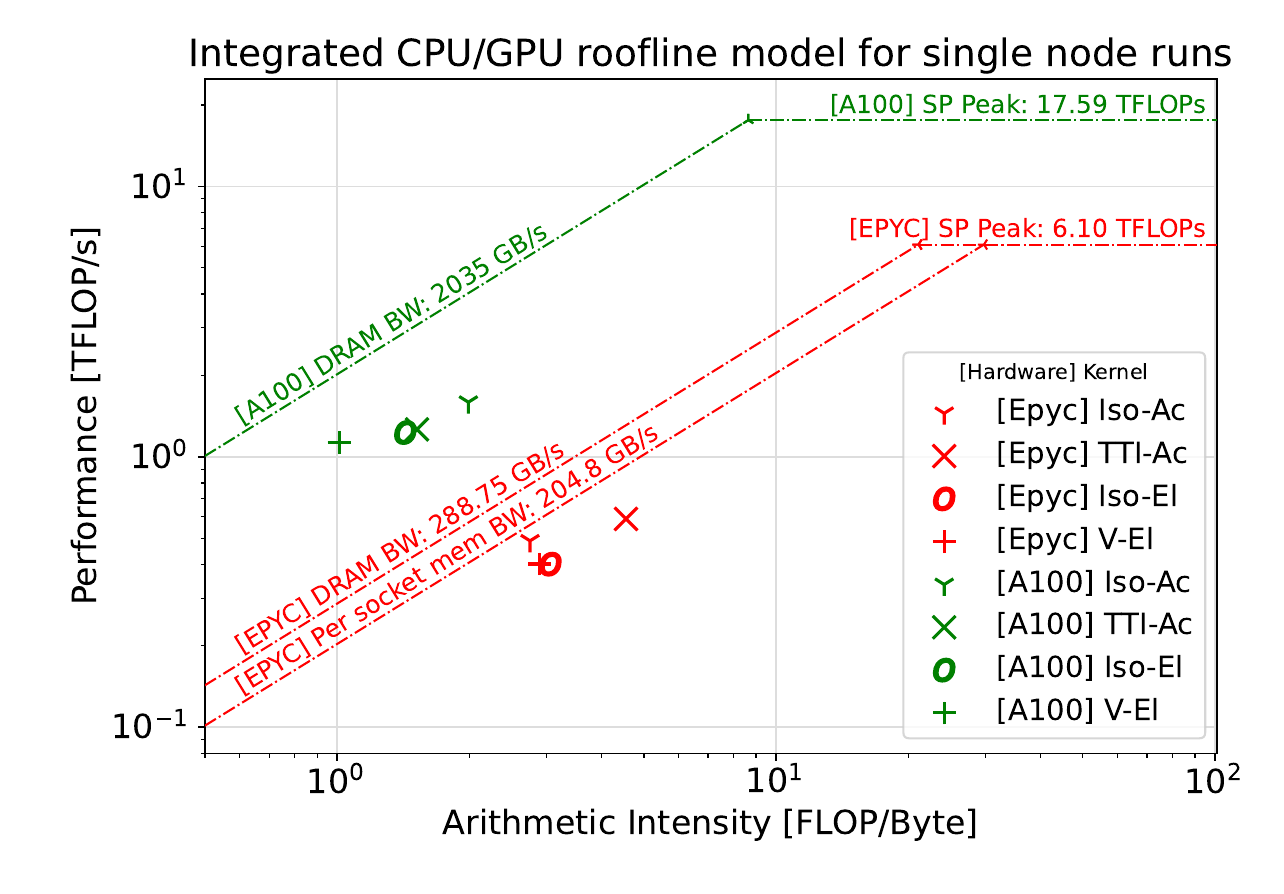}
        \caption{Flop optimized kernels are mainly DRAM BW bound}
        \label{fig:roofline}
\end{figure}

It is important to note that an OOM system issue was encountered in one particular experiment
\footnote{\href{https://docs.archer2.ac.uk/faq/}{OOM error on ARCHER2}},
when weak scaling the viscoelastic model on 128 nodes, having a big memory footprint per core.
For this experiment, we adjusted the MPI/OpenMP balance configuration to 4 ranks and 32 OpenMP cores,
scaling to the same number of cores.
The results for experiments that required further reducing the number of MPI ranks were left empty.
For the \emph{full} mode, we present results obtained after manual tuning for custom
topology in the decomposition of the computational grid (refer to \autoref{sec:dmp}).

\subsection{Strong scaling performance evaluation}\label{sec:strong-scaling}

Ideal line numbers show the percentage of the achieved ideal efficiency using
$(GPts/s$ for\ $N$ nodes)$/$($(GPts/s$ for $1\ node)*N$) for the CPU runs while using
$(GPts/s$ for\ $N$ GPUs)$/$($(GPts/s$ for $1\ GPU)*N$) for the GPU runs.

The isotropic acoustic stencil kernel is a relatively cheap memory-bound kernel with low communication requirements.
\autoref{fig:ac-stencil-scaling} shows that this kernel's strong scaling benefits more from the \emph{basic} and \emph{diagonal} modes.
\emph{full} mode's remainder computations cost more than the cheap issued communications.
However, efficiency is still on par, around 68\% of the ideal.
While GPUs are superior to CPUs in low number of nodes, they tend to be less efficient as we scale and the problem to solve becomes smaller.
While computation is way faster in GPUs, the communication overhead is comparable to CPUs, and the problem size is not big enough to hide this overhead. Using 128 A100-80 GPUs, the throughput is around 1470 GPts/s (37\% of the ideal), while using 128 EPYC 7742 CPUs, the throughput is around 1050 GPts/s (64\% of the ideal).

\begin{figure}[!htbp]
	\centering
	\begin{subfigure}[b]{0.24\textwidth}
        \includegraphics[width=\textwidth]{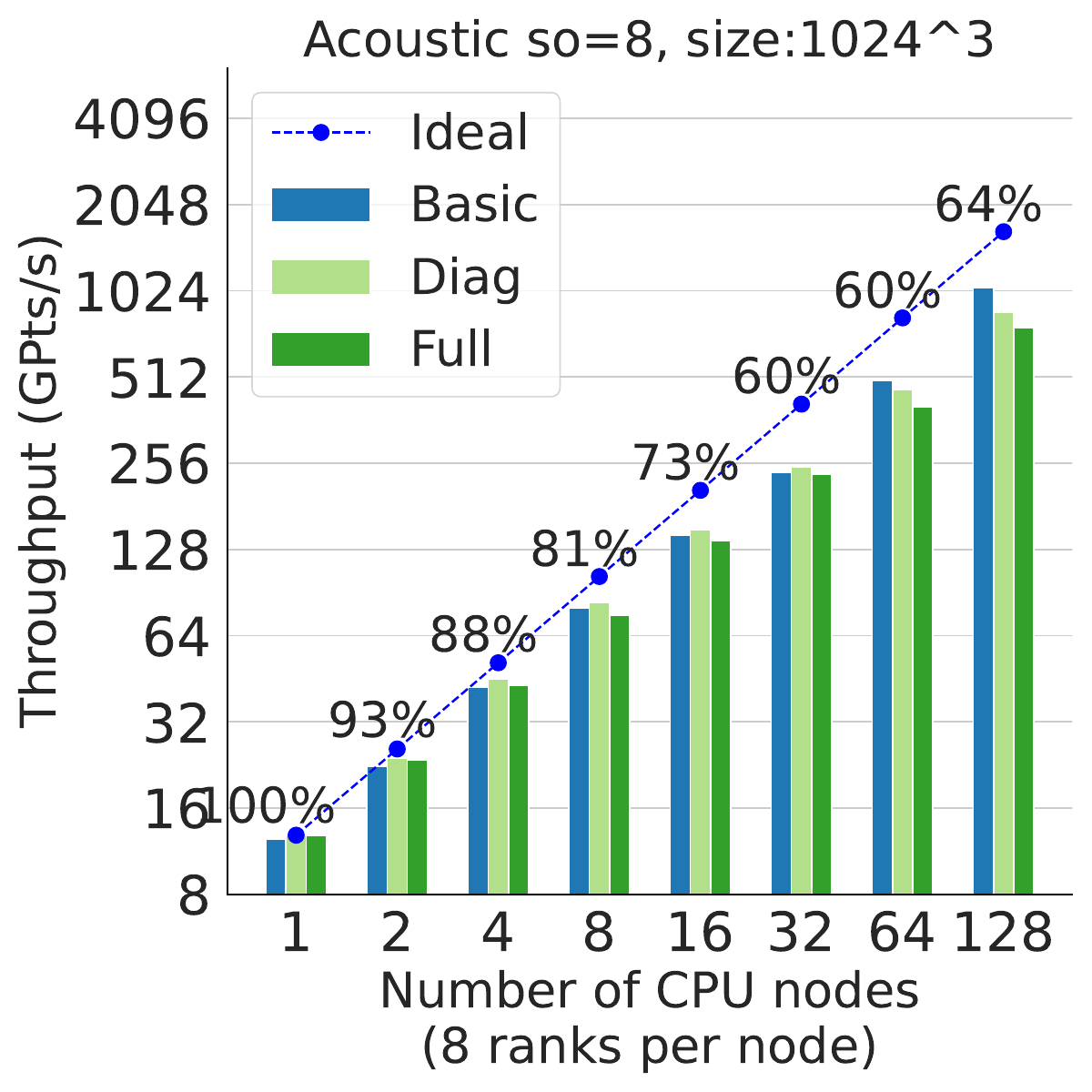}
        \caption{CPU scaling for Acoustic}
        \label{fig:ac_so08_platform}
        \end{subfigure}
	\begin{subfigure}[b]{0.24\textwidth}
        \includegraphics[width=\textwidth]{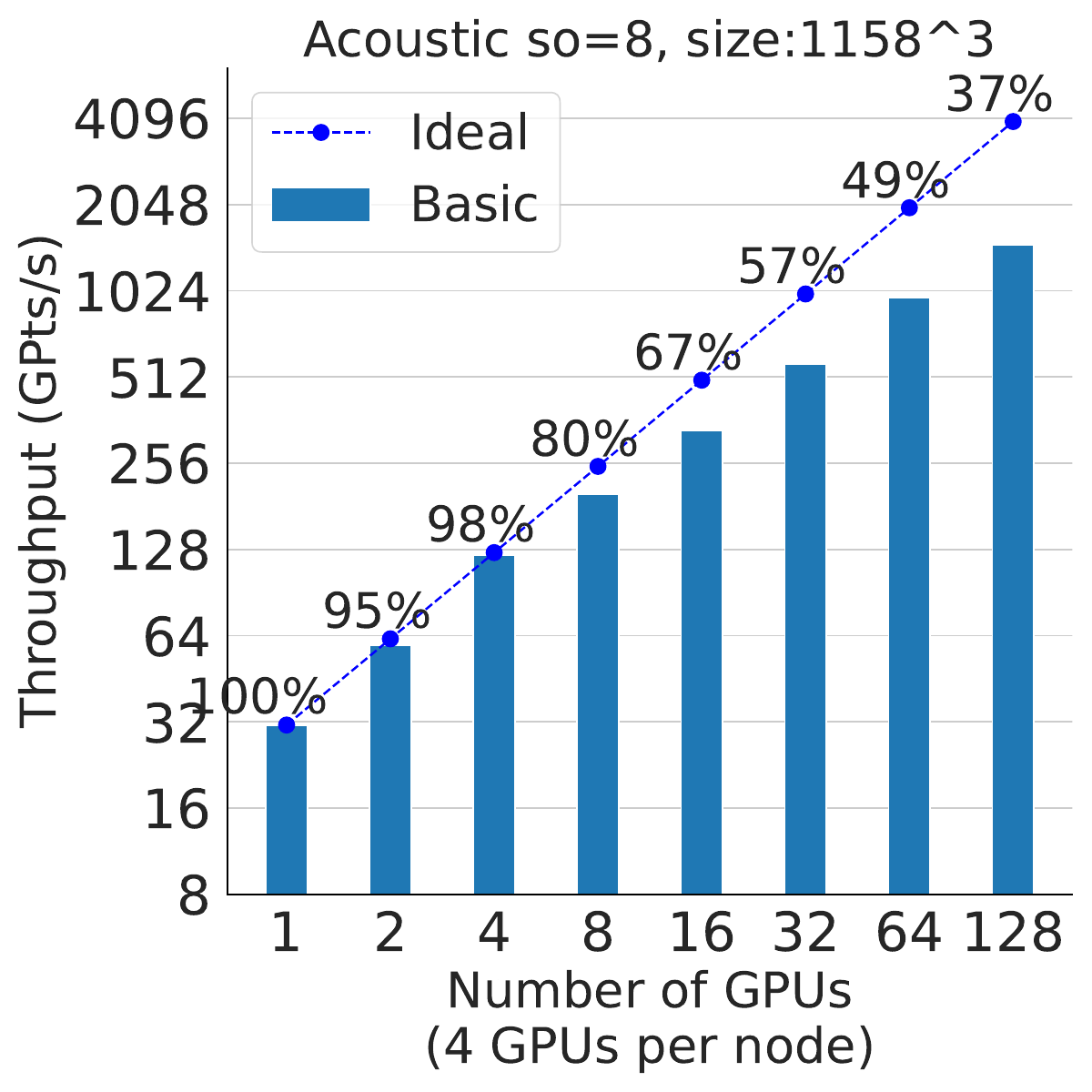}
        \caption{GPU scaling for Acoustic}
        \label{fig:ac_so8_gpu_platform}
	\end{subfigure}
        \caption{Strong scaling for the \textbf{isotropic acoustic} kernel.}
    \label{fig:ac-stencil-scaling}
\end{figure}

The elastic model's computation cost is around 5 times higher than iso-acoustic,
and its communication demands are almost 4.4 times higher, as 22 fields are needed instead of 5.
\autoref{fig:el-stencil-scaling} shows that elastic is benefiting more from \emph{diagonal}
compared to iso-acoustic.
\emph{full} mode shows improved throughput for a number of experiments,
but it tends to be less efficient at scale than the acoustic.
This trend is also present in GPUs, as we observe high communication demands while computing less and updating fewer points.
The exhibited efficiency is around 46\% of the ideal for CPUs, $\approx$106 GPts/s, while we achieve $\approx$164 GPts/s on GPUs with 25\% efficiency.

\begin{figure}[!htbp]
        \centering
        \begin{subfigure}[b]{0.24\textwidth}
        \includegraphics[width=\textwidth]{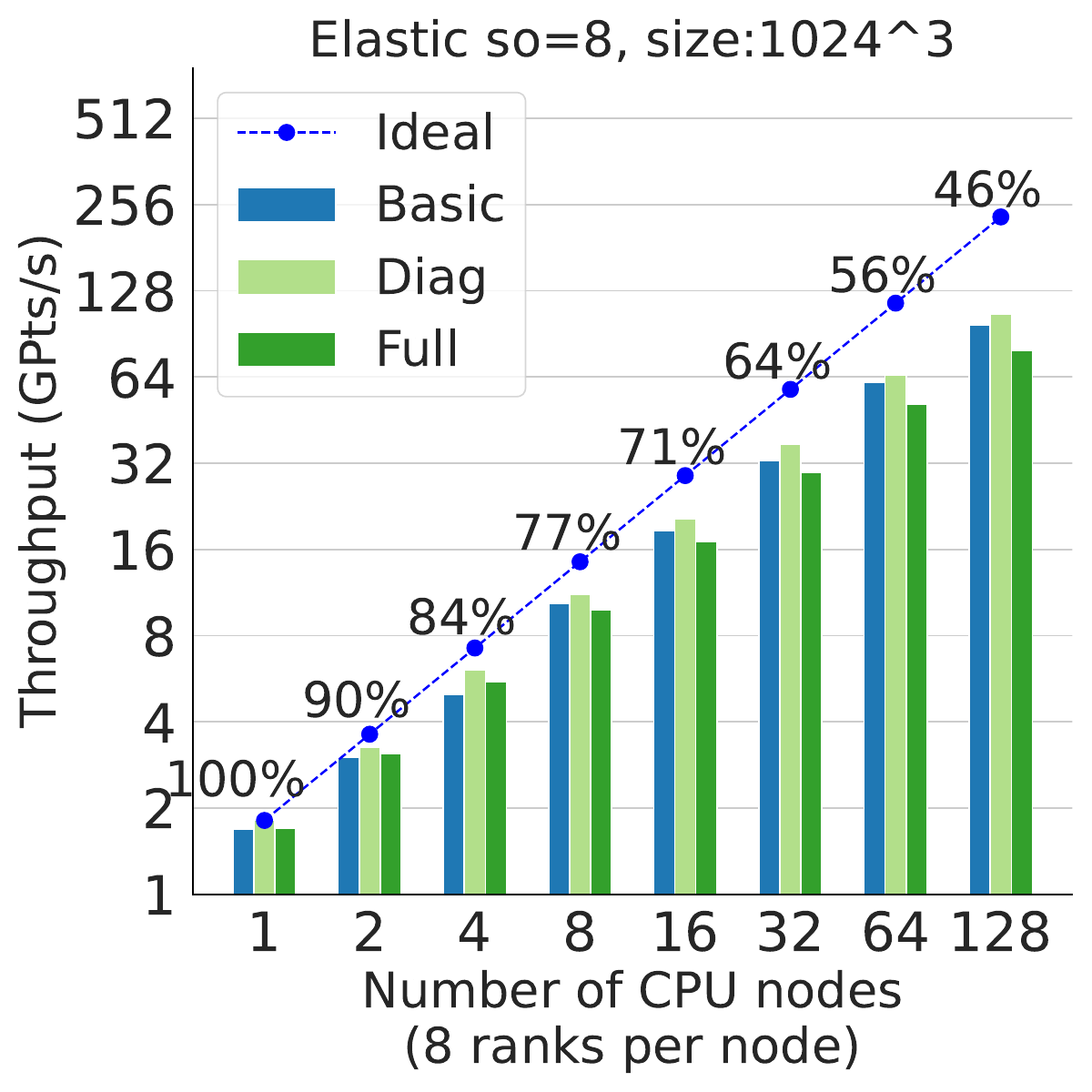}
        \caption{CPU scaling for Elastic}
        \label{fig:el_so08_platform}
        \end{subfigure}
        \begin{subfigure}[b]{0.24\textwidth}
        \includegraphics[width=\textwidth]{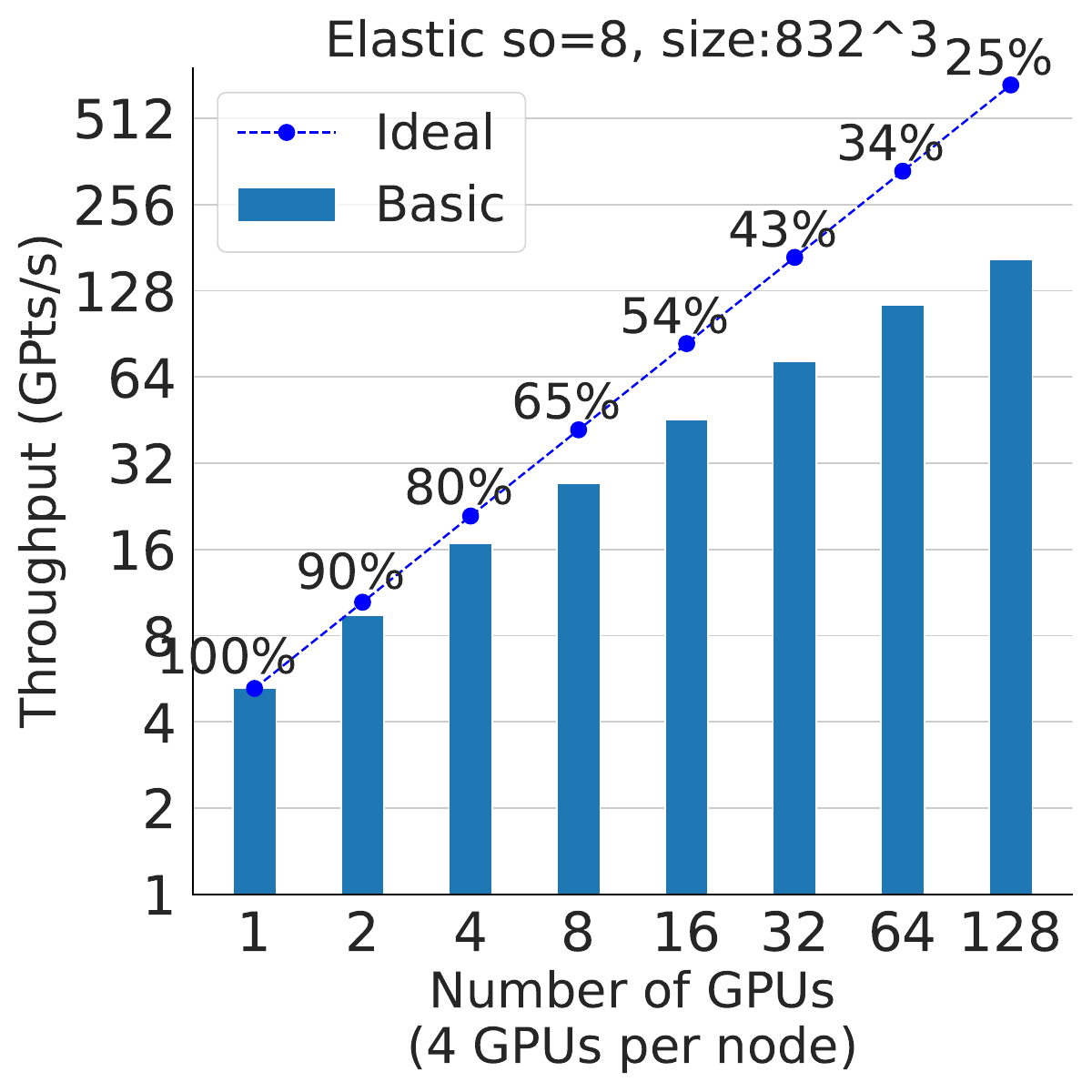}
        \caption{GPU scaling for Elastic}
        \label{fig:el_so8_gpu_platform}
        \end{subfigure}
        \caption{Strong scaling for the \textbf{isotropic elastic} kernel.}
    \label{fig:el-stencil-scaling}
\end{figure}

The TTI kernel is by far the computationally more intensive kernel and has the highest computation-to-communication ratio,
therefore exhibiting the highest scaling efficiency, as shown in \autoref{fig:tti-stencil-scaling}.
Kernels with these characteristics clearly benefit from not sacrificing computational resources over faster
communications. The remainders' computation cost is higher than the communication hidden away.
Consequently, there are better candidates than \emph {full} mode for TTI kernels.
\emph{diagonal} is the best performing mode for most of the cases, with \emph{basic} following.
It is important to note that benefits from the enhanced locality for this kernel outperform the time spent in communications, resulting in near-perfect scaling up to 32 and 64 nodes.
This computationally heavy kernel is an ideal candidate for GPU strong scaling, achieving 460 GPts/s (42\% efficiency) and around 314 GPts/s (69\% efficiency) on CPUs.

\begin{figure}[!htbp]
	\centering
	\begin{subfigure}[b]{0.24\textwidth}
        \includegraphics[width=\textwidth]{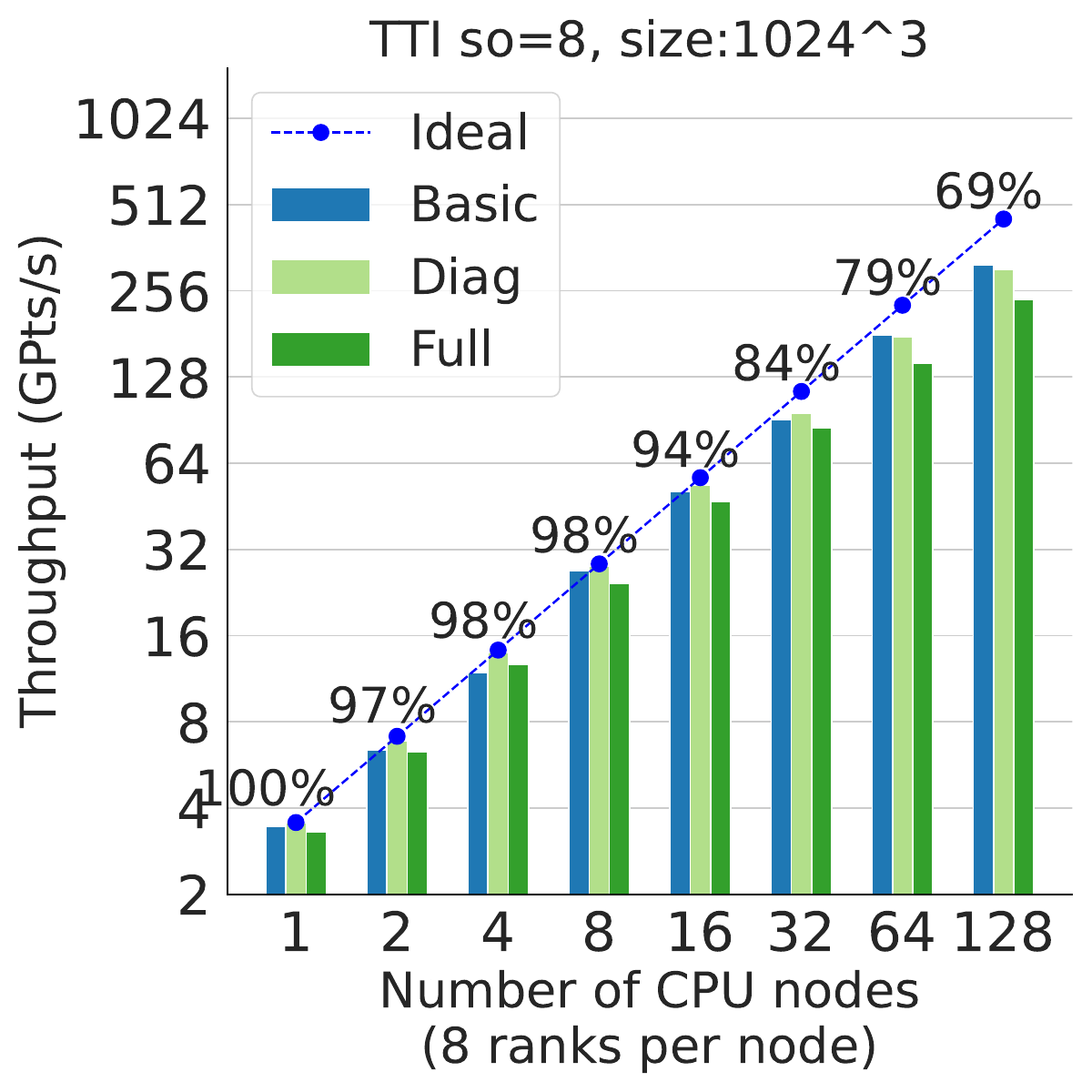}
        \caption{CPU scaling for TTI}
        \label{fig:tti_so08_platform}
        \end{subfigure}
	\begin{subfigure}[b]{0.24\textwidth}
        \includegraphics[width=\textwidth]{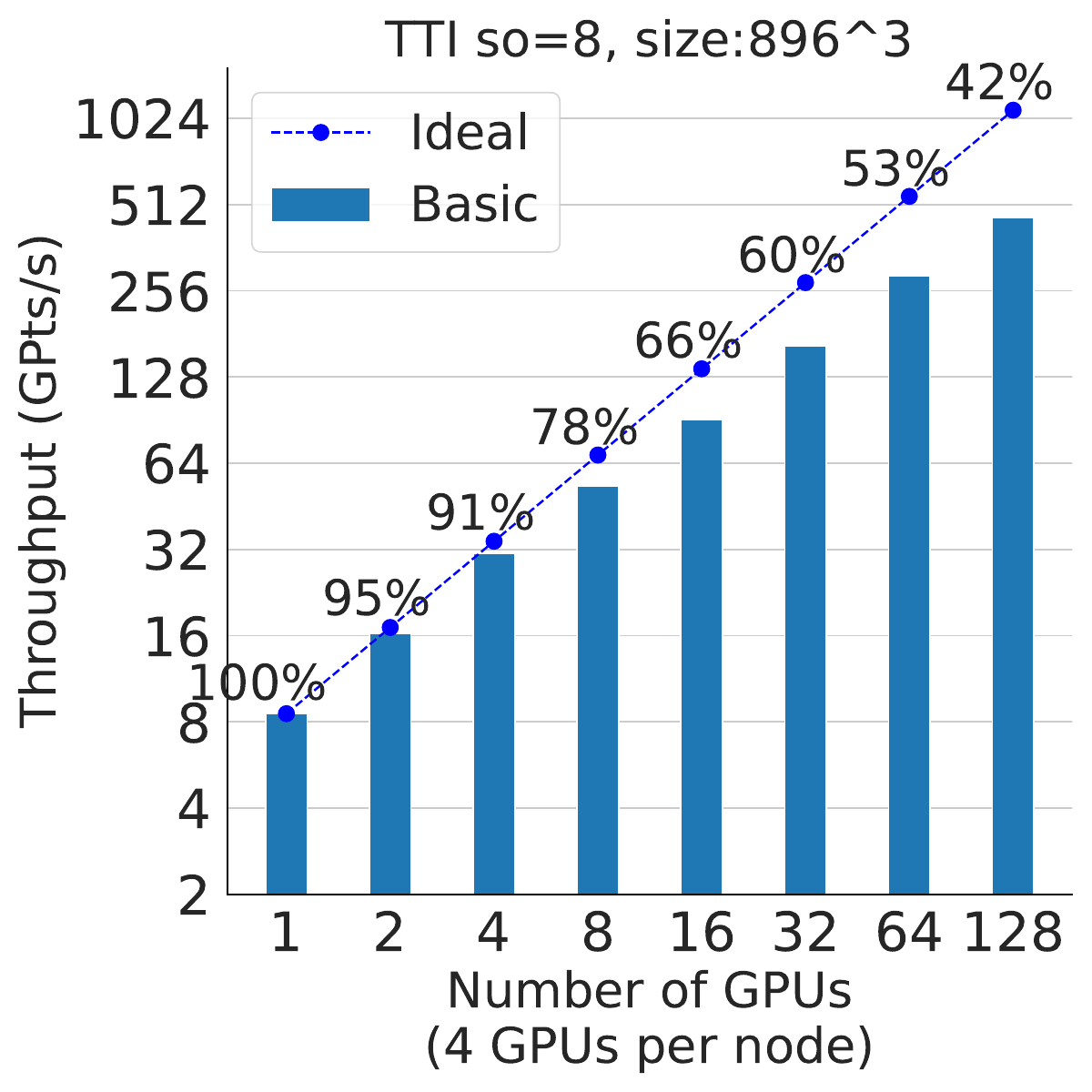}  
        \caption{GPU scaling for TTI}
        \label{fig:tti_so08_gpu_platform}
	\end{subfigure}
        \caption{Strong scaling for the \textbf{TTI} kernel.}
    \label{fig:tti-stencil-scaling}
\end{figure}

\autoref{fig:vel-stencil-scaling} shows that the viscoelastic stencil kernel benefits more from \emph{basic} and \emph{diagonal} modes.
This kernel has the highest memory footprint and, therefore, the most expensive communication cost.
The computation cost is similar to elastic, but the communication cost is around 65\% higher (36 vs. 22 fields).
The evaluation shows that this extra cost is handled slightly better from the \emph{basic} mode.
The exhibited efficiency is 46\% on CPUs ($\approx$73 GPts/s) and 30\% on GPUs ($\approx$107 GPts/s).
It is evident in all GPU experiments that there is a decrease in efficiency after 4 GPUs, owing to using not only the intra-node NVLink but also the Infiniband network.

\begin{figure}[!htbp]
	\centering
	\begin{subfigure}[b]{0.24\textwidth}
        \includegraphics[width=\textwidth]{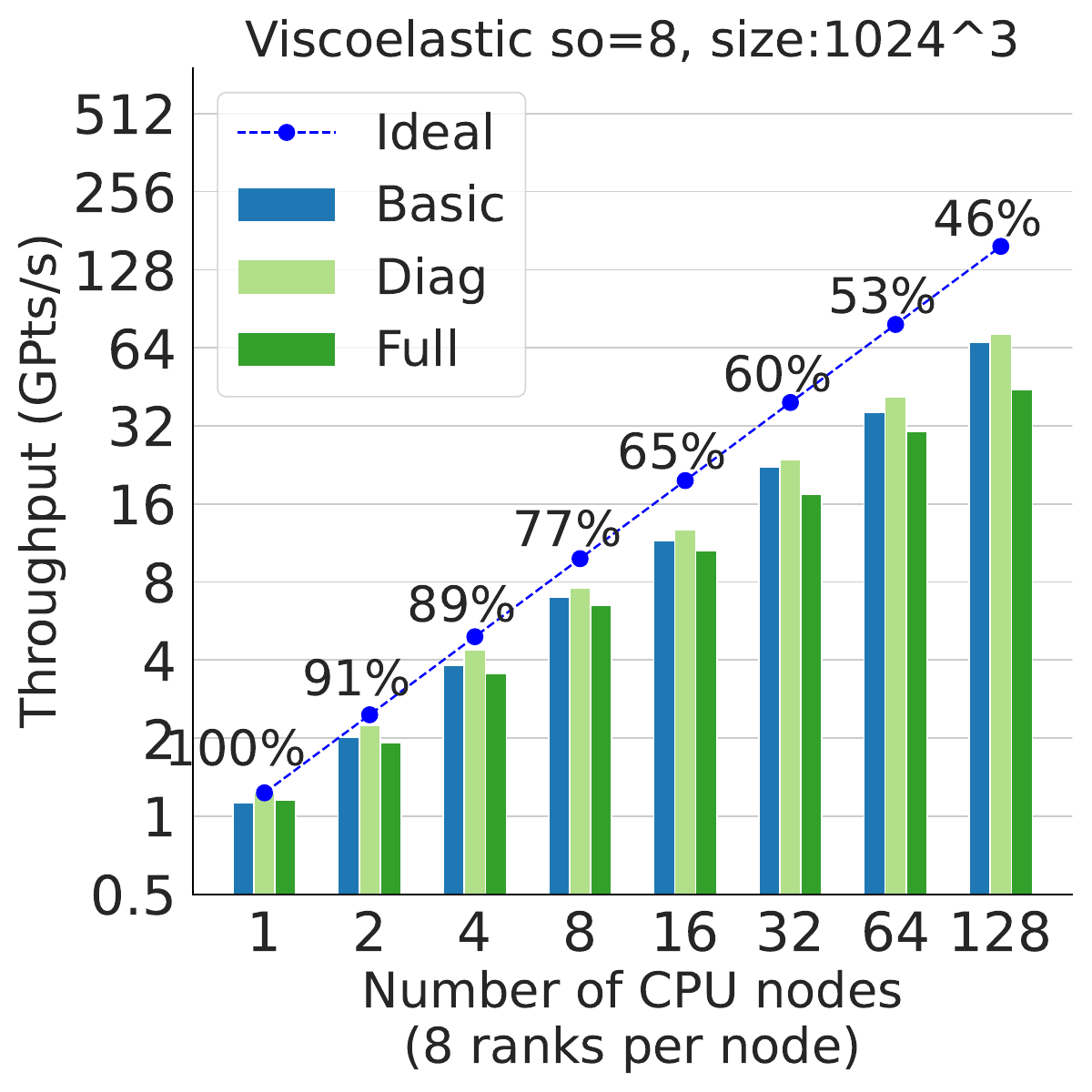}
        \caption{CPU scaling for Viscoelastic}
        \label{fig:vel_so08_platform}
        \end{subfigure}
	\begin{subfigure}[b]{0.24\textwidth}
        \includegraphics[width=\textwidth]{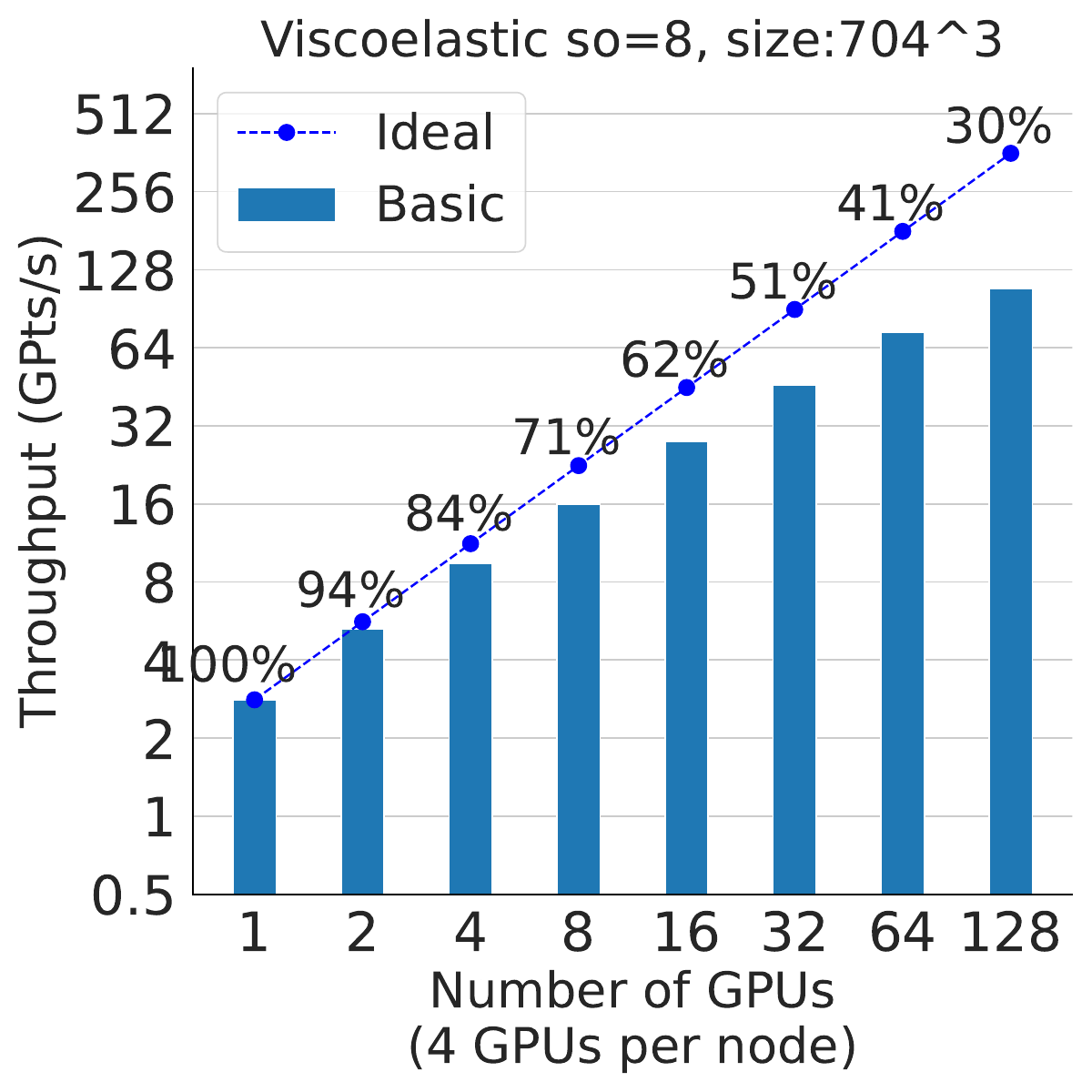}
        \caption{GPU scaling for Viscoelastic}
        \label{fig:vel_so08_gpu_platform}
	\end{subfigure}
        \caption{Strong scaling for the \textbf{viscoelastic} kernel.}
        \label{fig:vel-stencil-scaling}
\end{figure}

\subsection{Weak scaling performance evaluation}\label{sec:weak-scaling}

We select a constant per rank and node size of $256^3$ for weak scaling for all the benchmarked models.
We cyclically double the number of points per dimension as we double the number of CPU nodes/GPU devices (e.g., 512*256*256 on 2 nodes, up to 2048*1024*1024 on 128 nodes). 

\begin{figure}[!htbp]
        \centering
        \includegraphics[width=0.42\textwidth]{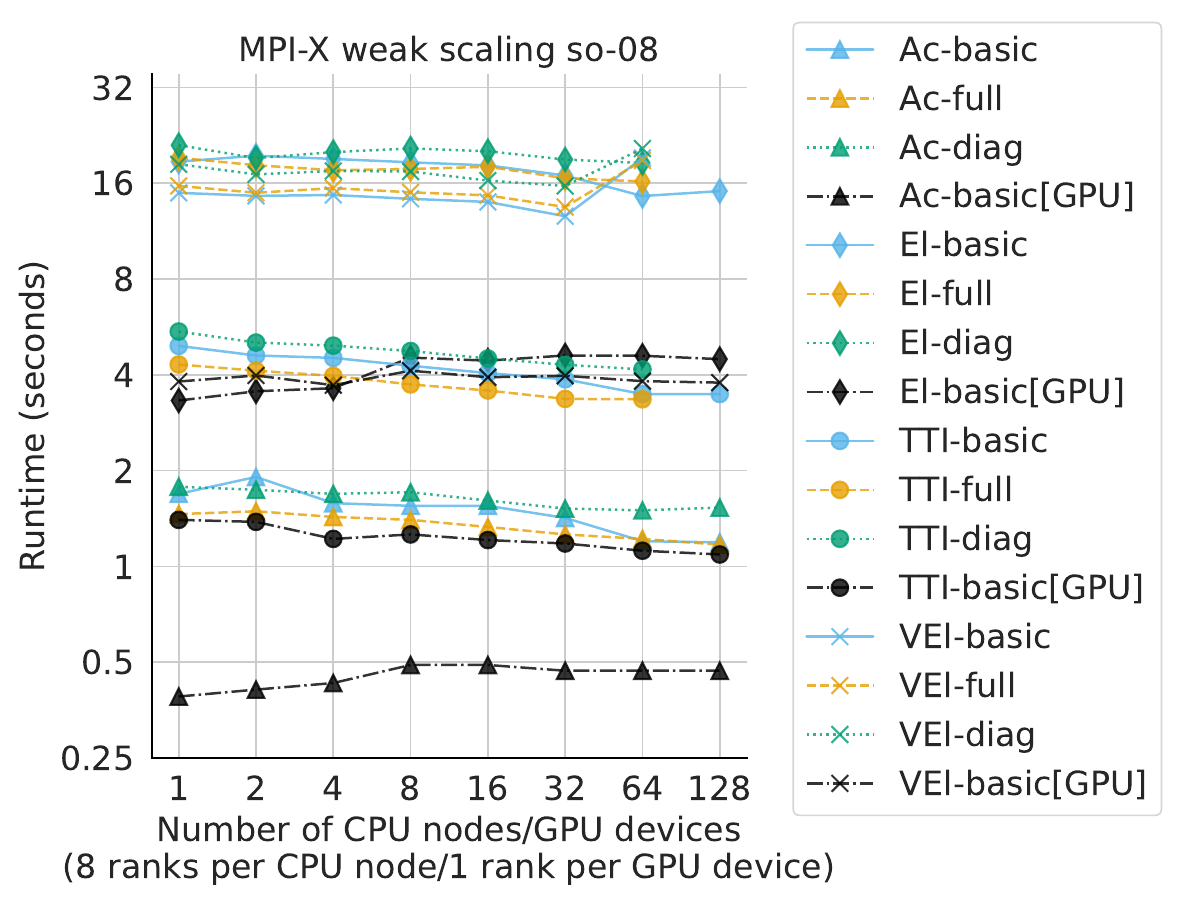}
        \label{fig:weak_scaling}
        \caption{MPI-X weak scaling (runtime). GPU is constantly ~4 times faster.}
        \label{fig:weak-scaling}
\end{figure}

\autoref{fig:weak-scaling} demonstrates a nearly constant runtime as we scale to more CPU nodes and GPU devices.
As expected, we observe a negligible runtime decrease in all experiments.
This minor improvement occurs because each time we double the size of a dimension and decompose along it, the ratio of halo regions that require communication to the computational domain is asymptotically reduced.
We observe that for the acoustic and TTI models, \emph{full} mode performs better when it is superior for one node.
This is expected since the core-to-remainder ratio stays the same and is in accordance with the claims in the strong scaling on how the core-to-remainder ratio affects the performance.
The trends for elastic and viscoelastic are similarly consistent, having only the outlier of 64 and 128 nodes for viscoelastic, owing to the OOM issue.

Weak scaling unlocks the GPU's potential, allowing for a more efficient time-to-solution.
\autoref{fig:weak-scaling} shows that all kernels are almost 4 times faster on GPUs compared to CPUs for the same amount of processed points.
GPUs evidently help solve big problems at scale faster and more efficiently.

\subsection{Discussion}\label{sec:discussion}

Regarding scaling on CPUs, in the context of the \emph{full} mode, local decomposed domains exhibit a lower core-to-remainder ratio as we scale.
Naturally, this ratio lowers more when using higher SDOs.
Consequently, as we scale, more time is spent in the computationally less efficient remainder areas.
In the ideal scenario of perfect computation/communication overlap, the total execution cost converges to the sum of the $\emph{core}$ and the less efficient $\emph{remainder}$ computation.
To establish \emph{full} mode as superior to others, the additional cost associated with inefficient strides in the remainder areas must be outweighed by the communication cost incurred in \emph{basic} and
\emph{diagonal} modes.
This is not easy, especially on systems equipped with high-speed InfiniBand interconnects, such as Archer2 and Tursa.
One obvious solution to improve the performance of the \emph{full} mode is customizing the decomposition to only split in $x$ and $y$ dimensions.
Message sizes are bigger, but inefficient strides over the z dimension are less of a problem. 
Therefore, a performance boost is observed, but continuous decomposition across $x$ and $y$ may lead to early shrinking of the decomposed domains.
In summary, we should tune topology for a golden spot in \emph{full} mode's throughput.

Naturally, kernels like elastic, with a higher communication-to-computation ratio,
tend to benefit from the \emph{full}, especially in kernels with high SDOs.
On the other hand, kernels like TTI do not leverage this benefit, as the computation-to-communication ratio is very high, even with higher SDOs.

Regarding scaling on GPUs, distributing the problem to smaller pieces makes scaling less efficient; however, it still performs better than CPUs in strong scaling.
At full scale, using 128 GPUs, acoustic and TTI are 1.6x faster on GPUs, while elastic and viscoelastic are 1.8x faster.
However, GPUs need to be more utilized and benefit from their full potential.
Bigger problems are run, as seen in weak scaling, where GPUs are 4x faster compared to GPUs at all numbers of nodes.

Limitations of this work that could further be addressed or alleviated in future work are:
support for preallocated on-device buffers, an automated tuning system for selecting the best-performing MPI pattern without exploring all three options manually, and another level of automated tuning for custom decompositions for the \emph{full} mode.
Further ideas to be explored, including, among others, data layout transformations that could benefit the computation of remainder areas \cite{zhao2018} and more advanced MPI patterns that could benefit from advanced cache-blocking optimizations, like temporal blocking \cite{pencil-sc20, malas2015}.

\section{Related work}\label{sec:related-work}

The acceleration of practical applications through domain decomposition methods
find, among others, applications in finite-difference stencil computations
\cite{physis2011, malas2015, yount2016, gridtools2021, omlin2024distributed, omlin2024xpu, bisbas2024},
unstructured mesh solvers \cite{firedrake2017, Betteridge2021b, mudalige-gas2022, pyfr2016, lfric2019},
image processing \cite{halide-dist2016}, magnetohydrodynamics \cite{pekkila-cudampi-2022}, 
Lattice-Boltzmann simulations for blood flow \cite{zacharoudiou2023}.
However, manual implementation of codes with DMP is error-prone and requires significant effort.
Several DSL and compiler frameworks aim to automatically generate code using MPI domain decomposition from higher-level abstractions to address these challenges.
Our work differentiates from the current state of the art in offering a fully automated workflow that starts from a symbolic DSL and can seamlessly support various MPI computation and communication schemes with zero changes in code. Notably, it achieves peak performance thanks to advanced flop-reduction optimizations.
Furthermore, this framework extends its supports beyond stencils commonly used for benchmarking in literature and provides the means for additional operators necessary for practical applications focused on geophysics.
This inclusive and automated approach distinguishes our work in addressing the challenges associated with DMP for application development.
OpenSBLI \cite{opensbli2017,opensbli2021} expresses PDEs using Einstein notation
and the C/C++-based OPS library \cite{reguly2018} for automatically leveraging SMP, DMP, and other optimizations.
However, the main application focus of OpenSBLI is hydrodynamics, gas turbines \cite{mudalige2019}, and supersonic-related applications.
In \cite{pencil-sc20}, the authors combine communication and computation overlap with temporal blocking, evaluating stencils from the CFD community.
Firedrake \cite{firedrake2017, Betteridge2021b, firedrake-manual2023} has long been
employing automated MPI code generation from high-level abstractions using
the UFL language from the FENICS project \cite{logg2012automated}.
PyFR \cite{pyfr2016} automatically generates MPI code targeting CPUs and GPUs,
mainly employing template-based JIT code generation.
LFRic \cite{lfric2019} for weather and climate modeling processes and ports Fortran
code to PSyclone, where hybrid MPI-OpenMP code is automatically generated.
Gridtools \cite{gridtools2021}, operating at a stencil-level DSL, can also
automatically generate hybrid MPI-CUDA code.
Additionally, frameworks such as OP2 \cite{mudalige2012} for unstructured mesh
solvers and OPS \cite{reguly2018} for structured mesh solvers support automated
MPI code generation for large CPU and GPU runs.
Finally, Saiph \cite{saiph2023} also provides a high-level language, focusing more on CFD applications, without support for more advanced tools like sparse operators.
Stencil workloads have recently been targeting newer architectures, such as the Cerebras Wafer-Scale engine.
In \cite{cerebras-sc22}, the authors focus on the iso-acoustic stencil,
presenting near-peak single-node efficiency, while in \cite{ltaief2023}, authors explore the scaling of seismic applications up to 48 CS-2 devices.

\section{Conclusions}\label{sec:conclusions}

This paper introduces a compiler approach to automate MPI code generation for solving PDEs targetting diverse hardware at scale.
The contributed methodology and implementation are integrated within the Devito DSL and compiler framework, but the concepts generally apply to any DSL and compiler frameworks. 
All the contributed machinery is abstracted from the user side, offering seamless portability and migration to HPC clusters of CPUs and GPUs.
Extending previous works in this scientific area, users can benefit from expressing non-trivial PDE simulations with complex physics in high-level symbolic mathematics and harnessing the power of DMP via MPI via HPC-ready automatic code generation.
We benchmarked three variants of the automatically generated hybrid MPI-OpenMP code on 16,384 AMD EPYC 7742 cores and hybrid MPI-OpenACC code on 128 A100-80s for four wave propagator kernels of academic and industrial interest (used in FWI, RTM).
Weak and strong scaling results show highly competitive throughput. Performance analysis highlights the strengths and limitations of these DMP patterns for computation kernels with different operational intensity and memory footprint.

\section*{Acknowledgments}
This research is funded by the Engineering and Physical Sciences Research Council (EPSRC) grants
EP/\-L016796/\-1, EP/\-R029423/\-1, EP/\-V001493/\-1, EP/\-W007789/\-1, EP/W007940/\-1.
This work used the ARCHER2 UK National Supercomputing Service (https://www.archer2.ac.uk), with an allocation provided by NERC,
and TURSA supercomputer.
We would also like to thank Amik St-Cyr for the helpful discussions on this topic.
\bibliographystyle{IEEEtran}
\bibliography{bibliography}

\newpage
\appendices

\clearpage

\input{appendix.tex}

\end{document}

%% file: appendix.tex
\appendix

\subsection{More details on the evaluated wave propagators}\label{sec:app:waves}

This subsection aims to provide more details on the benchmarked wave propagators used in this paper.

\subsubsection{Isotropic acoustic}\label{sec:app:waves:iso}
The acoustic wave equation is formulated in terms of the square slowness parameter,
This equation represents a single-scalar PDE resulting in a Jacobi-like stencil.
The acoustic wave equation is formulated in terms of the square slowness parameter,
denoted as $m$, where $m=1/c^2$ and $c$ is the speed of pressure waves in the given physical medium. 
A source term $q$ is also included to account for external excitations.
The isotropic acoustic wave equation is expressed as

\begin{equation}
    \begin{cases}
     m(x) \frac{\partial^2 u(t, x)}{\partial t^2} - \Delta u(t, x) = \delta(x_s) q(t), \\
     u(0, .) = \frac{\partial u(t, x)}{\partial t}(0, .) = 0, \\
     d(t, .) = u(t, x_r),
     \end{cases}
    \label{acou}
\end{equation}
where $u(t, x)$ is the pressure wavefield, $x_s$ is the point source position, 
$q(t)$ is the source time signature, $d(t, .)$ is the measured data at positions $x_r$
and $m(x)$ is the squared slowness.
This equation can be expressed concisely using the Devito symbolic API as
shown in \autoref{lst:wave-equation}.

\begin{python}[label=app:lst:wave-equation, caption=Symbolic definition of the wave-equation]
from devito import solve, Eq, Operator
eq = m * u.dt2 - u.laplace
stencil_eq = Eq(u.forward, solve(eq, u.forward))
\end{python}

\subsubsection{Anisotropic acoustic (TTI)}\label{sec:app:waves:tti}

This equation is widely employed in industrial applications such
as subsurface imaging, including FWI and RTM \cite{zhang2011stable, louboutin2018effects, duveneck, Alkhalifah2000AnAW, Bube2016self}.
This model captures the complexities of layered geological strata.

It is a pseudo-acoustic anisotropic equation consisting of a coupled system of two scalar PDEs.
More details on the mathematical formulation for the anisotropic acoustic are presented in 
Unlike the more straightforward isotropic acoustic, this formulation incorporates direction-dependent
propagation speeds. As a result, the discretised equation involves a rotated anisotropic Laplacian
kernel, significantly increasing the number of operations \cite{Louboutin2017a}.

For instance, the first dimension $x$ component of the Laplacian is defined as:

\begin{equation}
    \begin{aligned}
      G_{\bar{x}\bar{x}} &= D_{\bar{x}}^T D_{\bar{x}} \\
      D_{\bar{x}} &= \cos({\theta})\cos({\phi})\frac{\partial}{\partial x} + \cos({\theta})\sin({\phi})\frac{\partial}{\partial y} - \sin({\theta})\frac{\partial}{\partial z}.
    \end{aligned}
\label{rot}
\end{equation}

where $\theta$ is the (spatially dependent) tilt angle (rotation around $z$),
$\phi$ is the (spatially dependent) azimuth angle (rotation around $y$).
A more detailed description of the physics and discretisation can be found
in \cite{zhang2011stable, louboutin2018effects}. This model requires 12 grids.

\subsubsection{Isotropic elastic}\label{sec:app:waves:elastic}

The next is the isotropic elastic equation, which encapsulates the
complete physics of elastic waves, including compressional and shear waves.
Unlike the previous acoustic approximations, this equation possesses two
notable characteristics. Firstly, it is a first-order system in time,
allowing us to extend our work to a smaller range of local data dependencies over time.
Compared to second-order in time systems requiring three buffers to store
dependencies, first-order systems only require one buffer. 
Consequently, we demonstrate that the benefits of time-blocking and our
implementation thereof are not limited to a single pattern along the time
dimension. Secondly, this equation constitutes a coupled system consisting
of a vectorial and a tensorial PDE, resulting in drastically increased data
movement. While the isotropic acoustic equation has one or two state parameters,
the isotropic elastic equation involves nine parameters in the wavefield and
non-scalar expressions in the source and receiver expressions that involve
multiple wavefields.
The isotropic elastic model has a bigger working set compared to the other models (22 grids)
while also having a high number of floating point operations even at its arithmetically
optimised version.
The isotropic elastic wave equation, parameterised by the Lamé parameters
$\lambda$ and $\mu$, along with the density $\rho$, is defined as follows
\cite{virieux1986p}:

\begin{equation}
  \begin{aligned}
    &\frac{1}{\rho}\frac{\partial v}{\partial t} = \nabla . \tau \\
    &\frac{\partial \tau}{\partial t} = \lambda \mathrm{tr}(\nabla v) {I}  + \mu (\nabla v + (\nabla v)^T)
  \end{aligned}
  \label{elas1}
\end{equation}

where $v$ is a vector-valued function with one component per cartesian direction,
and the stress $\tau$ is a symmetric second-order tensor-valued function.

\subsubsection{Visco-elastic}\label{sec:app:waves:visco-elastic}

Finally, we introduce the visco-elastic wave propagation model, modelled upon the work
introduced in~\cite{Robertson:Viscoelastic94}. 
In three dimensions, using a single relaxation mode, the governing equations can be written as

\begin{subequations}\label{eq:model-equations}
\begin{alignat}{4}
 &v_i = &&\frac{1}{\rho}\partial_j\sigma_{ij}, &&&\\
 &\dot{\sigma}_{ij} = &&\pi\frac{\tau^p_\varepsilon}{\tau_\sigma}\partial_k v_k  \nonumber &&&\\
 & &&+2\mu\frac{\tau^s_\varepsilon}{\tau_\sigma}(\partial_i v_j-\partial_k v_k)+r_{ij}, \; &&& i=j, \\
 &\dot{\sigma}_{ij} = &&\mu\frac{\tau^s_\varepsilon}{\tau_\sigma}(\partial_i v_j+\partial_j v_i) + r_{ij}, \; &&& i\ne j, \\
 &\dot{r}_{ij} = &&-\frac{1}{\tau_\sigma}\biggl(r_{ij}+\left(\pi\frac{\tau^p_\varepsilon}{\tau_\sigma}
  -2\mu\frac{\tau^s_\varepsilon}{\tau_\sigma}\right)\partial_k v_k \nonumber &&&\\
 & &&+2\mu\frac{\tau^s_\varepsilon}{\tau_\sigma}\partial_i v_j\biggr), \; &&& i=j, \\
 &\dot{r}_{ij} = &&-\frac{1}{\tau_\sigma}\left(r_{ij} +
   \mu\frac{\tau^s_\varepsilon}{\tau_\sigma}(\partial_i v_j+\partial_j v_i)\right), \; &&& i\ne j, 
\end{alignat}
\end{subequations}
where suffix notation has been employed. In \autoref{eq:model-equations}, $v$ represents
the particle velocity, $\sigma$ the stress tensor and $r$ the memory variable. The physical parameters
appearing in \autoref{eq:model-equations} are detailed in \autoref{tab:model_parameters}.
This model has a first-order accurate discretisation in time and is computationally very interesting,
as it has a higher memory footprint (36 grids) and peak operation intensity and uses a staggered grid with
highly increased communication costs.
Furthermore, it is used for high-fidelity modelling - closer to the physical models people are interested in.

\begin{table}[!htbp]
        \centering
        \caption{Description of the parameters appearing in \autoref{eq:model-equations}.}
        \begin{tabular}{ll}
         $\tau^p_\varepsilon$ & Strain relaxation time for P-waves \\
         $\tau^s_\varepsilon$ & Strain relaxation time for SV-waves \\
         $\tau_\sigma$ & Stress relaxation time for both P- and SV-waves \\
         $\mu$ & Relaxation modulus corresponding to SV-waves \\
               & (analog of the Lam\'{e} constant $\mu$ in the elastic case) \\
         $\pi$ & Relaxation modulus corresponding to P-waves \\
               & (analog of $\lambda+2\mu$ in the elastic case)
        \end{tabular}
        \label{tab:model_parameters}
\end{table}

\clearpage

\newpage

\subsection{Automatically generated code excerpt}\label{sec:app:code.c}

\autoref{lst:code.c} shows the generated code for the input in \autoref{lst:dsl}.
We show only the computational kernel, excluding e.g. headers.

\begin{clang}[label=lst:code.c, caption={The Devito compiler automatically applied optimisations
        to the equations modelled in Listing~\ref{lst:dsl} and generated the C-code for the stencil update pattern.}]
float r0 = 1.0F/dt;
float r1 = 1.0F/(h_x*h_x);
float r2 = 1.0F/(h_y*h_y);

for (int time = time_m, t0 = (time)%
{
    for (int x = x_m; x <= x_M; x += 1)
    {
    #pragma omp simd aligned(u:32)
        for (int y = y_m; y <= y_M; y += 1)
        {
            float r3 = -2.0F*u[t0][x + 2][y + 2];
            u[t1][x + 2][y + 2] =
                         dt*(r0*u[t0][x + 2][y + 2]
                   + r1*r3 + r1*u[t0][x + 1][y + 2]
                           + r1*u[t0][x + 3][y + 2]
                   + r2*r3 + r2*u[t0][x + 2][y + 1]
                           + r2*u[t0][x + 2][y + 3]);
        }
    }
}
        
\end{clang}

\clearpage

\newpage

\subsection{Example scripts}\label{sec:app:example_scripts}

All the CPU experiments were executed and benchmarked on Archer2.
An example of the \emph{slurm} script and the parameters used is illustrated in \autoref{lst:cpu-slurm-script}.

\begin{python}[breaklines=True, label=lst:cpu-slurm-script,
                caption={An example of the SLURM file and the parameters submitted to Archer2 and used to execute the experiments.
                         This listing shows a 4-nodes example, using the \emph{diagonal} method}]
#!/bin/bash

# Slurm job options
#SBATCH --job-name=
#SBATCH --time=1:20:00
#SBATCH --nodes=4
#SBATCH --ntasks-per-node=8
#SBATCH --cpus-per-task=16
#SBATCH --switches=1@360 # Each group has 128 nodes

# Replace [budget code] below with your project code (e.g. t01)
#SBATCH --account=xxxxxx
#SBATCH --partition=standard
#SBATCH --qos=standard
#SBATCH -o ./jobs-diag2/output-4-diag2.%

# Propagate the cpus-per-task setting from script to srun commands
# By default, Slurm does not propagate this setting from the sbatch
# options to srun commands in the job script. If this is not done,
# process/thread pinning may be incorrect leading to poor performance
export SRUN_CPUS_PER_TASK=$SLURM_CPUS_PER_TASK

module load cray-python
module load cray-mpich
# Activate environment devito is installed
source environments/python3-env/bin/activate
cd devito

# Set the number of threads to $SLURM_CPUS_PER_TASK
# We want one thread per physical core
export OMP_NUM_THREADS=$SLURM_CPUS_PER_TASK
export OMP_PLACES=cores

# Devito-specific env variables
export DEVITO_ARCH=cray
export DEVITO_LANGUAGE=openmp
export DEVITO_LOGGING=DEBUG
# Select your MPI mode
export DEVITO_MPI=diag2
export DEVITO_PROFILING=advanced2

# Archer specific
export FI_OFI_RXM_SAR_LIMIT=524288
export FI_OFI_RXM_BUFFER_SIZE=131072
export MPICH_SMP_SINGLE_COPY_SIZE=16384
export CRAY_OMP_CHECK_AFFINITY=TRUE
export SLURM_CPU_FREQ_REQ=2250000

\end{python}

All the GPU experiments were executed and benchmarked on TURSA.
An example of the \emph{slurm} script and the parameters used is illustrated in \autoref{lst:gpu-slurm-script}.

\begin{python}[breaklines=True, label=lst:gpu-slurm-script,
                caption={An example of a SLURM file and the parameters submitted to TURSA and used to execute the experiments.
                         This listing shows a 2-nodes example, using the \emph{basic} method}]
#!/bin/bash

# Slurm job options
#SBATCH --job-name=GPU-2-job
#SBATCH --time=00:50:00
#SBATCH --partition=gpu-a100-80
#SBATCH --qos=standard
# Replace [budget code] below with your budget code (e.g. t01)
#SBATCH --account=xxxxxx

# Request right number of full nodes (48 cores by node for A100-80 GPU nodes))
#SBATCH --nodes=2
#SBATCH --ntasks-per-node=48
#SBATCH --cpus-per-task=1
#SBATCH --gres=gpu:4
#SBATCH --gpu-freq=1410

# Add our Python to PATH
cd /home/xxxxxx/xxxxxx/dc-bisb2/Python-3.12.6/
export PATH=${PWD}:$PATH
cd /home/xxxxxx/xxxxxx/dc-bisb2/devito

# Load needed modules: WARNING: You need other modules to BUILD mpi4py
module load nvhpc/23.5-nompi
export PATH=/home/y07/shared/utils/core/nvhpc/23.5/Linux_x86_64/23.5/comm_libs/mpi/bin:$PATH
mpicxx --version
module list

# Devito environment
export DEVITO_MPI=1
export DEVITO_LANGUAGE=openacc
export DEVITO_LOGGING=BENCH
export DEVITO_PROFILING=advanced2
export DEVITO_PLATFORM=nvidiaX
export DEVITO_COMPILER=nvc

\end{python}

The following commands are part of the job files used to run strong scaling on ARCHER2.
The benchmarks are open source and available online at \cite{devito_zenodo_4.8.10}.

\begin{python}[breaklines=True, label=lst:jobs-python, caption={Commands used to run the benchmarks on Archer2, shape, timesteps and space order can be seen for each problem executed.}]
srun --distribution=block:block --hint=nomultithread python examples/seismic/acoustic/acoustic_example.py -d 1024 1024 1024 --tn 512 -so 4 -a aggressive
srun --distribution=block:block --hint=nomultithread python examples/seismic/acoustic/acoustic_example.py -d 1024 1024 1024 --tn 512 -so 8 -a aggressive
srun --distribution=block:block --hint=nomultithread python examples/seismic/acoustic/acoustic_example.py -d 1024 1024 1024 --tn 512 -so 12 -a aggressive
srun --distribution=block:block --hint=nomultithread python examples/seismic/acoustic/acoustic_example.py -d 1024 1024 1024 --tn 512 -so 16 -a aggressive

srun --distribution=block:block --hint=nomultithread python examples/seismic/elastic/elastic_example.py -d 1024 1024 1024 --tn 512 -so 4 -a aggressive
srun --distribution=block:block --hint=nomultithread python examples/seismic/elastic/elastic_example.py -d 1024 1024 1024 --tn 512 -so 8 -a aggressive
srun --distribution=block:block --hint=nomultithread python examples/seismic/elastic/elastic_example.py -d 1024 1024 1024 --tn 512 -so 12 -a aggressive
srun --distribution=block:block --hint=nomultithread python examples/seismic/elastic/elastic_example.py -d 1024 1024 1024 --tn 512 -so 16 -a aggressive

srun --distribution=block:block --hint=nomultithread python examples/seismic/tti/tti_example.py -d 1024 1024 1024 --tn 512 -so 4 -a aggressive
srun --distribution=block:block --hint=nomultithread python examples/seismic/tti/tti_example.py -d 1024 1024 1024 --tn 512 -so 8 -a aggressive
srun --distribution=block:block --hint=nomultithread python examples/seismic/tti/tti_example.py -d 1024 1024 1024 --tn 512 -so 12 -a aggressive
srun --distribution=block:block --hint=nomultithread python examples/seismic/tti/tti_example.py -d 1024 1024 1024 --tn 512 -so 16 -a aggressive

srun --distribution=block:block --hint=nomultithread python examples/seismic/viscoelastic/viscoelastic_example.py -d 768 768 768 --tn 512 -so 4 -a aggressive
srun --distribution=block:block --hint=nomultithread python examples/seismic/viscoelastic/viscoelastic_example.py -d 768 768 768 --tn 512 -so 8 -a aggressive
srun --distribution=block:block --hint=nomultithread python examples/seismic/viscoelastic/viscoelastic_example.py -d 768 768 768 --tn 512 -so 12 -a aggressive
srun --distribution=block:block --hint=nomultithread python examples/seismic/viscoelastic/viscoelastic_example.py -d 768 768 768 --tn 512 -so 16 -a aggressive

\end{python}

\clearpage

\newpage

\subsection{CPU strong scaling benchmarks}\label{sec:app:cpu_benchmarks}

\subsubsection{Extended strong scaling results for space discretisation orders 4, 8, 12, 16}\label{sec:app:extended_results}

\begin{figure}[htbp]
	\centering
	\begin{subfigure}[b]{0.24\textwidth}
        \includegraphics[width=\textwidth]{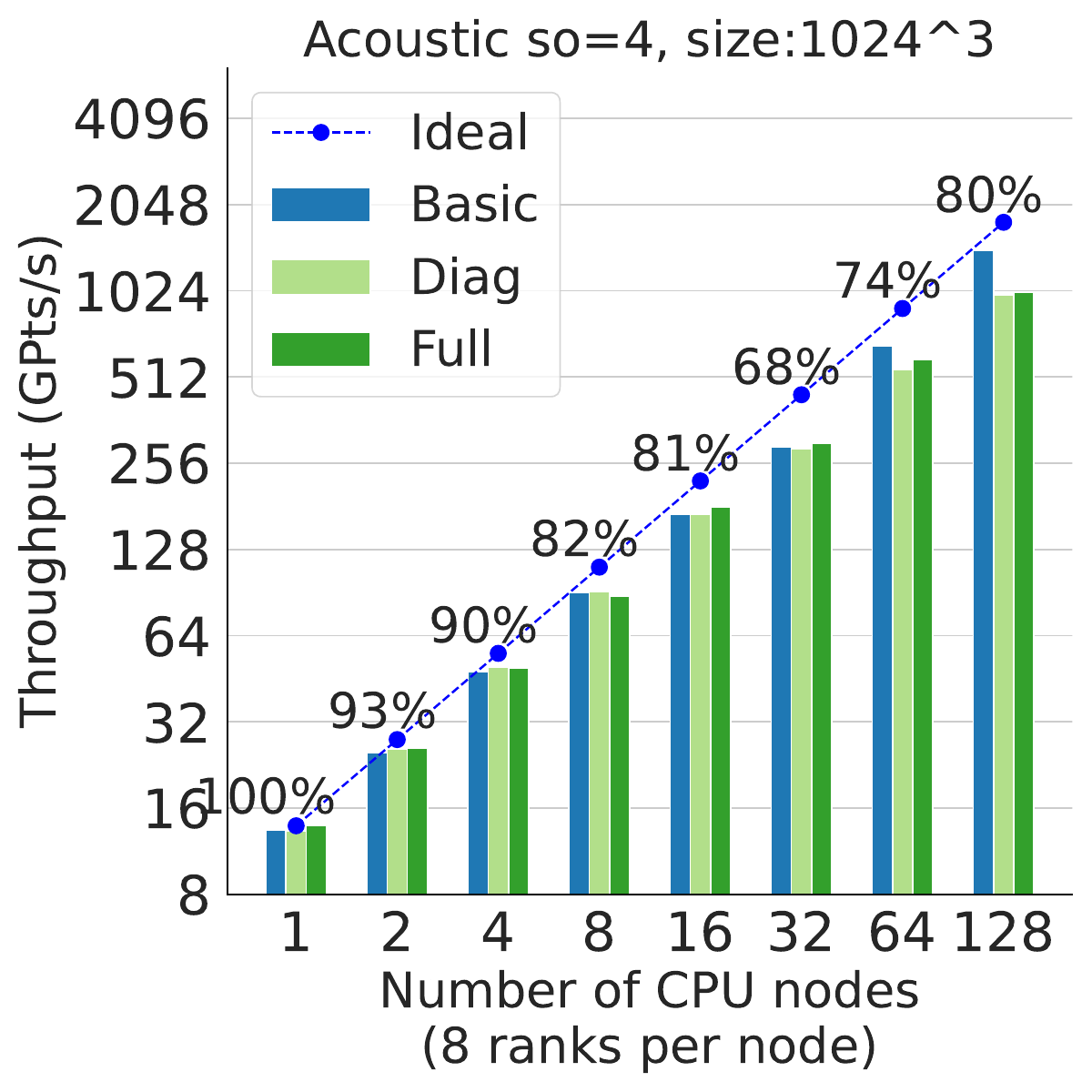}
        \caption{SDO 4}
        \label{fig:ac_so04_platform-a}
	\end{subfigure}
	\begin{subfigure}[b]{0.24\textwidth}
        \includegraphics[width=\textwidth]{figures/Acoustic_throughput_so-08SH.pdf}
        \caption{SDO 8}
        \label{fig:ac_so08_platform-a}
        \end{subfigure}
	\begin{subfigure}[b]{0.24\textwidth}
        \includegraphics[width=\textwidth]{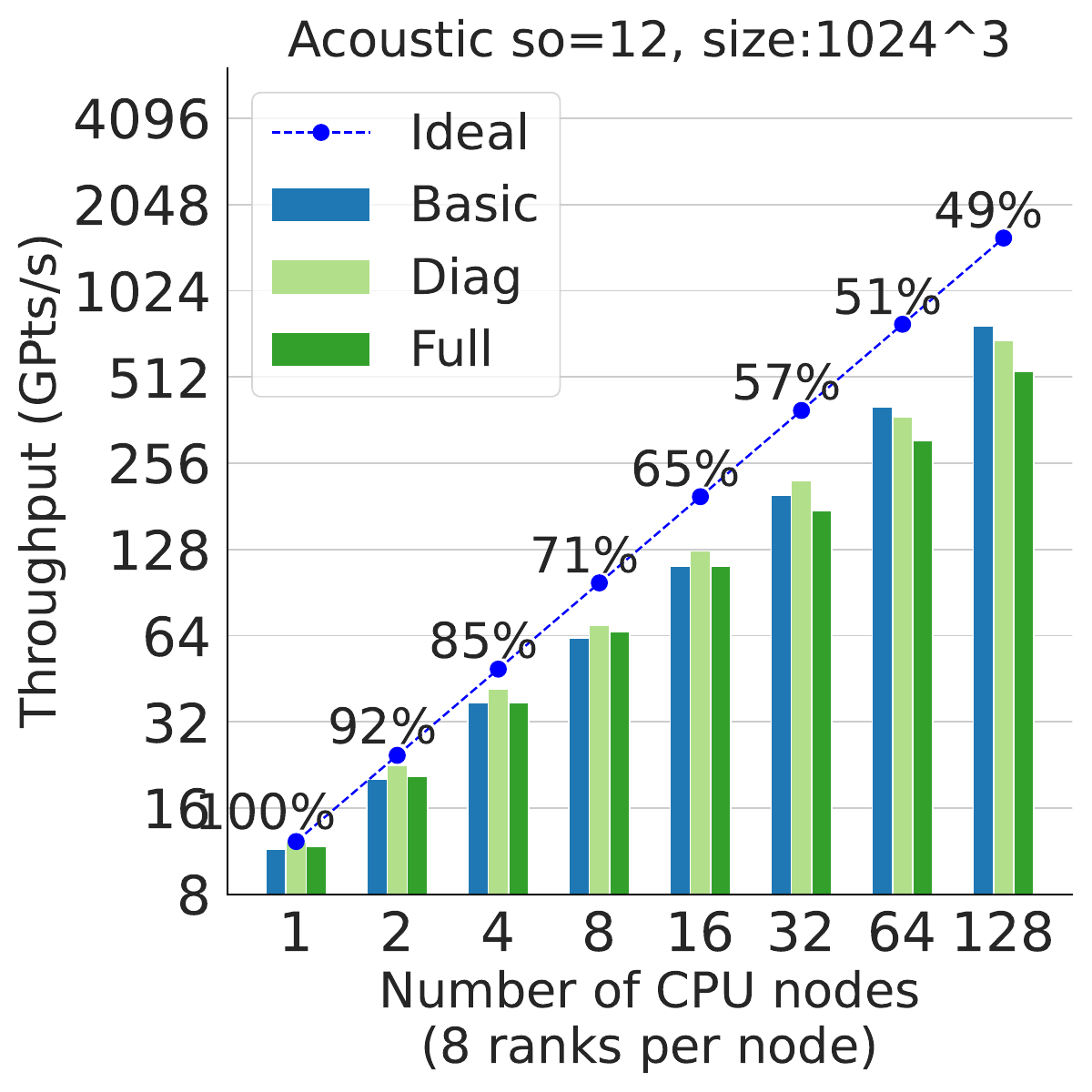}
        \caption{SDO 12}
        \label{fig:ac_so12_platform-a}
	\end{subfigure}
	\begin{subfigure}[b]{0.24\textwidth}
        \includegraphics[width=\textwidth]{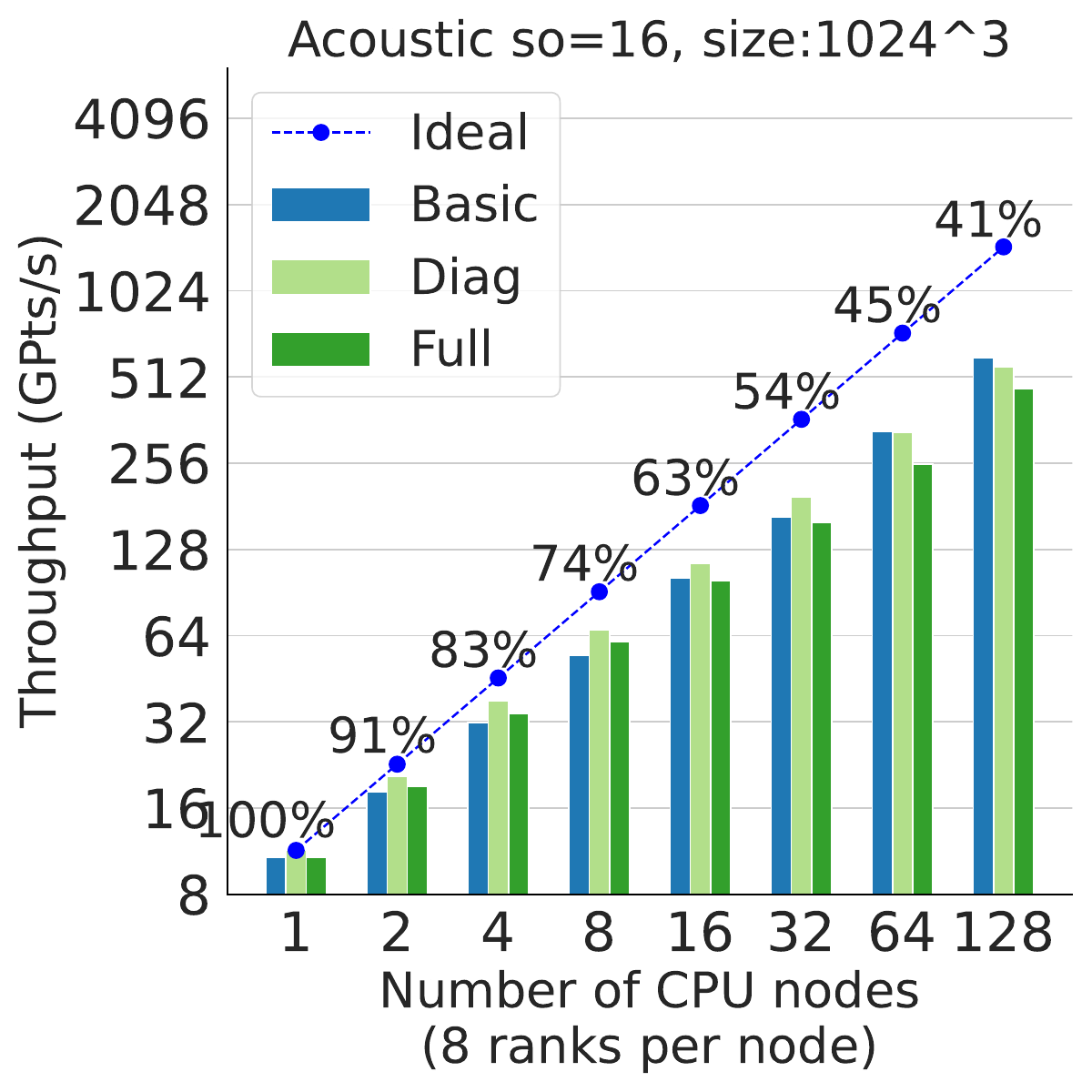}
        \caption{SDO 16}
        \label{fig:ac_so16_platform-a}
	\end{subfigure}
        \caption{Strong scaling for the \textbf{iso-acoustic} kernel for SDOs 4, 8, 12 and 16.}
        \label{fig:ac-stencil-scaling-a}
\end{figure}

This section includes extended results, presenting SDOs 4, 12 and 16 in addition to the paper's results on SDO 8.
\autoref{fig:ac-stencil-scaling-a} and tables \autoref{tab:ac:so-04}, \autoref{tab:ac:so-08}, \autoref{tab:ac:so-12}, \autoref{tab:ac:so-16} present the extended results for the acoustic stencil kernel for space orders 4, 8, 12 and 16 respectively.

\input{tables_raw/acoustic_raw.tex}

\begin{figure}[!htbp]
	\centering
	\begin{subfigure}[b]{0.24\textwidth}
        \includegraphics[width=\textwidth]{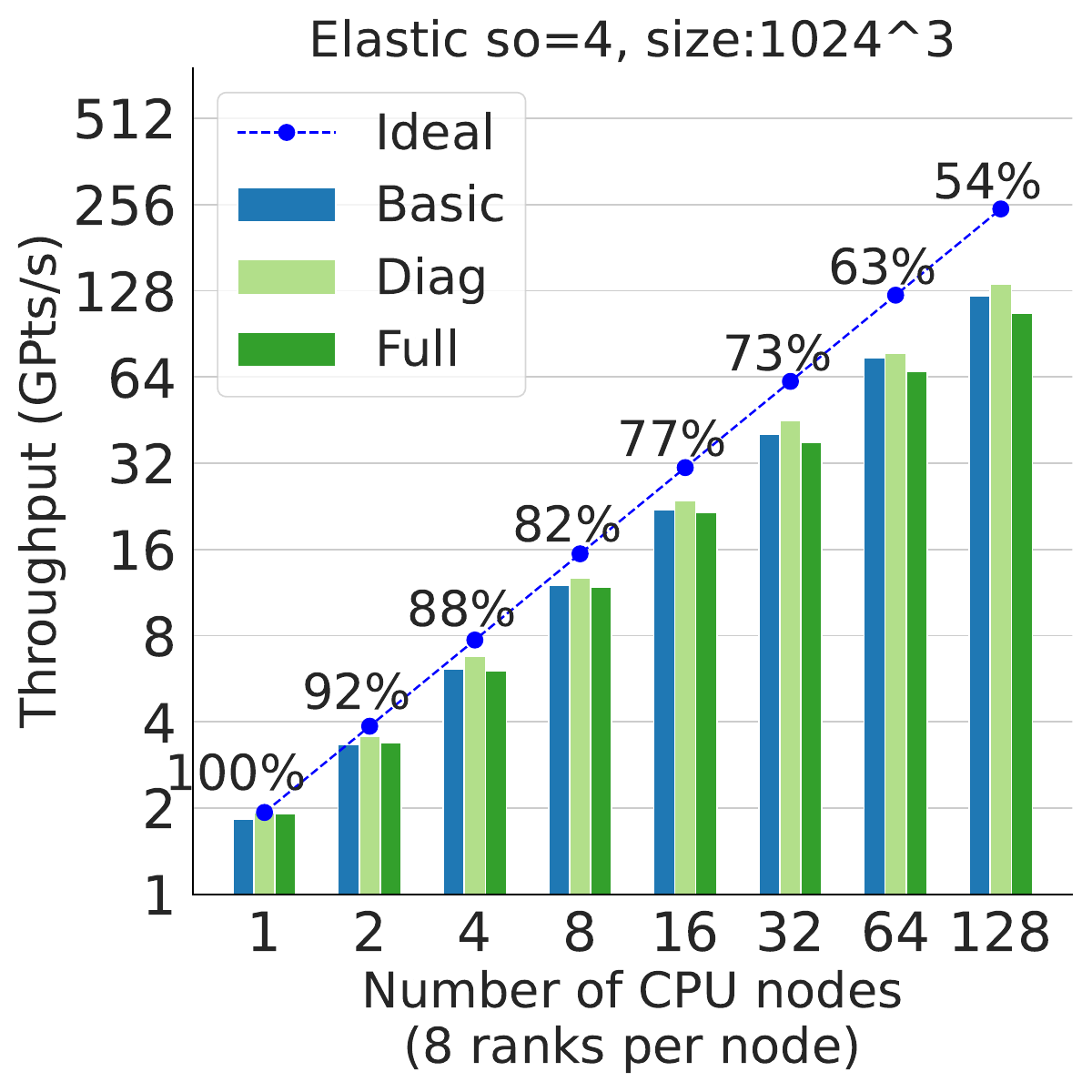}
        \caption{SDO 4}
        \label{fig:el_so04_platform-a}
	\end{subfigure}
	\begin{subfigure}[b]{0.24\textwidth}
        \includegraphics[width=\textwidth]{figures/Elastic_throughput_so-08SH.pdf}
        \caption{SDO 8}
        \label{fig:el_so08_platform-a}
        \end{subfigure}
	\begin{subfigure}[b]{0.24\textwidth}
        \includegraphics[width=\textwidth]{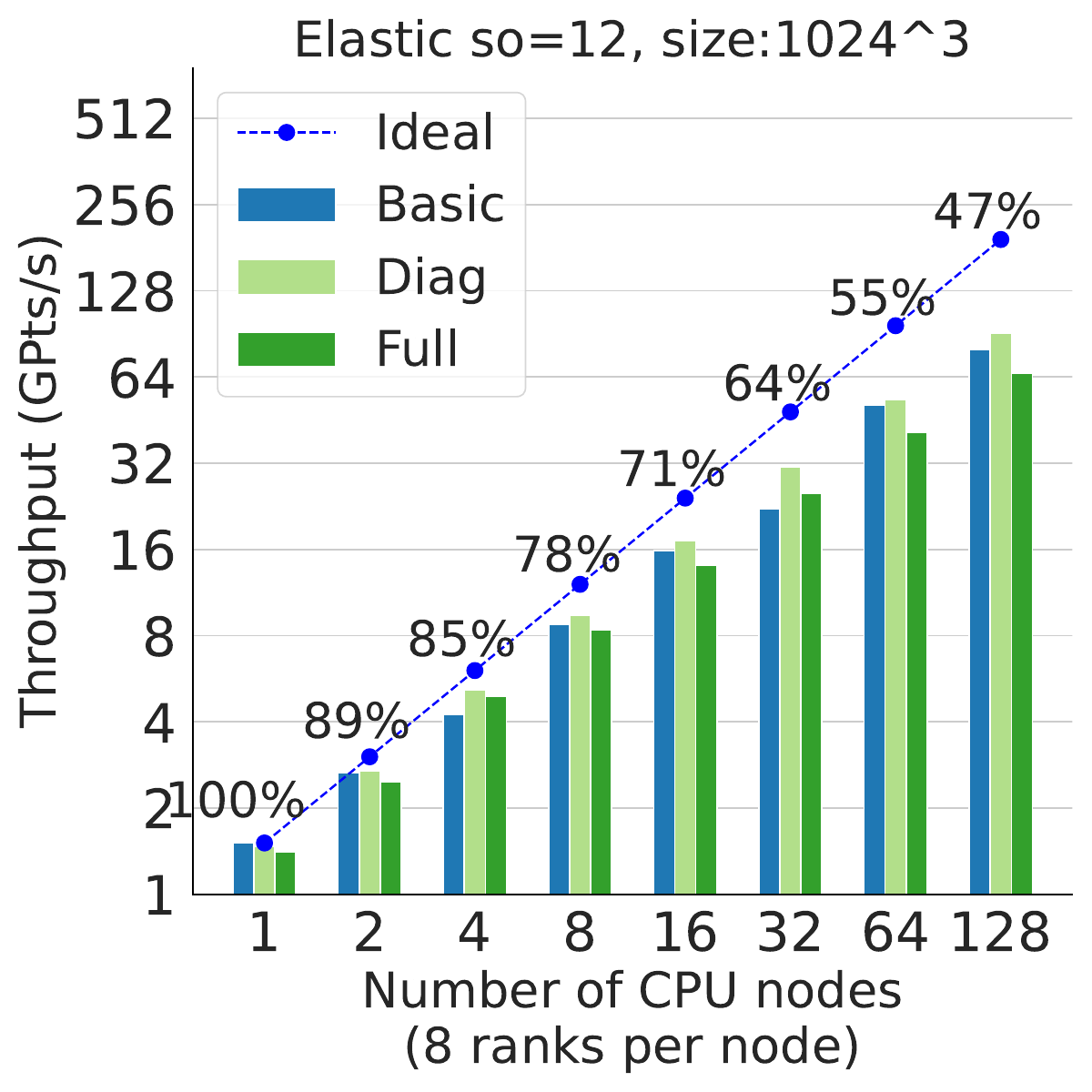}
        \caption{SDO 12}
        \label{fig:el_so12_platform-a}
	\end{subfigure}
	\begin{subfigure}[b]{0.24\textwidth}
        \includegraphics[width=\textwidth]{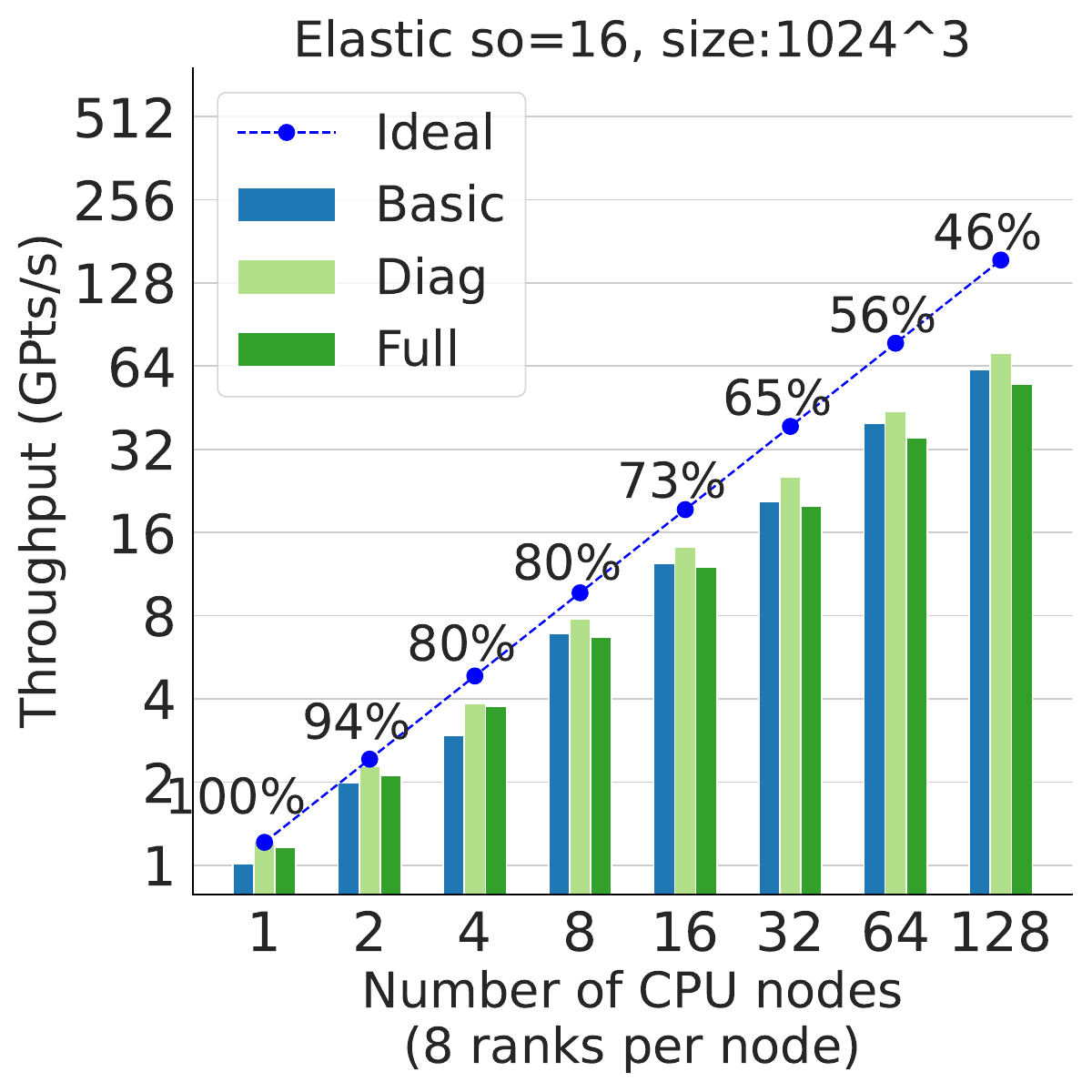}
        \caption{SDO 16}
        \label{fig:el_so16_platform-a}
	\end{subfigure}
        \caption{Strong scaling for the \textbf{elastic} kernel for SDOs 4, 8, 12 and 16.}
        \label{fig:el-stencil-scaling-a}
\end{figure}

\autoref{fig:el-stencil-scaling-a} and tables \ref{tab:el:so-04}, \ref{tab:el:so-08}, \ref{tab:el:so-12}, \ref{tab:el:so-16}
present the extended results for the elastic stencil kernel for various space orders.

\input{tables_raw/elastic_raw.tex}

\begin{figure}[!htbp]
	\centering
	\begin{subfigure}[b]{0.24\textwidth}
        \includegraphics[width=\textwidth]{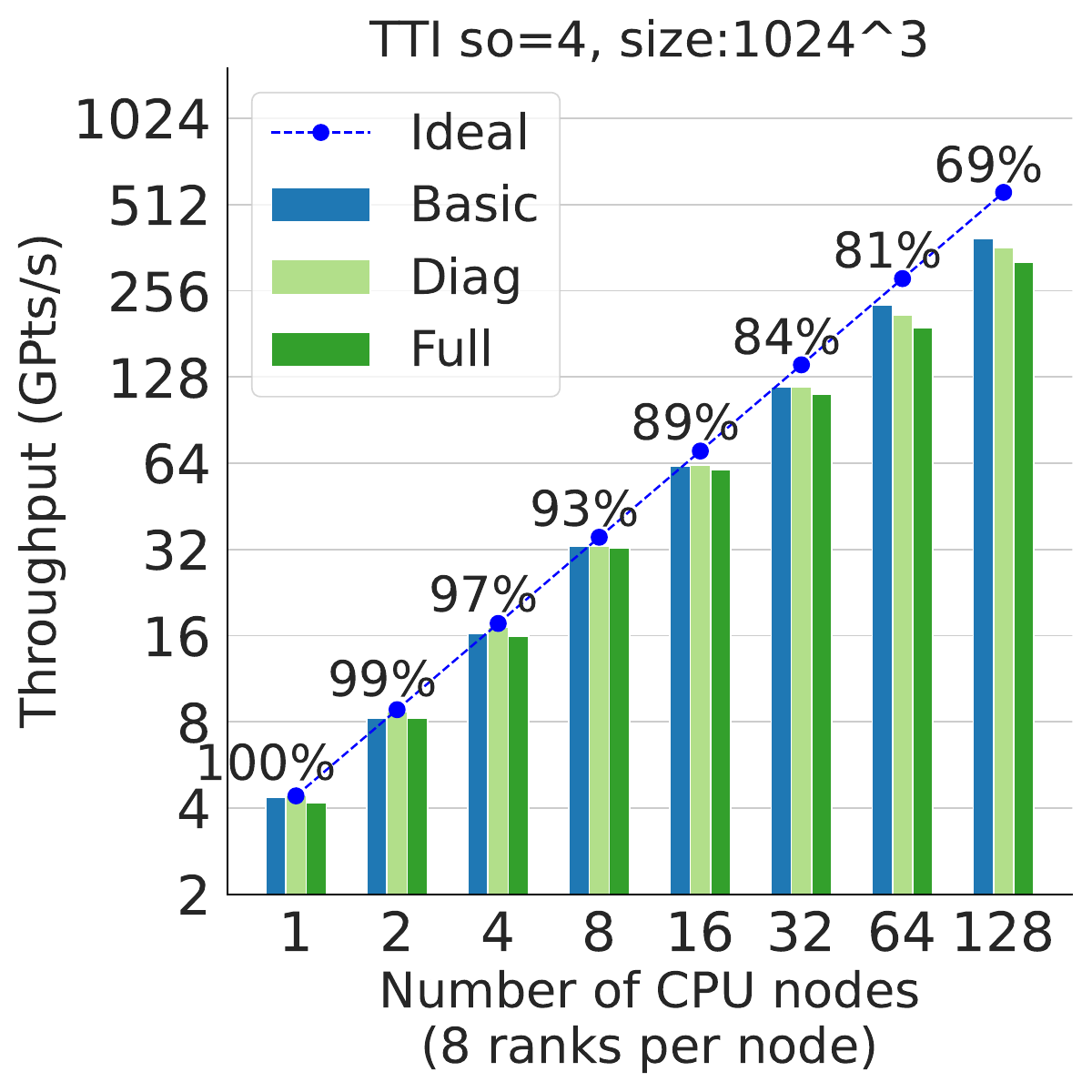}
        \caption{SDO 4}
        \label{fig:tti_so04_platform-a}
	\end{subfigure}
	\begin{subfigure}[b]{0.24\textwidth}
        \includegraphics[width=\textwidth]{figures/TTI_throughput_so-08SH.pdf}
        \caption{SDO 8}
        \label{fig:tti_so08_platform-a}
        \end{subfigure}
	\begin{subfigure}[b]{0.24\textwidth}
        \includegraphics[width=\textwidth]{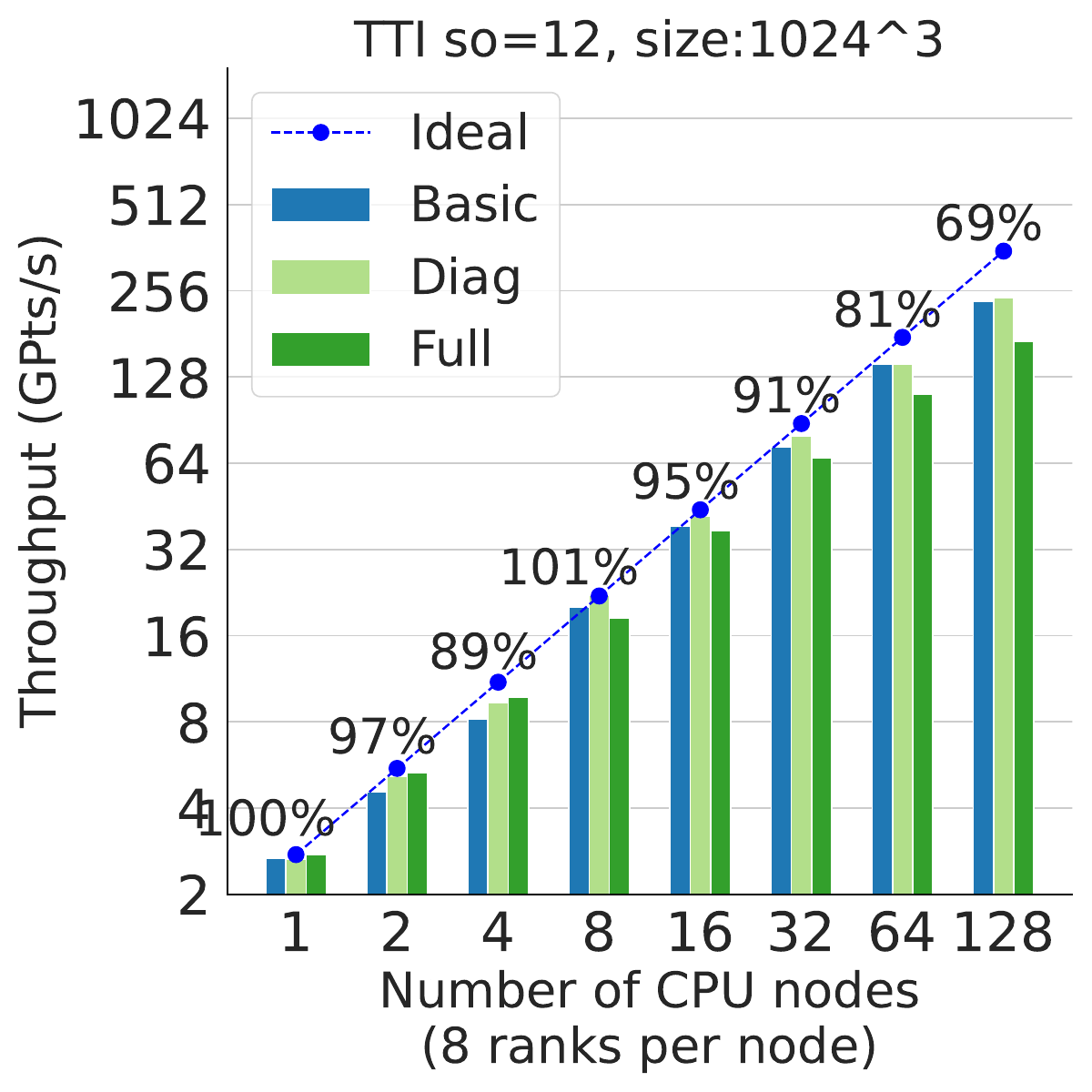}
        \caption{SDO 12}
        \label{fig:tti_so12_platform-a}
	\end{subfigure}
	\begin{subfigure}[b]{0.24\textwidth}
        \includegraphics[width=\textwidth]{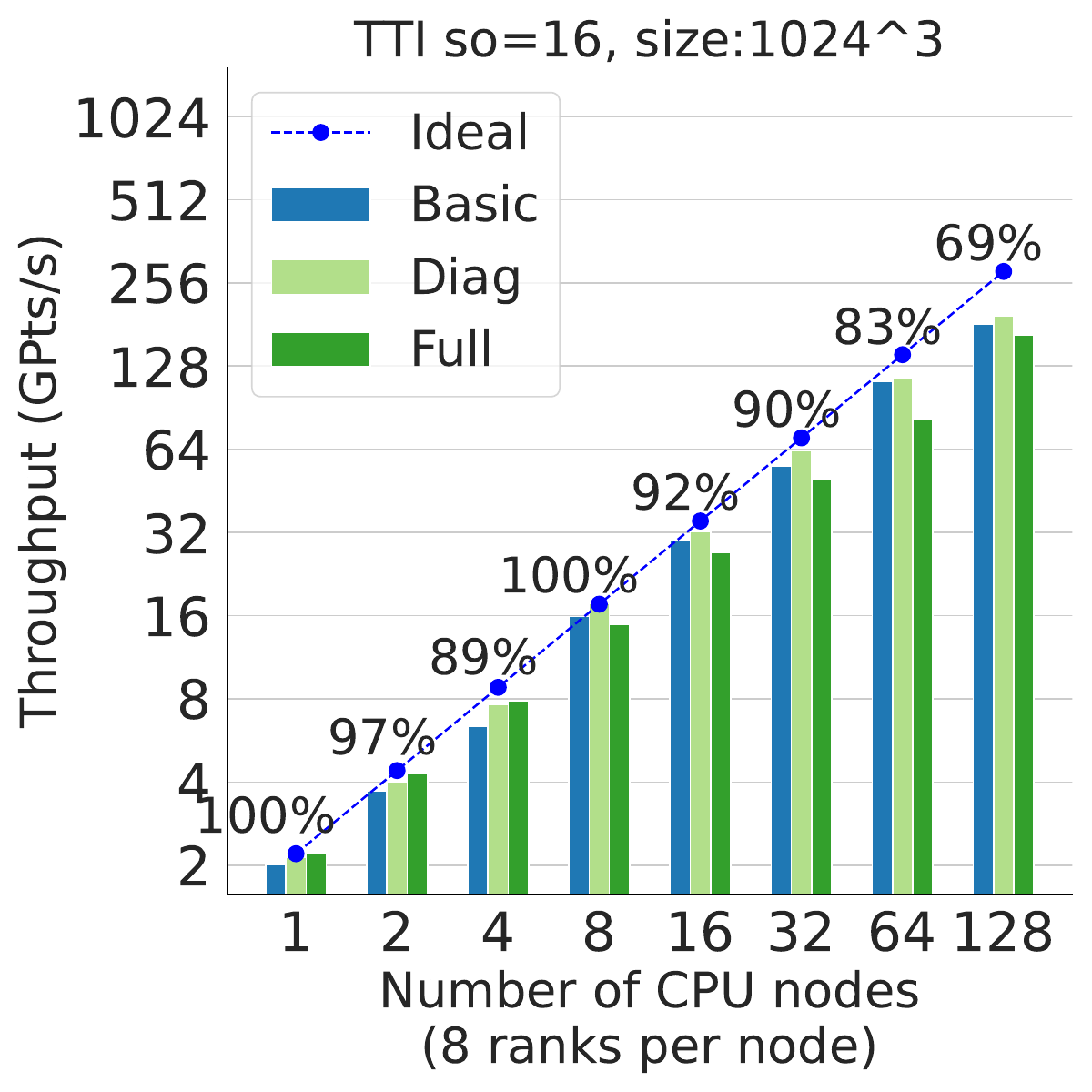}
        \caption{SDO 16}
        \label{fig:tti_so16_platform-a}
	\end{subfigure}
        \caption{Strong scaling for the \textbf{TTI} kernel for SDOs 4, 8, 12 and 16.}
        \label{fig:tti-stencil-scaling-a}
\end{figure}

\autoref{fig:tti-stencil-scaling-a} and tables \ref{tab:tti:so-04}, \ref{tab:tti:so-08}, \ref{tab:tti:so-12}, \ref{tab:tti:so-16}
present the extended results for the TTI stencil kernel for various space orders.

\input{tables_raw/tti_raw.tex}

\begin{figure}[!htbp]
	\centering
	\begin{subfigure}[b]{0.24\textwidth}
        \includegraphics[width=\textwidth]{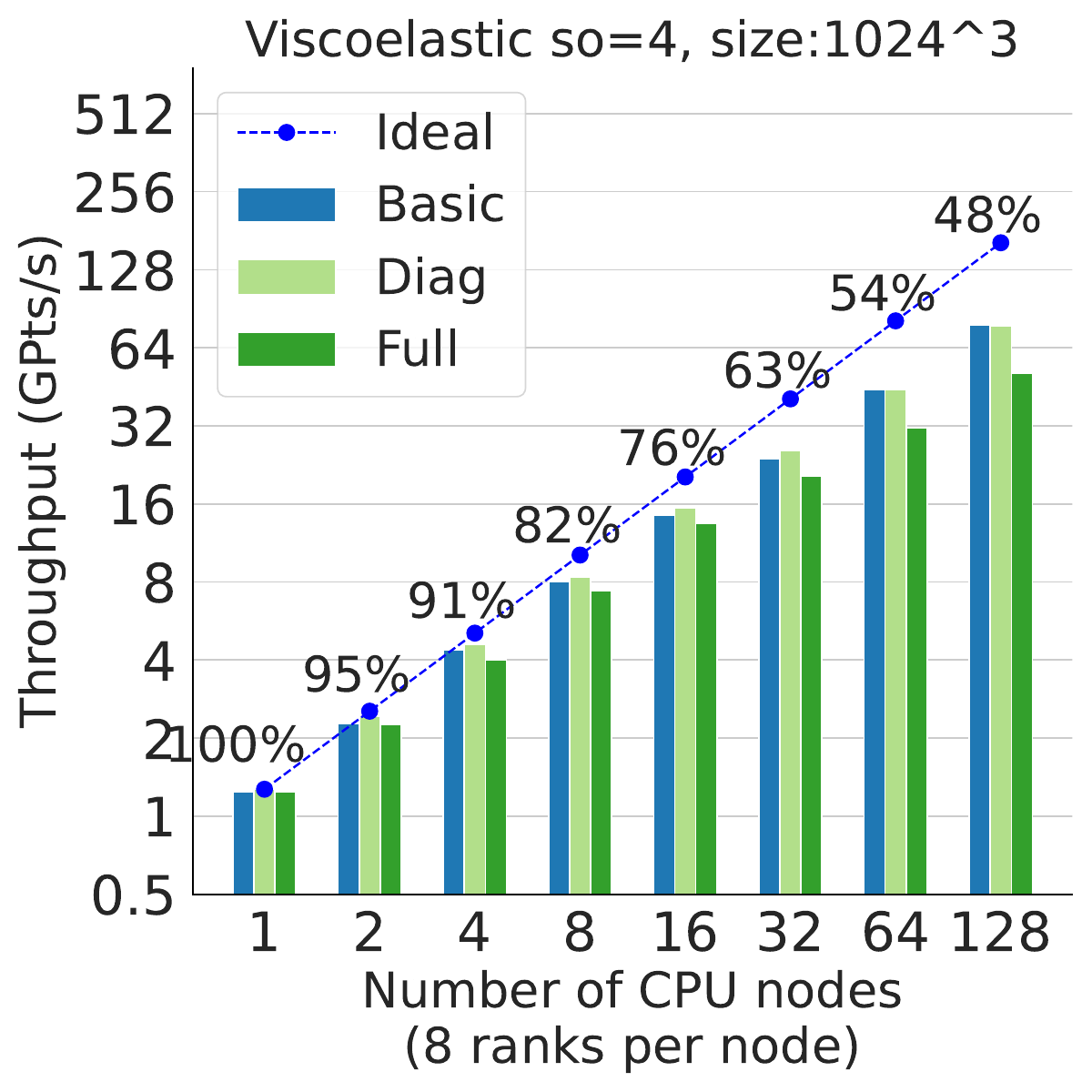}
        \caption{SDO 4}
        \label{fig:vel_so04_platform-a}
	\end{subfigure}
	\begin{subfigure}[b]{0.24\textwidth}
        \includegraphics[width=\textwidth]{figures/Viscoelastic_throughput_so-08SH.pdf}
        \caption{SDO 8}
        \label{fig:vel_so08_platform-a}
        \end{subfigure}
	\begin{subfigure}[b]{0.24\textwidth}
        \includegraphics[width=\textwidth]{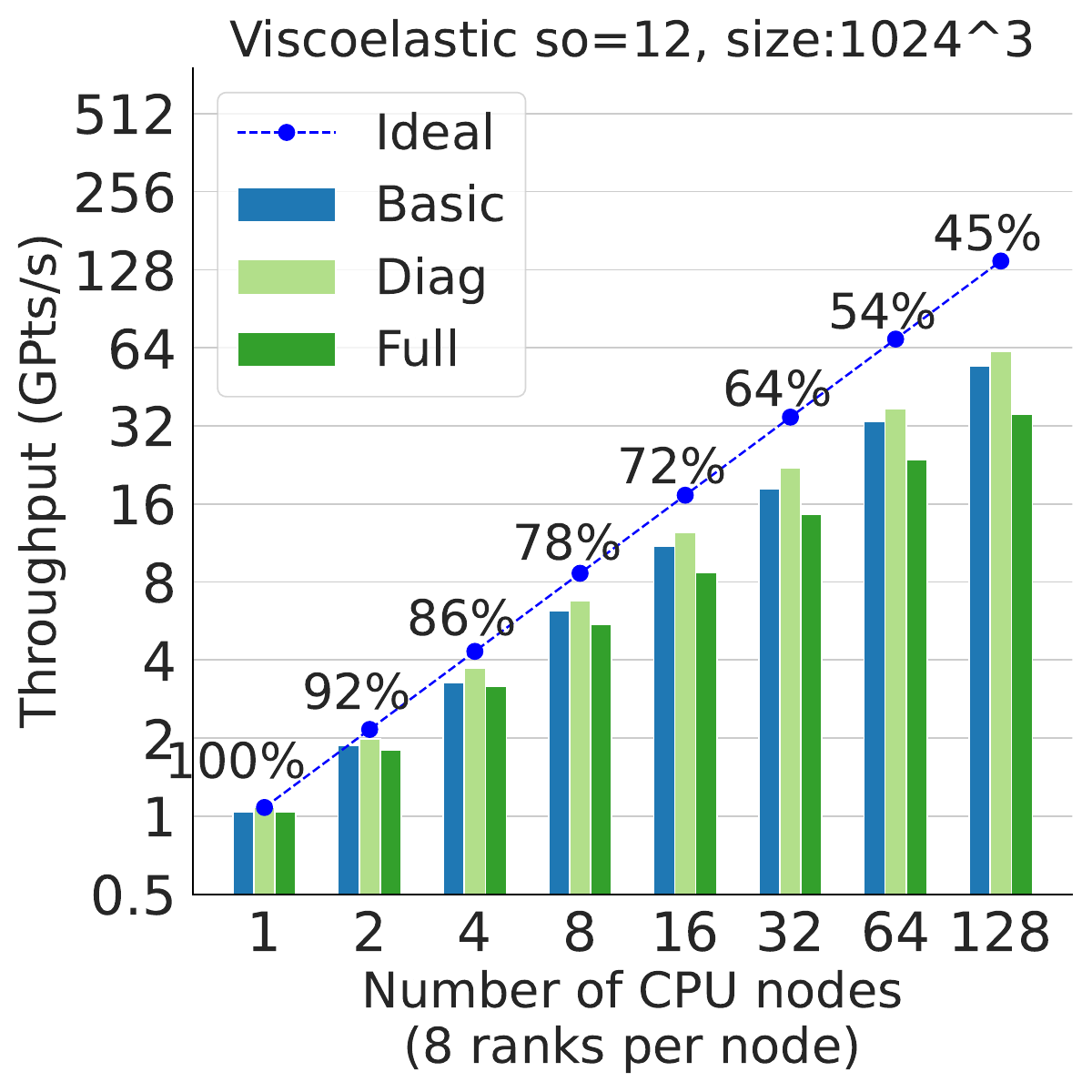}
        \caption{SDO 12}
        \label{fig:vel_so12_platform-a}
	\end{subfigure}
	\begin{subfigure}[b]{0.24\textwidth}
        \includegraphics[width=\textwidth]{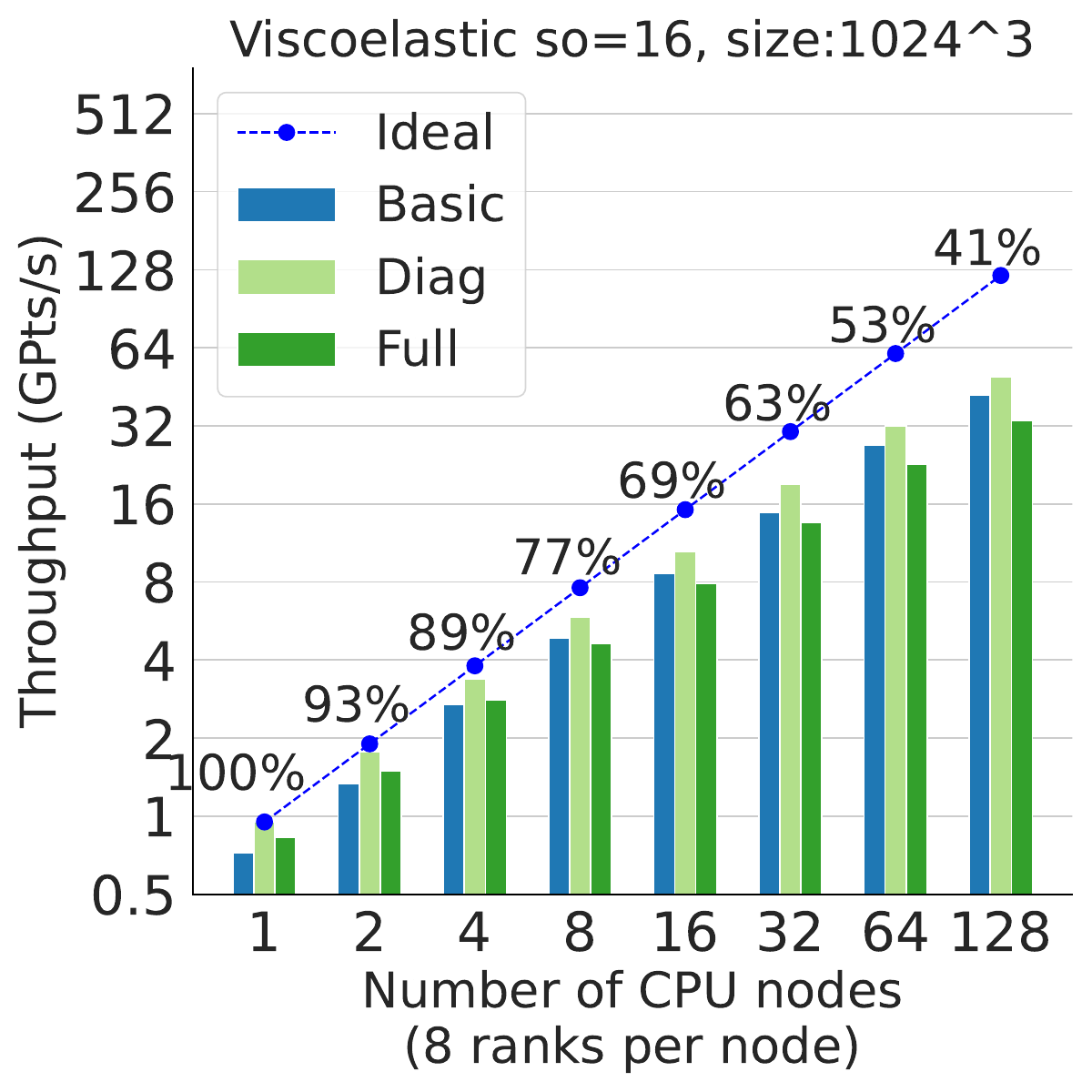}
        \caption{SDO 16}
        \label{fig:vel_so16_platform-a}
	\end{subfigure}
        \caption{Strong scaling for the \textbf{viscoelastic} kernel for SDOs 4, 8, 12 and 16.}
        \label{fig:vel-stencil-scaling-a}
\end{figure}

\autoref{fig:vel-stencil-scaling-a} and tables \ref{tab:vel:so-04}, \ref{tab:vel:so-08}, \ref{tab:vel:so-12}, \ref{tab:vel:so-16}
present the extended results for the viscoelastic stencil kernel for various space orders.

\input{tables_raw/v_elastic_raw.tex}

\newpage
\subsection{GPU strong scaling benchmarks}\label{sec:app:gpu_benchmarks}

\subsubsection{Extended GPU strong scaling results for space discretisation orders 4, 8, 12, 16}\label{sec:app:extended_results_gpu}

This section includes extended results, presenting SDOs 4, 12 and 16 in addition to the paper's results on SDOs 8.

\begin{figure}[htbp]
	\centering
	\begin{subfigure}[b]{0.24\textwidth}
        \includegraphics[width=\textwidth]{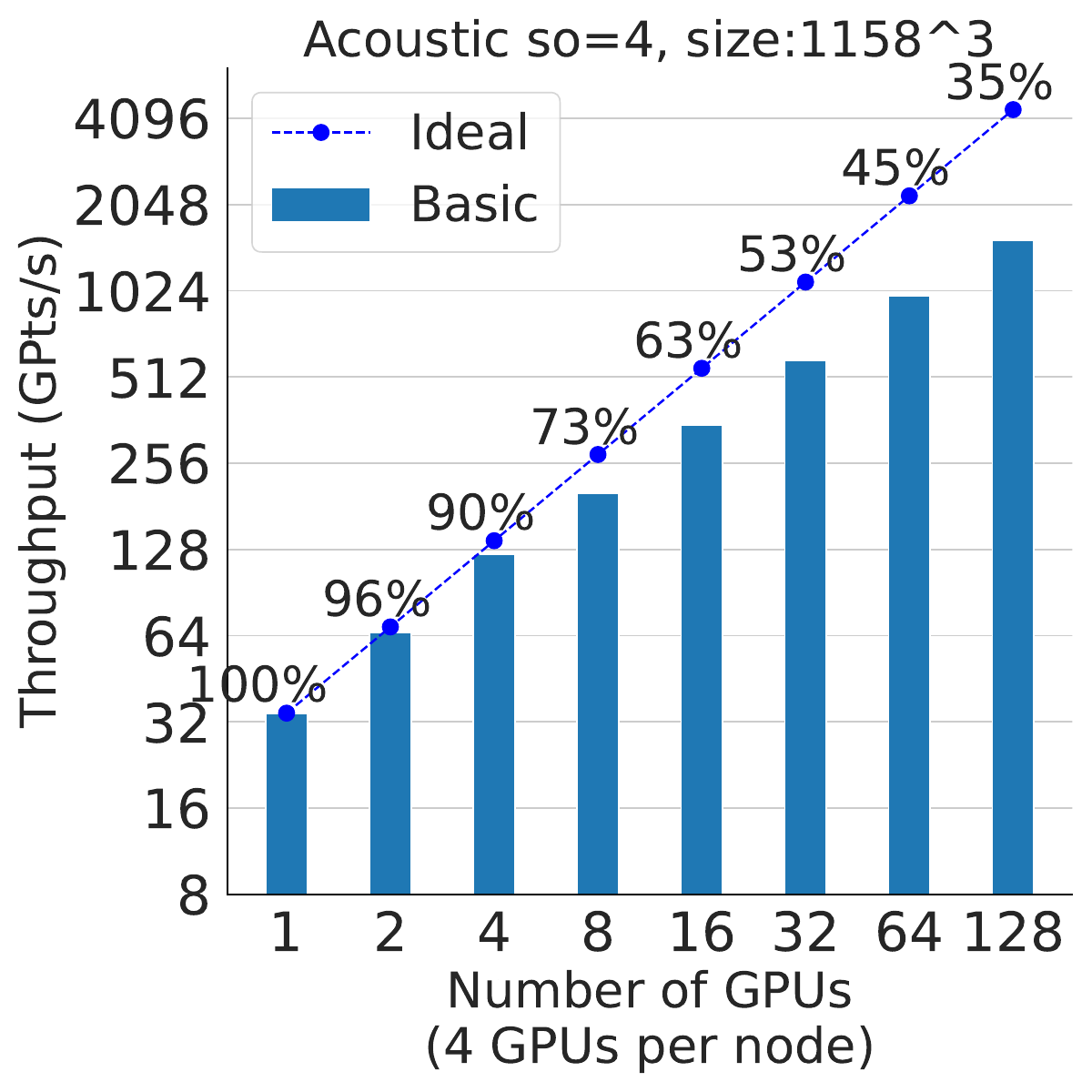}
        \caption{SDO 4}
        \label{fig:ac_so04_gpu_platform-a}
	\end{subfigure}
	\begin{subfigure}[b]{0.24\textwidth}
        \includegraphics[width=\textwidth]{figures/Acoustic_throughput_Basic_so-08GPU_SH.pdf}
        \caption{SDO 8}
        \label{fig:ac_so08_gpu_platform-a}
        \end{subfigure}
	\begin{subfigure}[b]{0.24\textwidth}
        \includegraphics[width=\textwidth]{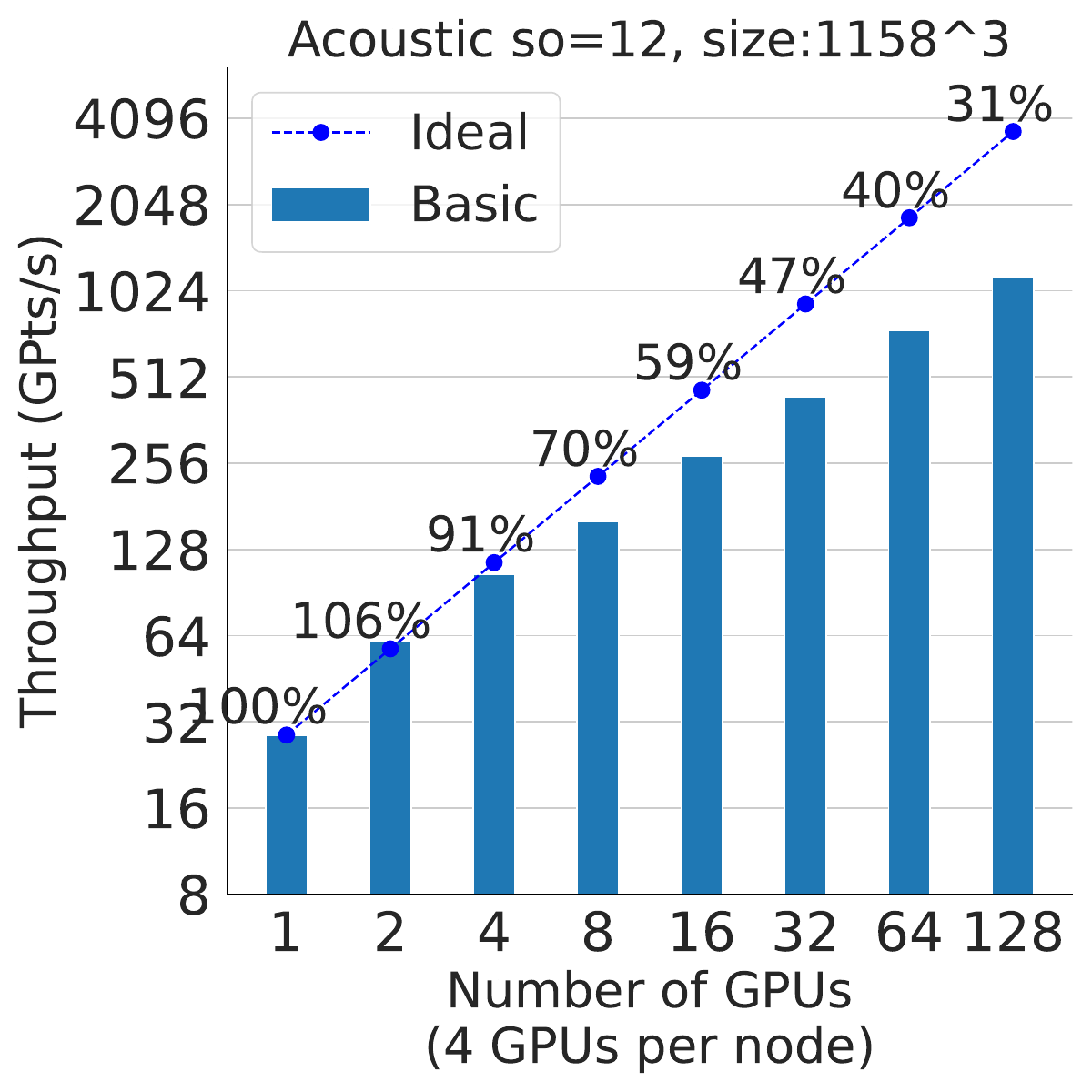}
        \caption{SDO 12}
        \label{fig:ac_so12_gpu_platform-a}
	\end{subfigure}
	\begin{subfigure}[b]{0.24\textwidth}
        \includegraphics[width=\textwidth]{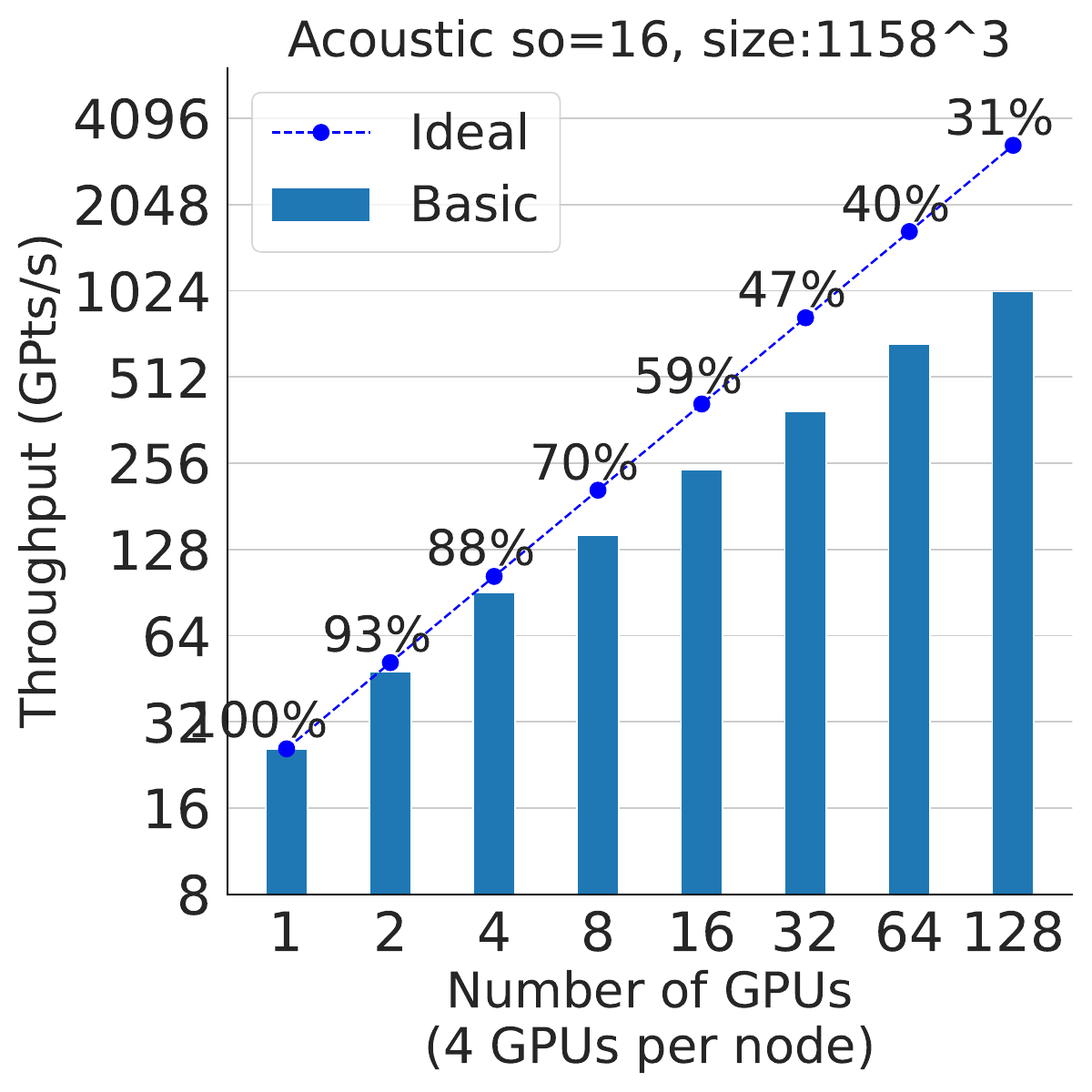}
        \caption{SDO 16}
        \label{fig:ac_so16_gpu_platform-a}
	\end{subfigure}
        \caption{GPU Strong scaling for the \textbf{iso-acoustic} kernel for SDOs 4, 8, 12 and 16.}
        \label{fig:ac-stencil-gpu-scaling-a}
\end{figure}

\autoref{fig:ac-stencil-gpu-scaling-a} and tables \autoref{tab:ac_gpu:so-04}, \autoref{tab:ac_gpu:so-08}, \autoref{tab:ac_gpu:so-12}, \autoref{tab:ac_gpu:so-16}
present the extended results for the acoustic stencil kernel for various space orders.

\input{tables_raw/acoustic_gpu_raw.tex}

\begin{figure}[htbp]
	\centering
	\begin{subfigure}[b]{0.24\textwidth}
        \includegraphics[width=\textwidth]{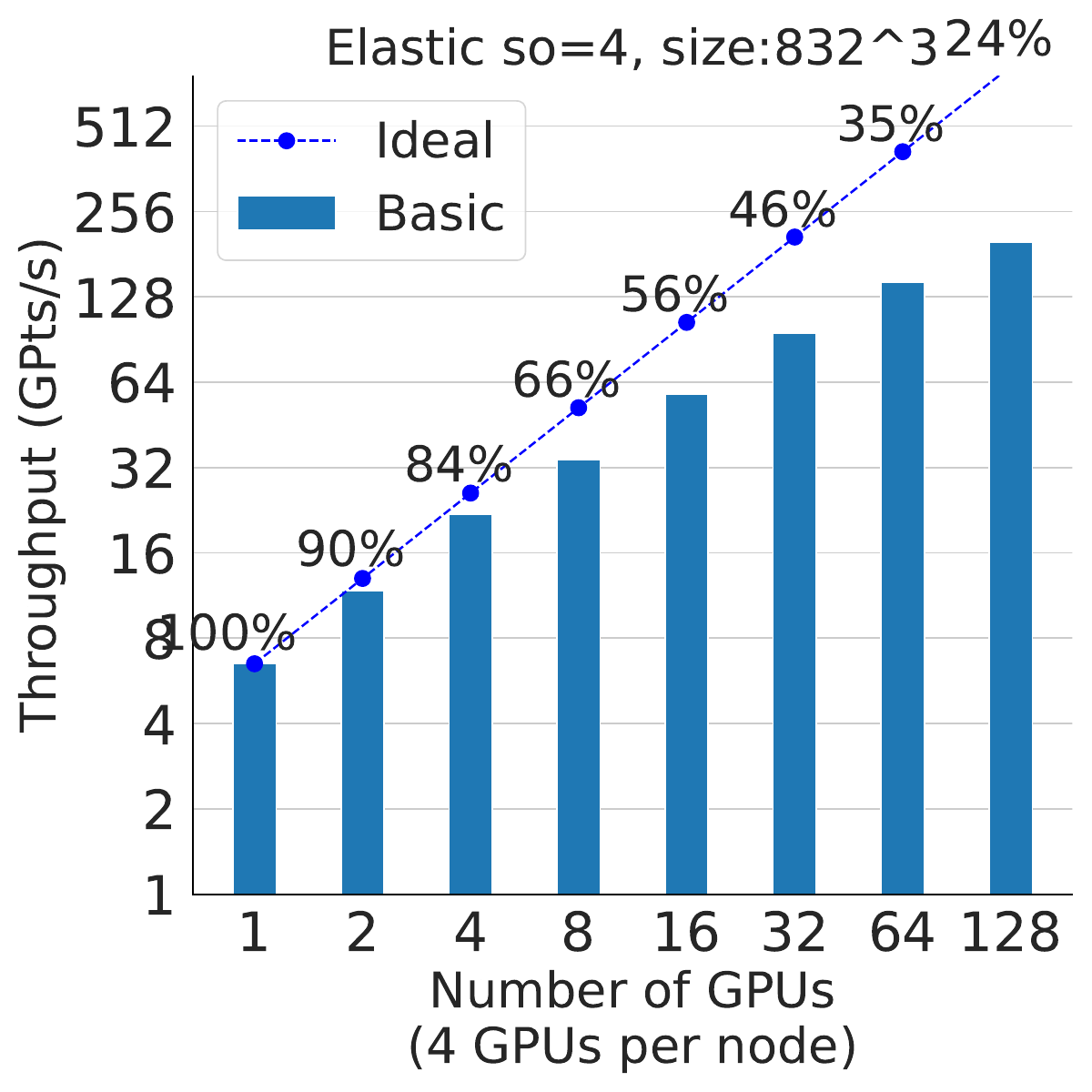}
        \caption{SDO 4}
        \label{fig:el_so04_gpu_platform-a}
	\end{subfigure}
	\begin{subfigure}[b]{0.24\textwidth}
        \includegraphics[width=\textwidth]{figures/Elastic_throughput_Basic_so-08GPU_SH.pdf}
        \caption{SDO 8}
        \label{fig:el_so08_gpu_platform-a}
        \end{subfigure}
	\begin{subfigure}[b]{0.24\textwidth}
        \includegraphics[width=\textwidth]{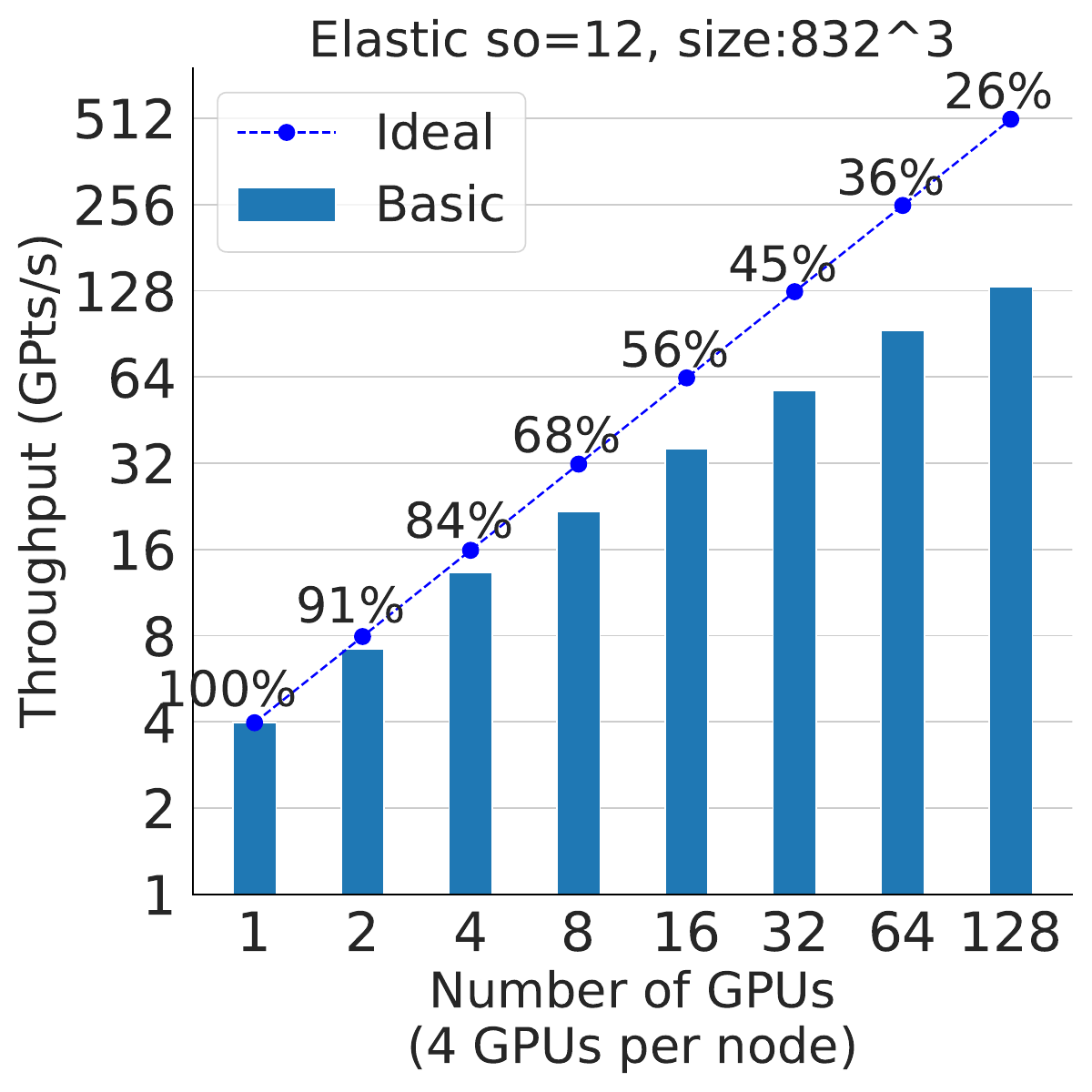}
        \caption{SDO 12}
        \label{fig:el_so12_gpu_platform-a}
	\end{subfigure}
	\begin{subfigure}[b]{0.24\textwidth}
        \includegraphics[width=\textwidth]{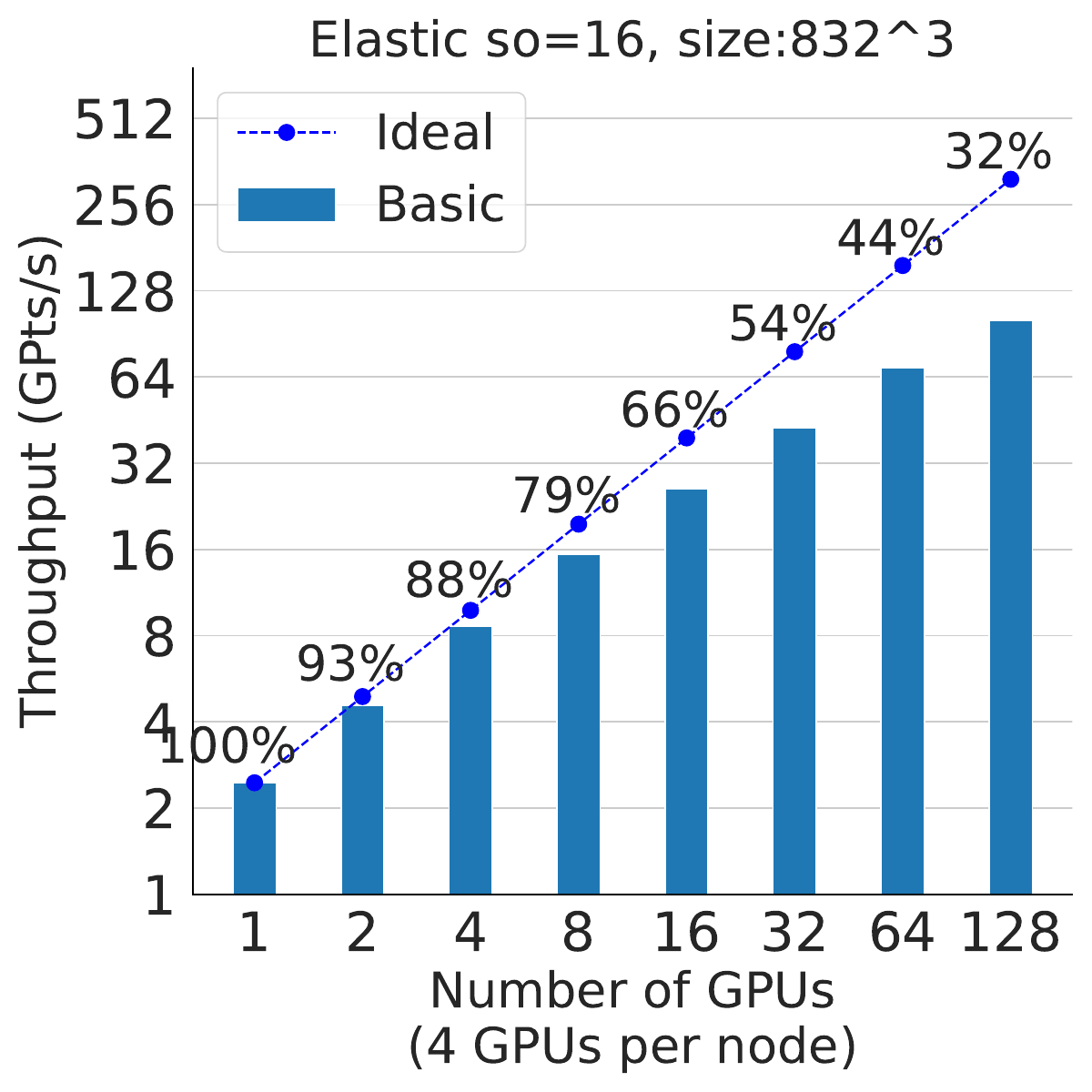}
        \caption{SDO 16}
        \label{fig:el_so16_gpu_platform-a}
	\end{subfigure}
        \caption{GPU Strong scaling for the \textbf{iso-elastic} kernel for SDOs 4, 8, 12 and 16.}
        \label{fig:el-stencil-gpu-scaling-a}
\end{figure}

\autoref{fig:el-stencil-gpu-scaling-a} and tables \ref{tab:el_gpu:so-04}, \ref{tab:el_gpu:so-08}, \ref{tab:el_gpu:so-12}, \ref{tab:el_gpu:so-16}
present the extended results for the elastic stencil kernel for various space orders.

\input{tables_raw/elastic_gpu_raw.tex}

\newpage

\begin{figure}[htbp]
    \centering
    \begin{subfigure}[b]{0.24\textwidth}
        \includegraphics[width=\textwidth]{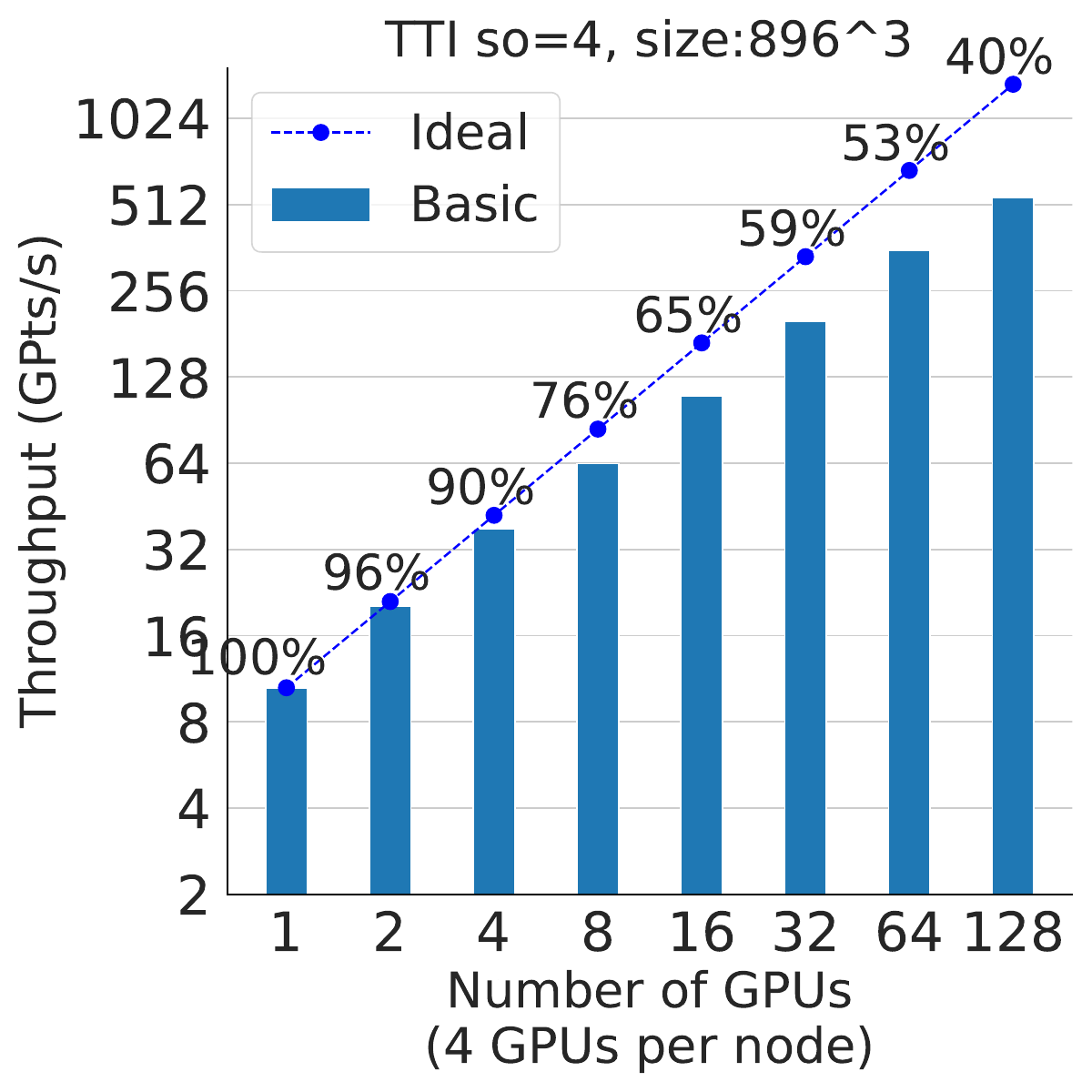}
        \caption{SDO 4}
        \label{fig:tti_so04_gpu_platform-a}
    \end{subfigure}
    \begin{subfigure}[b]{0.24\textwidth}
        \includegraphics[width=\textwidth]{figures/TTI_throughput_Basic_so-08GPU_SH.pdf}
        \caption{SDO 8}
        \label{fig:tti_so08_gpu_platform-a}
        \end{subfigure}
    \begin{subfigure}[b]{0.24\textwidth}
        \includegraphics[width=\textwidth]{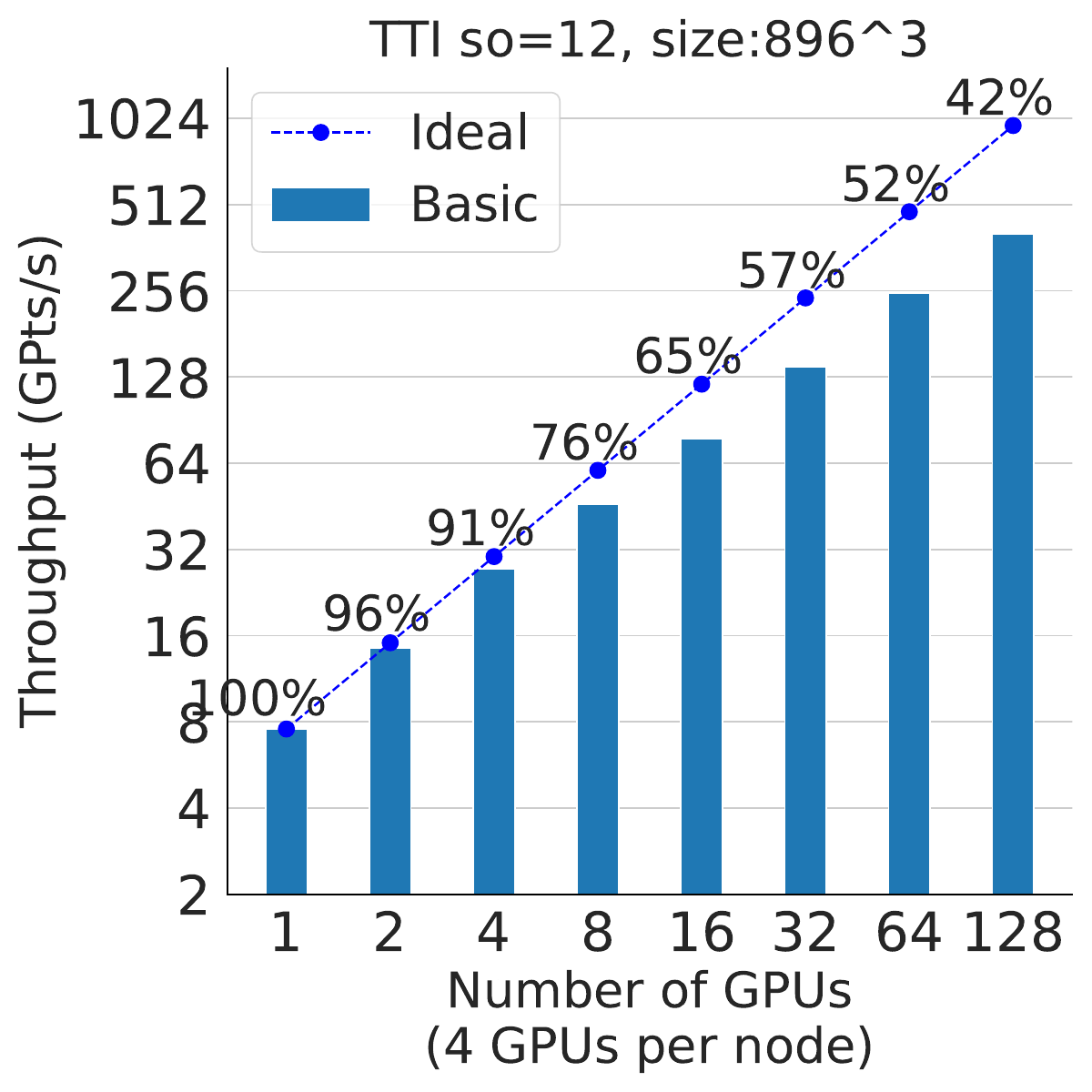}
        \caption{SDO 12}
        \label{fig:tti_so12_gpu_platform-a}
    \end{subfigure}
    \begin{subfigure}[b]{0.24\textwidth}
        \includegraphics[width=\textwidth]{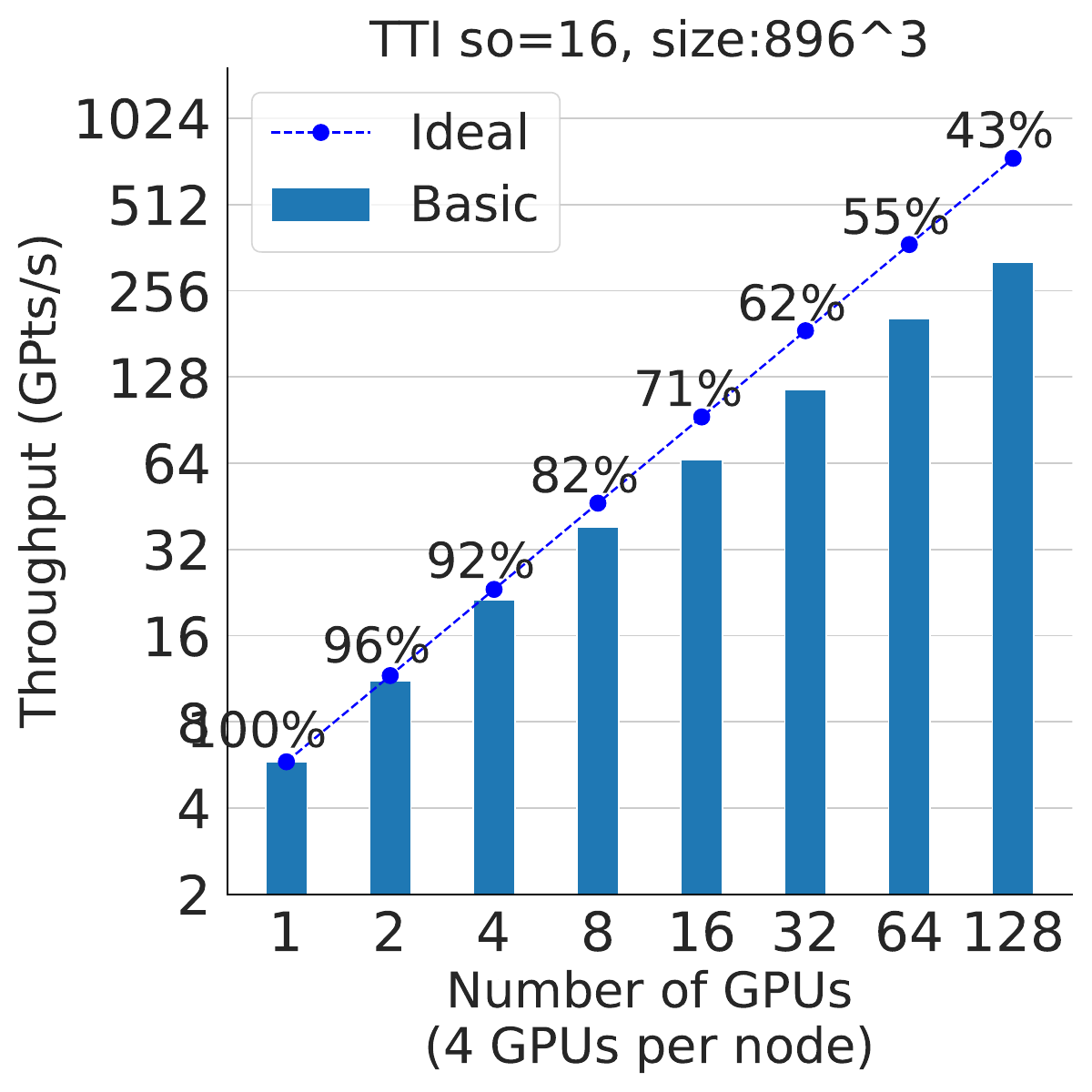}
        \caption{SDO 16}
        \label{fig:tti_so16_gpu_platform-a}
    \end{subfigure}
        \caption{GPU Strong scaling for the \textbf{TTI} kernel for SDOs 4, 8, 12 and 16.}
        \label{fig:tti-stencil-gpu-scaling-a}
\end{figure}

\autoref{fig:tti-stencil-gpu-scaling-a} and tables \ref{tab:tti_gpu:so-04}, \ref{tab:tti_gpu:so-08}, \ref{tab:tti_gpu:so-12}, \ref{tab:tti_gpu:so-16}
present the extended results for the TTI stencil kernel for various space orders.

\input{tables_raw/tti_gpu_raw.tex}

\begin{figure}
    \centering
    \begin{subfigure}[b]{0.24\textwidth}
        \includegraphics[width=\textwidth]{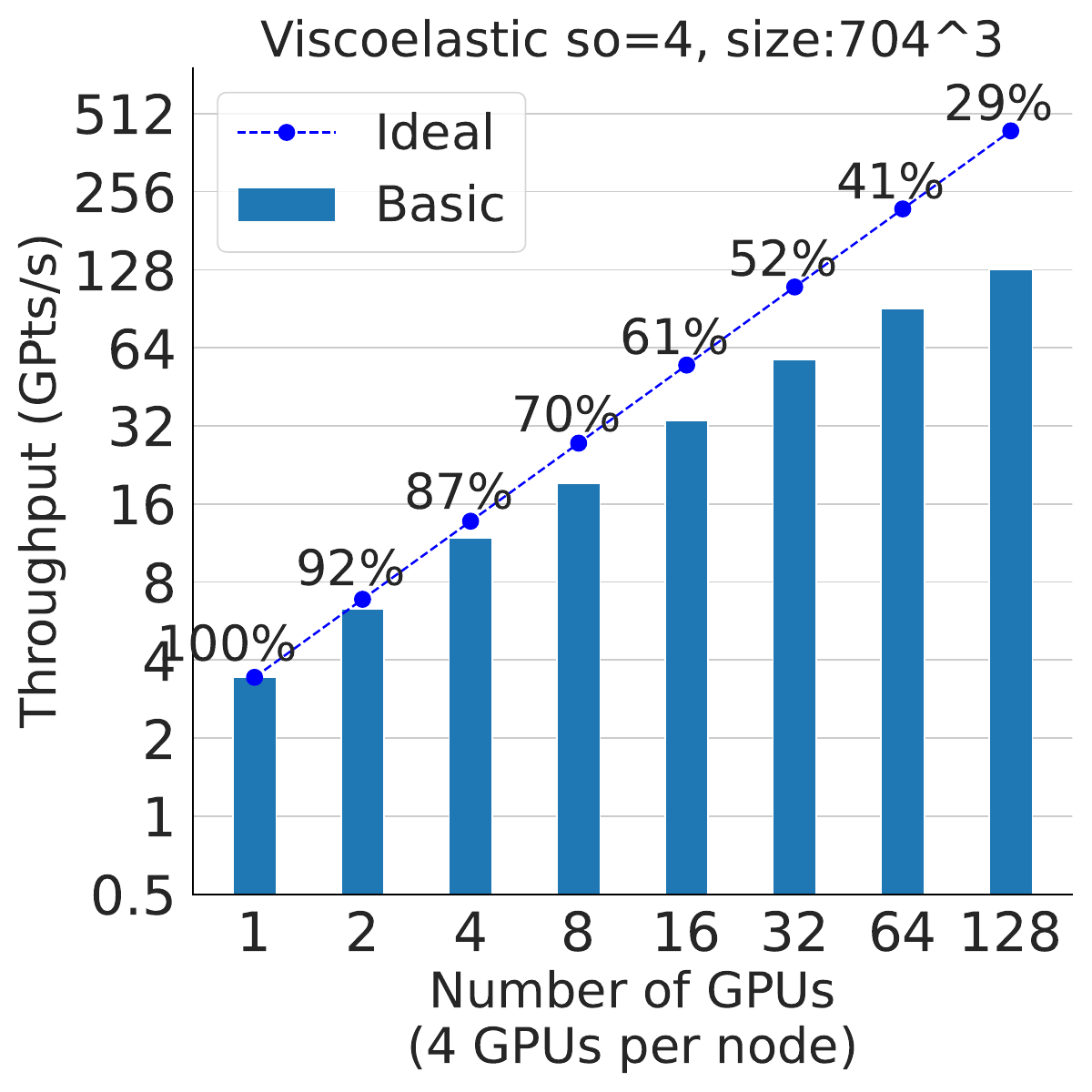}
        \caption{SDO 4}
        \label{fig:vel_so04_gpu_platform-a}
    \end{subfigure}
    \begin{subfigure}[b]{0.24\textwidth}
        \includegraphics[width=\textwidth]{figures/Viscoelastic_throughput_Basic_so-08GPU_SH.pdf}
        \caption{SDO 8}
        \label{fig:vel_so08_gpu_platform-a}
        \end{subfigure}
    \begin{subfigure}[b]{0.24\textwidth}
        \includegraphics[width=\textwidth]{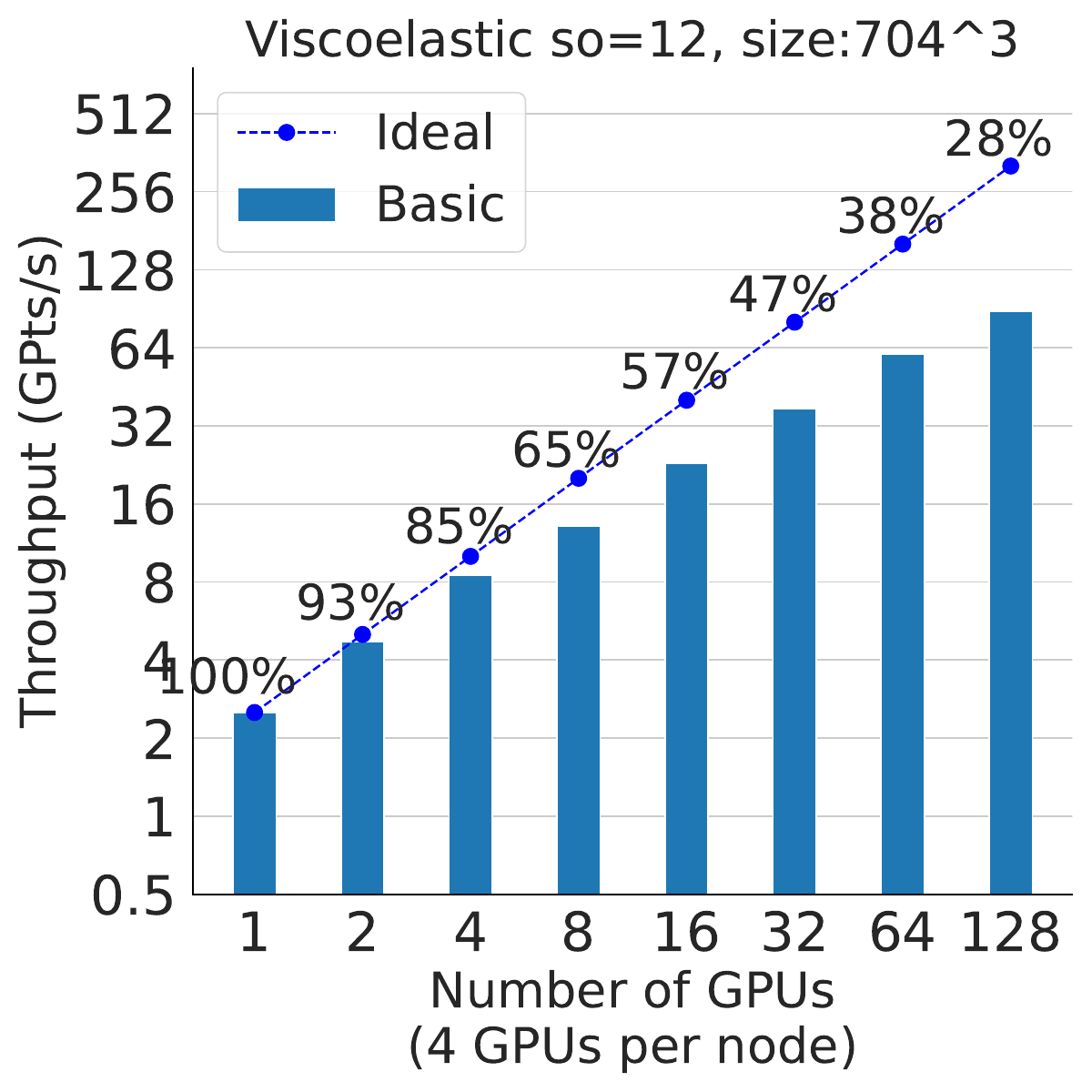}
        \caption{SDO 12}
        \label{fig:vel_so12_gpu_platform-a}
    \end{subfigure}
    \begin{subfigure}[b]{0.24\textwidth}
        \includegraphics[width=\textwidth]{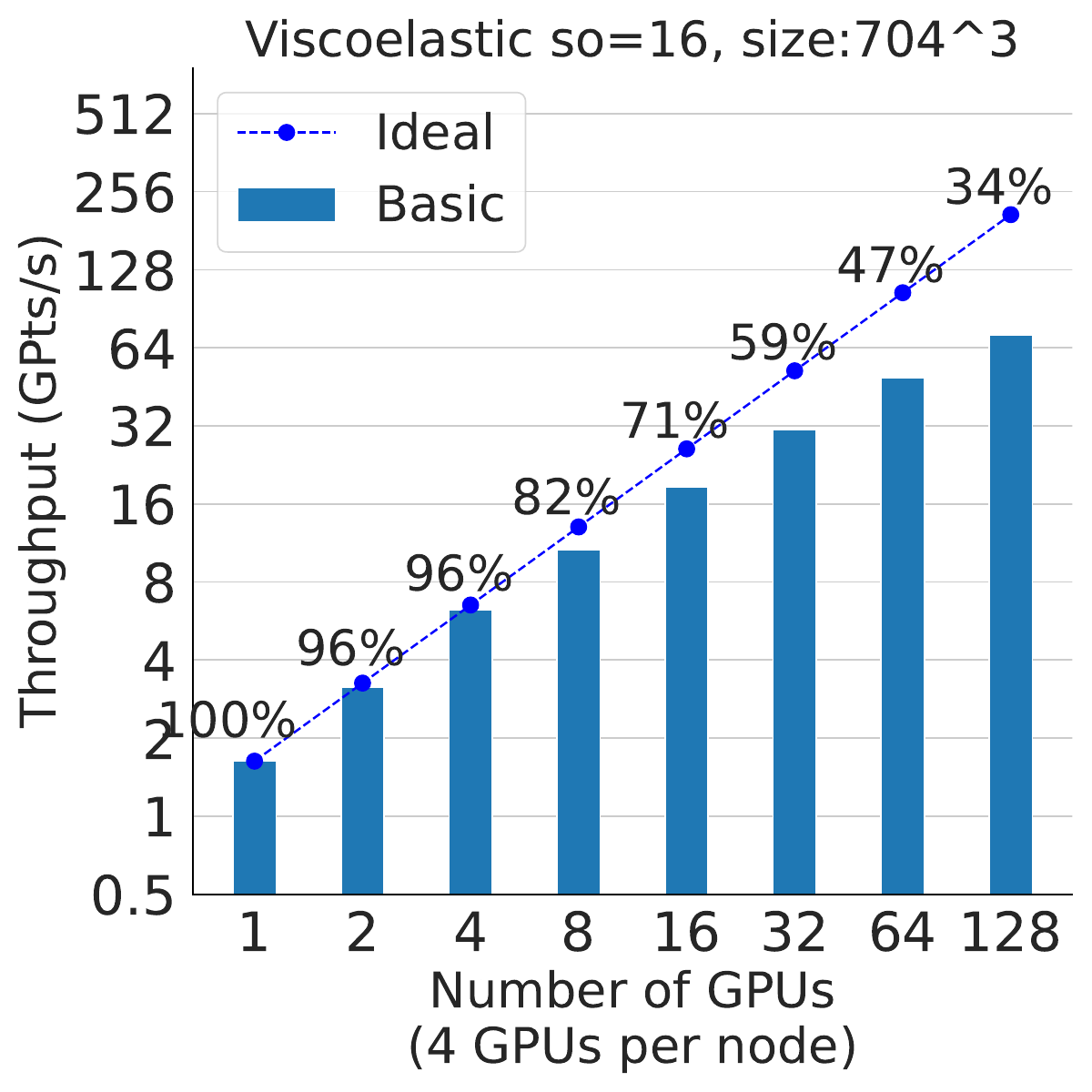}
        \caption{SDO 16}
        \label{fig:vel_so16_gpu_platform-a}
    \end{subfigure}
        \caption{GPU Strong scaling for the \textbf{viscoelastic} kernel for SDOs 4, 8, 12 and 16.}
        \label{fig:vel-stencil-gpu-scaling-a}
\end{figure}

\autoref{fig:vel-stencil-gpu-scaling-a} and tables \ref{tab:vel_gpu:so-04}, \ref{tab:vel_gpu:so-08}, \ref{tab:vel_gpu:so-12}, \ref{tab:vel_gpu:so-16}
present the extended results for the viscoelastic stencil kernel for various space orders.

\input{tables_raw/v_elastic_gpu_raw.tex}

\begin{figure}[!htbp]
        \centering
        \includegraphics[width=0.42\textwidth]{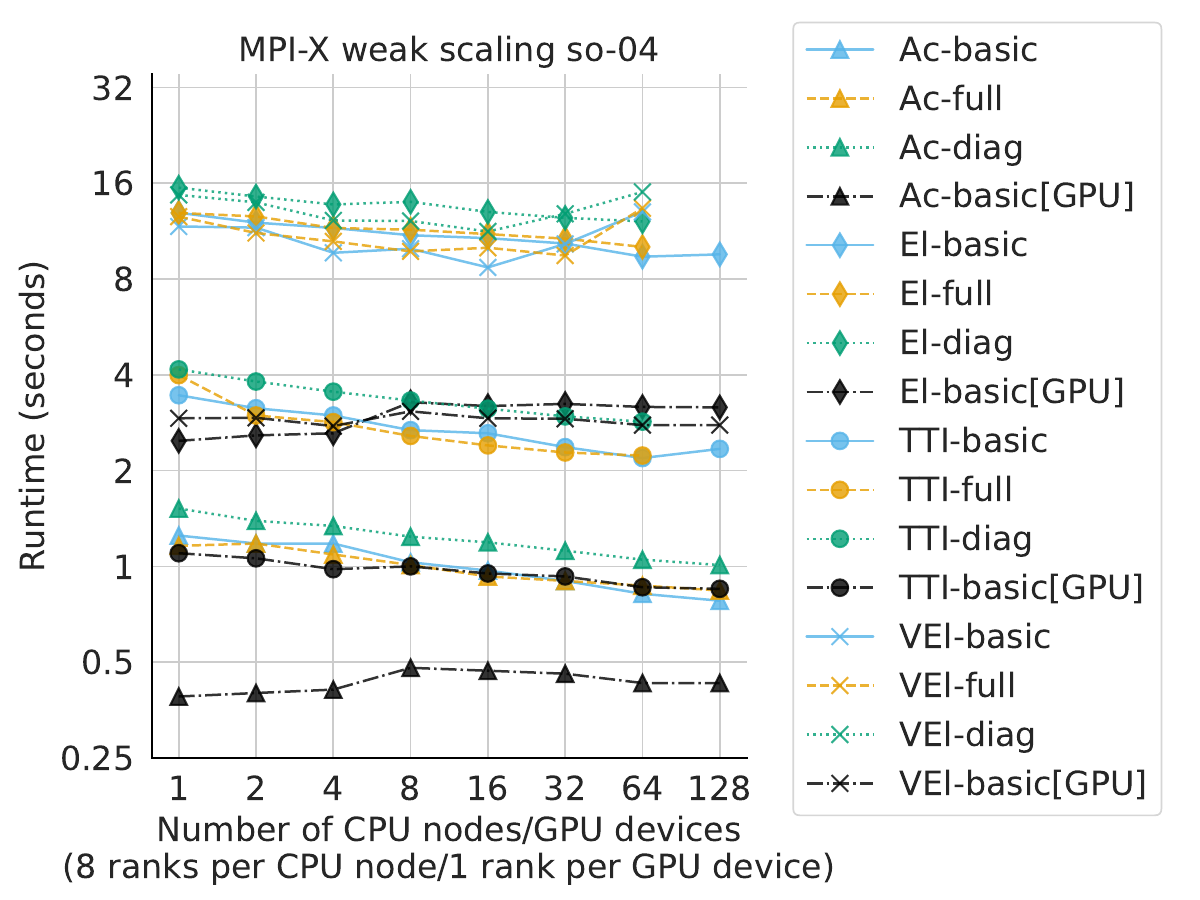}
        \caption{MPI-X Weak scaling runtime for SDO 4}
        \label{fig:weak_scaling-so4}
\end{figure}

\begin{figure}[!htbp]
        \centering
        \includegraphics[width=0.42\textwidth]{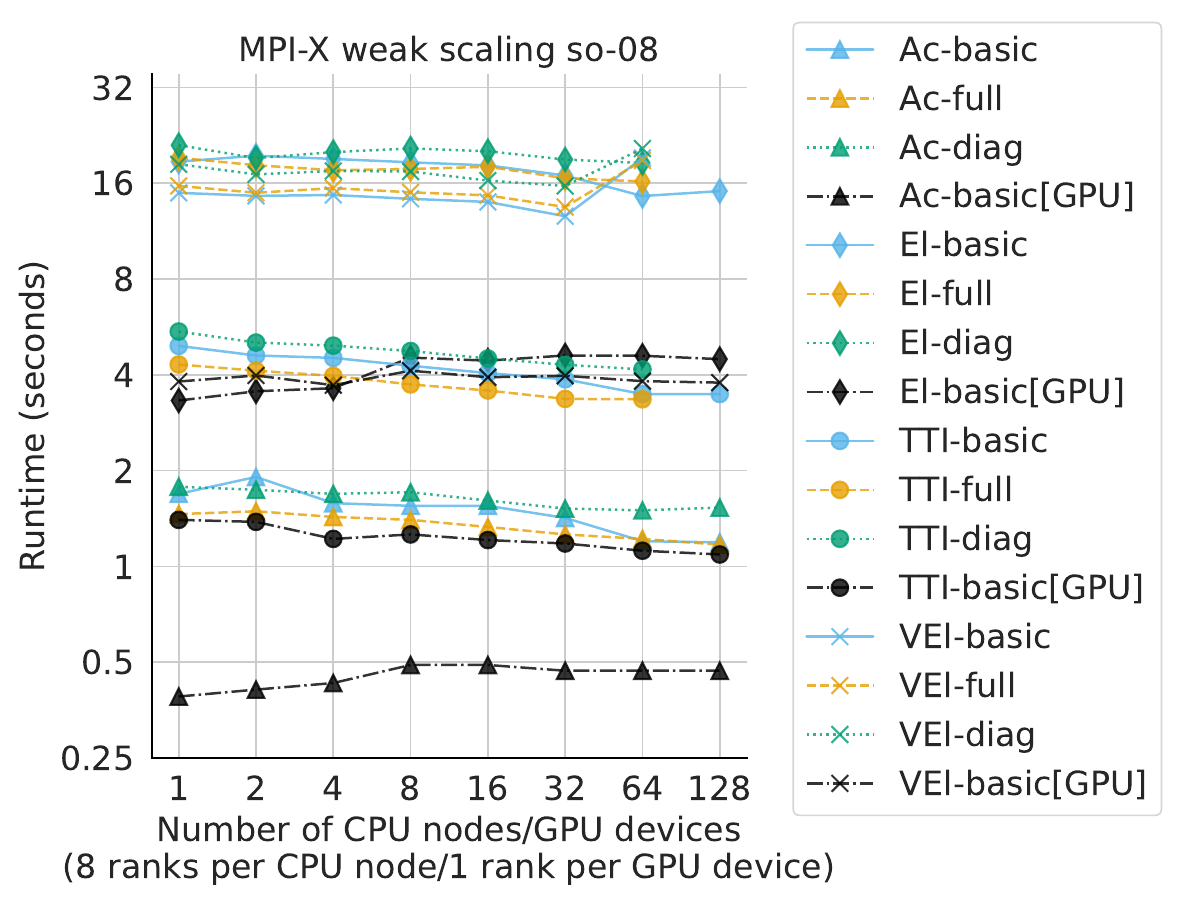}
        \caption{MPI-X Weak scaling runtime for SDO 8}
        \label{fig:weak_scaling-so8}
\end{figure}

\begin{figure}[!htbp]
        \centering
        \includegraphics[width=0.42\textwidth]{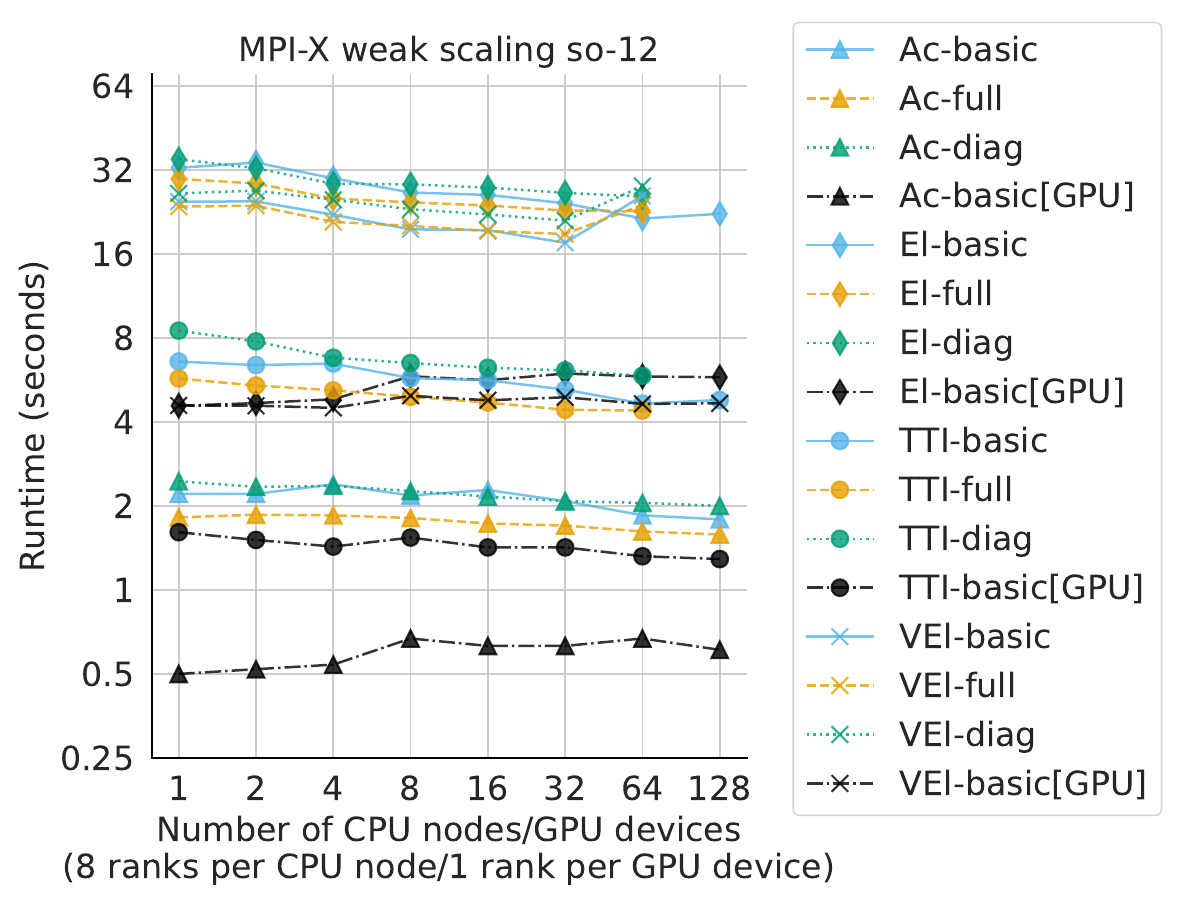}
        \caption{MPI-X Weak scaling runtime for SDO 12}
        \label{fig:weak_scaling-so12}
\end{figure}

\begin{figure}[!htbp]
        \centering
        \includegraphics[width=0.42\textwidth]{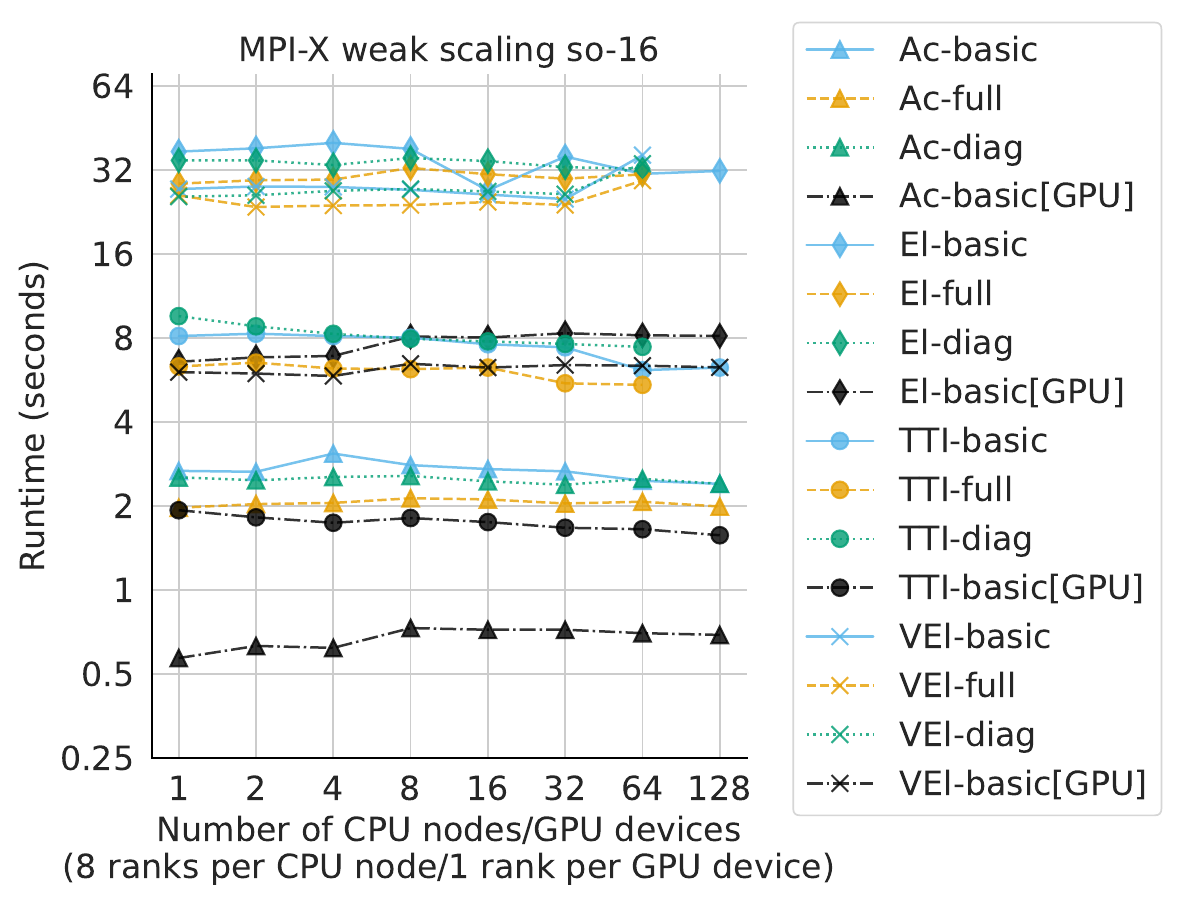}
        \caption{MPI-X Weak scaling runtime for SDO 16}
        \label{fig:weak_scaling-so16}
\end{figure}

\autoref{fig:weak_scaling-so4}, \autoref{fig:weak_scaling-so8}, \autoref{fig:weak_scaling-so12} and \autoref{fig:weak_scaling-so16}
present the weak scaling extended results for the stencil kernels for various space orders.

%% file: tables_raw/acoustic_raw.tex
\begin{table}[htbp]
\caption{Acoustic so-04 kernel throughput (Gpts/s)}
\label{tab:ac:so-04}
\begin{tabular}{m{0.5cm}m{0.5cm}m{0.5cm}m{0.6cm}m{0.6cm}m{0.6cm}m{0.6cm}m{0.6cm}m{0.6cm}m{0.6cm}}
\toprule
 & 1 & 2 & 4 & 8 & 16 & 32 & 64 & 128 \\
\midrule
Basic & 13.4 & \cellcolor[HTML]{FFFF00} \color{red} \textit{\textbf{25.0}} & \cellcolor[HTML]{FFFF00} \color{red} \textit{\textbf{48.0}} & 90.7 & 170.1 & 292.5 & \cellcolor[HTML]{FFFF00} \color{green} \textit{\textbf{655.4}} & \cellcolor[HTML]{FFFF00} \color{green} \textit{\textbf{1415.5}} \\
Diag & \cellcolor[HTML]{FFFF00} \color{red} \textit{\textbf{13.3}} & 25.7 & \cellcolor[HTML]{FFFF00} \color{green} \textit{\textbf{49.8}} & \cellcolor[HTML]{FFFF00} \color{green} \textit{\textbf{91.0}} & \cellcolor[HTML]{FFFF00} \color{red} \textit{\textbf{169.3}} & \cellcolor[HTML]{FFFF00} \color{red} \textit{\textbf{287.7}} & \cellcolor[HTML]{FFFF00} \color{red} \textit{\textbf{544.4}} & \cellcolor[HTML]{FFFF00} \color{red} \textit{\textbf{991.6}} \\
Full & \cellcolor[HTML]{FFFF00} \color{green} \textit{\textbf{13.9}} & \cellcolor[HTML]{FFFF00} \color{green} \textit{\textbf{25.8}} & 49.3 & \cellcolor[HTML]{FFFF00} \color{red} \textit{\textbf{88.0}} & \cellcolor[HTML]{FFFF00} \color{green} \textit{\textbf{180.0}} & \cellcolor[HTML]{FFFF00} \color{green} \textit{\textbf{299.9}} & 589.8 & 1011.1 \\
\bottomrule
\end{tabular}
\end{table}
\begin{table}[htbp]
\caption{Acoustic so-08 kernel throughput (Gpts/s)}
\label{tab:ac:so-08}
\begin{tabular}{m{0.5cm}m{0.5cm}m{0.5cm}m{0.6cm}m{0.6cm}m{0.6cm}m{0.6cm}m{0.6cm}m{0.6cm}m{0.6cm}}
\toprule
 & 1 & 2 & 4 & 8 & 16 & 32 & 64 & 128 \\
\midrule
Basic & \cellcolor[HTML]{FFFF00} \color{red} \textit{\textbf{12.4}} & \cellcolor[HTML]{FFFF00} \color{red} \textit{\textbf{22.3}} & \cellcolor[HTML]{FFFF00} \color{red} \textit{\textbf{42.4}} & 79.6 & 143.2 & 237.8 & \cellcolor[HTML]{FFFF00} \color{green} \textit{\textbf{497.5}} & \cellcolor[HTML]{FFFF00} \color{green} \textit{\textbf{1050.3}} \\
Diag & \cellcolor[HTML]{FFFF00} \color{green} \textit{\textbf{12.9}} & \cellcolor[HTML]{FFFF00} \color{green} \textit{\textbf{23.9}} & \cellcolor[HTML]{FFFF00} \color{green} \textit{\textbf{45.3}} & \cellcolor[HTML]{FFFF00} \color{green} \textit{\textbf{83.3}} & \cellcolor[HTML]{FFFF00} \color{green} \textit{\textbf{149.4}} & \cellcolor[HTML]{FFFF00} \color{green} \textit{\textbf{248.8}} & 461.1 & 859.3 \\
Full & 12.8 & 23.6 & 43.0 & \cellcolor[HTML]{FFFF00} \color{red} \textit{\textbf{75.5}} & \cellcolor[HTML]{FFFF00} \color{red} \textit{\textbf{137.0}} & \cellcolor[HTML]{FFFF00} \color{red} \textit{\textbf{233.4}} & \cellcolor[HTML]{FFFF00} \color{red} \textit{\textbf{402.2}} & \cellcolor[HTML]{FFFF00} \color{red} \textit{\textbf{759.2}} \\
\bottomrule
\end{tabular}
\end{table}
\begin{table}[htbp]
\caption{Acoustic so-12 kernel throughput (Gpts/s)}
\label{tab:ac:so-12}
\begin{tabular}{m{0.5cm}m{0.5cm}m{0.5cm}m{0.6cm}m{0.6cm}m{0.6cm}m{0.6cm}m{0.6cm}m{0.6cm}m{0.6cm}}
\toprule
 & 1 & 2 & 4 & 8 & 16 & 32 & 64 & 128 \\
\midrule
Basic & \cellcolor[HTML]{FFFF00} \color{red} \textit{\textbf{11.5}} & \cellcolor[HTML]{FFFF00} \color{red} \textit{\textbf{20.1}} & 37.3 & \cellcolor[HTML]{FFFF00} \color{red} \textit{\textbf{62.5}} & \cellcolor[HTML]{FFFF00} \color{red} \textit{\textbf{111.5}} & 198.1 & \cellcolor[HTML]{FFFF00} \color{green} \textit{\textbf{402.3}} & \cellcolor[HTML]{FFFF00} \color{green} \textit{\textbf{769.2}} \\
Diag & \cellcolor[HTML]{FFFF00} \color{green} \textit{\textbf{12.2}} & \cellcolor[HTML]{FFFF00} \color{green} \textit{\textbf{22.5}} & \cellcolor[HTML]{FFFF00} \color{green} \textit{\textbf{41.5}} & \cellcolor[HTML]{FFFF00} \color{green} \textit{\textbf{69.3}} & \cellcolor[HTML]{FFFF00} \color{green} \textit{\textbf{126.3}} & \cellcolor[HTML]{FFFF00} \color{green} \textit{\textbf{221.7}} & 371.6 & 686.6 \\
Full & 11.8 & 20.6 & \cellcolor[HTML]{FFFF00} \color{red} \textit{\textbf{37.2}} & 66.0 & 112.1 & \cellcolor[HTML]{FFFF00} \color{red} \textit{\textbf{175.0}} & \cellcolor[HTML]{FFFF00} \color{red} \textit{\textbf{307.3}} & \cellcolor[HTML]{FFFF00} \color{red} \textit{\textbf{534.5}} \\
\bottomrule
\end{tabular}
\end{table}
\begin{table}[htbp]
\caption{Acoustic so-16 kernel throughput (Gpts/s)}
\label{tab:ac:so-16}
\begin{tabular}{m{0.5cm}m{0.5cm}m{0.5cm}m{0.6cm}m{0.6cm}m{0.6cm}m{0.6cm}m{0.6cm}m{0.6cm}m{0.6cm}}
\toprule
 & 1 & 2 & 4 & 8 & 16 & 32 & 64 & 128 \\
\midrule
Basic & 10.8 & \cellcolor[HTML]{FFFF00} \color{red} \textit{\textbf{18.3}} & \cellcolor[HTML]{FFFF00} \color{red} \textit{\textbf{31.8}} & \cellcolor[HTML]{FFFF00} \color{red} \textit{\textbf{54.6}} & 101.4 & 165.5 & \cellcolor[HTML]{FFFF00} \color{green} \textit{\textbf{329.7}} & \cellcolor[HTML]{FFFF00} \color{green} \textit{\textbf{599.4}} \\
Diag & \cellcolor[HTML]{FFFF00} \color{green} \textit{\textbf{11.4}} & \cellcolor[HTML]{FFFF00} \color{green} \textit{\textbf{20.6}} & \cellcolor[HTML]{FFFF00} \color{green} \textit{\textbf{37.8}} & \cellcolor[HTML]{FFFF00} \color{green} \textit{\textbf{67.1}} & \cellcolor[HTML]{FFFF00} \color{green} \textit{\textbf{114.0}} & \cellcolor[HTML]{FFFF00} \color{green} \textit{\textbf{194.9}} & 326.9 & 557.2 \\
Full & \cellcolor[HTML]{FFFF00} \color{red} \textit{\textbf{10.7}} & 19.1 & 34.2 & 60.8 & \cellcolor[HTML]{FFFF00} \color{red} \textit{\textbf{99.7}} & \cellcolor[HTML]{FFFF00} \color{red} \textit{\textbf{158.9}} & \cellcolor[HTML]{FFFF00} \color{red} \textit{\textbf{253.6}} & \cellcolor[HTML]{FFFF00} \color{red} \textit{\textbf{465.7}} \\
\bottomrule
\end{tabular}
\end{table}

%% file: tables_raw/elastic_raw.tex
\begin{table}[htbp]
\caption{Elastic so-04 kernel throughput (Gpts/s)}
\label{tab:el:so-04}
\begin{tabular}{m{0.5cm}m{0.5cm}m{0.5cm}m{0.6cm}m{0.6cm}m{0.6cm}m{0.6cm}m{0.6cm}m{0.6cm}m{0.6cm}}
\toprule
 & 1 & 2 & 4 & 8 & 16 & 32 & 64 & 128 \\
\midrule
Basic & \cellcolor[HTML]{FFFF00} \color{red} \textit{\textbf{1.8}} & \cellcolor[HTML]{FFFF00} \color{red} \textit{\textbf{3.3}} & 6.1 & 12.0 & 22.0 & 40.5 & 74.6 & 123.0 \\
Diag & \cellcolor[HTML]{FFFF00} \color{green} \textit{\textbf{1.9}} & \cellcolor[HTML]{FFFF00} \color{green} \textit{\textbf{3.6}} & \cellcolor[HTML]{FFFF00} \color{green} \textit{\textbf{6.8}} & \cellcolor[HTML]{FFFF00} \color{green} \textit{\textbf{12.7}} & \cellcolor[HTML]{FFFF00} \color{green} \textit{\textbf{23.6}} & \cellcolor[HTML]{FFFF00} \color{green} \textit{\textbf{45.0}} & \cellcolor[HTML]{FFFF00} \color{green} \textit{\textbf{77.5}} & \cellcolor[HTML]{FFFF00} \color{green} \textit{\textbf{134.6}} \\
Full & 1.9 & 3.4 & \cellcolor[HTML]{FFFF00} \color{red} \textit{\textbf{6.0}} & \cellcolor[HTML]{FFFF00} \color{red} \textit{\textbf{11.8}} & \cellcolor[HTML]{FFFF00} \color{red} \textit{\textbf{21.4}} & \cellcolor[HTML]{FFFF00} \color{red} \textit{\textbf{37.7}} & \cellcolor[HTML]{FFFF00} \color{red} \textit{\textbf{66.7}} & \cellcolor[HTML]{FFFF00} \color{red} \textit{\textbf{106.9}} \\
\bottomrule
\end{tabular}
\end{table}
\begin{table}[htbp]
\caption{Elastic so-08 kernel throughput (Gpts/s)}
\label{tab:el:so-08}
\begin{tabular}{m{0.5cm}m{0.5cm}m{0.5cm}m{0.6cm}m{0.6cm}m{0.6cm}m{0.6cm}m{0.6cm}m{0.6cm}m{0.6cm}}
\toprule
 & 1 & 2 & 4 & 8 & 16 & 32 & 64 & 128 \\
\midrule
Basic & \cellcolor[HTML]{FFFF00} \color{red} \textit{\textbf{1.7}} & \cellcolor[HTML]{FFFF00} \color{red} \textit{\textbf{3.0}} & \cellcolor[HTML]{FFFF00} \color{red} \textit{\textbf{5.0}} & 10.3 & 18.6 & 32.8 & 61.3 & 97.3 \\
Diag & \cellcolor[HTML]{FFFF00} \color{green} \textit{\textbf{1.8}} & \cellcolor[HTML]{FFFF00} \color{green} \textit{\textbf{3.3}} & \cellcolor[HTML]{FFFF00} \color{green} \textit{\textbf{6.1}} & \cellcolor[HTML]{FFFF00} \color{green} \textit{\textbf{11.2}} & \cellcolor[HTML]{FFFF00} \color{green} \textit{\textbf{20.5}} & \cellcolor[HTML]{FFFF00} \color{green} \textit{\textbf{37.4}} & \cellcolor[HTML]{FFFF00} \color{green} \textit{\textbf{65.0}} & \cellcolor[HTML]{FFFF00} \color{green} \textit{\textbf{106.3}} \\
Full & 1.7 & 3.1 & 5.5 & \cellcolor[HTML]{FFFF00} \color{red} \textit{\textbf{9.8}} & \cellcolor[HTML]{FFFF00} \color{red} \textit{\textbf{17.0}} & \cellcolor[HTML]{FFFF00} \color{red} \textit{\textbf{29.6}} & \cellcolor[HTML]{FFFF00} \color{red} \textit{\textbf{51.4}} & \cellcolor[HTML]{FFFF00} \color{red} \textit{\textbf{79.3}} \\
\bottomrule
\end{tabular}
\end{table}
\begin{table}[htbp]
\caption{Elastic so-12 kernel throughput (Gpts/s)}
\label{tab:el:so-12}
\begin{tabular}{m{0.5cm}m{0.5cm}m{0.5cm}m{0.6cm}m{0.6cm}m{0.6cm}m{0.6cm}m{0.6cm}m{0.6cm}m{0.6cm}}
\toprule
 & 1 & 2 & 4 & 8 & 16 & 32 & 64 & 128 \\
\midrule
Basic & \cellcolor[HTML]{FFFF00} \color{green} \textit{\textbf{1.5}} & 2.7 & \cellcolor[HTML]{FFFF00} \color{red} \textit{\textbf{4.2}} & 8.8 & 15.8 & \cellcolor[HTML]{FFFF00} \color{red} \textit{\textbf{22.2}} & 50.9 & 80.0 \\
Diag & 1.5 & \cellcolor[HTML]{FFFF00} \color{green} \textit{\textbf{2.7}} & \cellcolor[HTML]{FFFF00} \color{green} \textit{\textbf{5.2}} & \cellcolor[HTML]{FFFF00} \color{green} \textit{\textbf{9.4}} & \cellcolor[HTML]{FFFF00} \color{green} \textit{\textbf{17.1}} & \cellcolor[HTML]{FFFF00} \color{green} \textit{\textbf{30.9}} & \cellcolor[HTML]{FFFF00} \color{green} \textit{\textbf{53.4}} & \cellcolor[HTML]{FFFF00} \color{green} \textit{\textbf{90.8}} \\
Full & \cellcolor[HTML]{FFFF00} \color{red} \textit{\textbf{1.4}} & \cellcolor[HTML]{FFFF00} \color{red} \textit{\textbf{2.5}} & 4.9 & \cellcolor[HTML]{FFFF00} \color{red} \textit{\textbf{8.4}} & \cellcolor[HTML]{FFFF00} \color{red} \textit{\textbf{14.1}} & 25.1 & \cellcolor[HTML]{FFFF00} \color{red} \textit{\textbf{41.0}} & \cellcolor[HTML]{FFFF00} \color{red} \textit{\textbf{65.7}} \\
\bottomrule
\end{tabular}
\end{table}
\begin{table}[htbp]
\caption{Elastic so-16 kernel throughput (Gpts/s)}
\label{tab:el:so-16}
\begin{tabular}{m{0.5cm}m{0.5cm}m{0.5cm}m{0.6cm}m{0.6cm}m{0.6cm}m{0.6cm}m{0.6cm}m{0.6cm}m{0.6cm}}
\toprule
 & 1 & 2 & 4 & 8 & 16 & 32 & 64 & 128 \\
\midrule
Basic & \cellcolor[HTML]{FFFF00} \color{red} \textit{\textbf{1.0}} & \cellcolor[HTML]{FFFF00} \color{red} \textit{\textbf{2.0}} & \cellcolor[HTML]{FFFF00} \color{red} \textit{\textbf{3.0}} & 6.9 & 12.4 & 20.7 & 39.9 & 62.3 \\
Diag & \cellcolor[HTML]{FFFF00} \color{green} \textit{\textbf{1.2}} & \cellcolor[HTML]{FFFF00} \color{green} \textit{\textbf{2.3}} & \cellcolor[HTML]{FFFF00} \color{green} \textit{\textbf{3.9}} & \cellcolor[HTML]{FFFF00} \color{green} \textit{\textbf{7.8}} & \cellcolor[HTML]{FFFF00} \color{green} \textit{\textbf{14.2}} & \cellcolor[HTML]{FFFF00} \color{green} \textit{\textbf{25.3}} & \cellcolor[HTML]{FFFF00} \color{green} \textit{\textbf{43.7}} & \cellcolor[HTML]{FFFF00} \color{green} \textit{\textbf{71.5}} \\
Full & 1.2 & 2.1 & 3.8 & \cellcolor[HTML]{FFFF00} \color{red} \textit{\textbf{6.7}} & \cellcolor[HTML]{FFFF00} \color{red} \textit{\textbf{12.0}} & \cellcolor[HTML]{FFFF00} \color{red} \textit{\textbf{19.9}} & \cellcolor[HTML]{FFFF00} \color{red} \textit{\textbf{35.2}} & \cellcolor[HTML]{FFFF00} \color{red} \textit{\textbf{55.2}} \\
\bottomrule
\end{tabular}
\end{table}

%% file: tables_raw/tti_raw.tex
\begin{table}[htbp]
\caption{TTI so-04 kernel throughput (Gpts/s)}
\label{tab:tti:so-04}
\begin{tabular}{m{0.5cm}m{0.5cm}m{0.5cm}m{0.6cm}m{0.6cm}m{0.6cm}m{0.6cm}m{0.6cm}m{0.6cm}m{0.6cm}}
\toprule
 & 1 & 2 & 4 & 8 & 16 & 32 & 64 & 128 \\
\midrule
Basic & 4.3 & 8.2 & 16.2 & 32.8 & 62.7 & \cellcolor[HTML]{FFFF00} \color{green} \textit{\textbf{118.4}} & \cellcolor[HTML]{FFFF00} \color{green} \textit{\textbf{228.2}} & \cellcolor[HTML]{FFFF00} \color{green} \textit{\textbf{388.7}} \\
Diag & \cellcolor[HTML]{FFFF00} \color{green} \textit{\textbf{4.4}} & \cellcolor[HTML]{FFFF00} \color{green} \textit{\textbf{8.7}} & \cellcolor[HTML]{FFFF00} \color{green} \textit{\textbf{17.1}} & \cellcolor[HTML]{FFFF00} \color{green} \textit{\textbf{32.8}} & \cellcolor[HTML]{FFFF00} \color{green} \textit{\textbf{63.0}} & 117.9 & 209.9 & 361.9 \\
Full & \cellcolor[HTML]{FFFF00} \color{red} \textit{\textbf{4.2}} & \cellcolor[HTML]{FFFF00} \color{red} \textit{\textbf{8.2}} & \cellcolor[HTML]{FFFF00} \color{red} \textit{\textbf{15.9}} & \cellcolor[HTML]{FFFF00} \color{red} \textit{\textbf{32.3}} & \cellcolor[HTML]{FFFF00} \color{red} \textit{\textbf{60.9}} & \cellcolor[HTML]{FFFF00} \color{red} \textit{\textbf{111.7}} & \cellcolor[HTML]{FFFF00} \color{red} \textit{\textbf{189.7}} & \cellcolor[HTML]{FFFF00} \color{red} \textit{\textbf{321.3}} \\
\bottomrule
\end{tabular}
\end{table}
\begin{table}[htbp]
\caption{TTI so-08 kernel throughput (Gpts/s)}
\label{tab:tti:so-08}
\begin{tabular}{m{0.5cm}m{0.5cm}m{0.5cm}m{0.6cm}m{0.6cm}m{0.6cm}m{0.6cm}m{0.6cm}m{0.6cm}m{0.6cm}}
\toprule
 & 1 & 2 & 4 & 8 & 16 & 32 & 64 & 128 \\
\midrule
Basic & 3.5 & 6.4 & \cellcolor[HTML]{FFFF00} \color{red} \textit{\textbf{11.8}} & 26.9 & 51.0 & 90.7 & \cellcolor[HTML]{FFFF00} \color{green} \textit{\textbf{178.9}} & \cellcolor[HTML]{FFFF00} \color{green} \textit{\textbf{314.4}} \\
Diag & \cellcolor[HTML]{FFFF00} \color{green} \textit{\textbf{3.6}} & \cellcolor[HTML]{FFFF00} \color{green} \textit{\textbf{6.9}} & \cellcolor[HTML]{FFFF00} \color{green} \textit{\textbf{13.9}} & \cellcolor[HTML]{FFFF00} \color{green} \textit{\textbf{27.9}} & \cellcolor[HTML]{FFFF00} \color{green} \textit{\textbf{53.6}} & \cellcolor[HTML]{FFFF00} \color{green} \textit{\textbf{95.6}} & 176.1 & 303.1 \\
Full & \cellcolor[HTML]{FFFF00} \color{red} \textit{\textbf{3.3}} & \cellcolor[HTML]{FFFF00} \color{red} \textit{\textbf{6.3}} & 12.7 & \cellcolor[HTML]{FFFF00} \color{red} \textit{\textbf{24.4}} & \cellcolor[HTML]{FFFF00} \color{red} \textit{\textbf{47.0}} & \cellcolor[HTML]{FFFF00} \color{red} \textit{\textbf{84.7}} & \cellcolor[HTML]{FFFF00} \color{red} \textit{\textbf{143.2}} & \cellcolor[HTML]{FFFF00} \color{red} \textit{\textbf{238.6}} \\
\bottomrule
\end{tabular}
\end{table}
\begin{table}[htbp]
\caption{TTI so-12 kernel throughput (Gpts/s)}
\label{tab:tti:so-12}
\begin{tabular}{m{0.5cm}m{0.5cm}m{0.5cm}m{0.6cm}m{0.6cm}m{0.6cm}m{0.6cm}m{0.6cm}m{0.6cm}m{0.6cm}}
\toprule
 & 1 & 2 & 4 & 8 & 16 & 32 & 64 & 128 \\
\midrule
Basic & 2.7 & \cellcolor[HTML]{FFFF00} \color{red} \textit{\textbf{4.6}} & \cellcolor[HTML]{FFFF00} \color{red} \textit{\textbf{8.2}} & 20.2 & 38.6 & 73.0 & 141.7 & 235.2 \\
Diag & \cellcolor[HTML]{FFFF00} \color{red} \textit{\textbf{2.7}} & 5.2 & 9.3 & \cellcolor[HTML]{FFFF00} \color{green} \textit{\textbf{22.2}} & \cellcolor[HTML]{FFFF00} \color{green} \textit{\textbf{41.7}} & \cellcolor[HTML]{FFFF00} \color{green} \textit{\textbf{79.9}} & \cellcolor[HTML]{FFFF00} \color{green} \textit{\textbf{142.3}} & \cellcolor[HTML]{FFFF00} \color{green} \textit{\textbf{241.8}} \\
Full & \cellcolor[HTML]{FFFF00} \color{green} \textit{\textbf{2.8}} & \cellcolor[HTML]{FFFF00} \color{green} \textit{\textbf{5.3}} & \cellcolor[HTML]{FFFF00} \color{green} \textit{\textbf{9.8}} & \cellcolor[HTML]{FFFF00} \color{red} \textit{\textbf{18.5}} & \cellcolor[HTML]{FFFF00} \color{red} \textit{\textbf{37.1}} & \cellcolor[HTML]{FFFF00} \color{red} \textit{\textbf{66.6}} & \cellcolor[HTML]{FFFF00} \color{red} \textit{\textbf{111.6}} & \cellcolor[HTML]{FFFF00} \color{red} \textit{\textbf{170.4}} \\
\bottomrule
\end{tabular}
\end{table}
\begin{table}[htbp]
\caption{TTI so-16 kernel throughput (Gpts/s)}
\label{tab:tti:so-16}
\begin{tabular}{m{0.5cm}m{0.5cm}m{0.5cm}m{0.6cm}m{0.6cm}m{0.6cm}m{0.6cm}m{0.6cm}m{0.6cm}m{0.6cm}}
\toprule
 & 1 & 2 & 4 & 8 & 16 & 32 & 64 & 128 \\
\midrule
Basic & \cellcolor[HTML]{FFFF00} \color{red} \textit{\textbf{2.0}} & \cellcolor[HTML]{FFFF00} \color{red} \textit{\textbf{3.7}} & \cellcolor[HTML]{FFFF00} \color{red} \textit{\textbf{6.4}} & 15.9 & 30.0 & 55.5 & 112.2 & 181.0 \\
Diag & 2.1 & 4.0 & 7.6 & \cellcolor[HTML]{FFFF00} \color{green} \textit{\textbf{17.7}} & \cellcolor[HTML]{FFFF00} \color{green} \textit{\textbf{32.2}} & \cellcolor[HTML]{FFFF00} \color{green} \textit{\textbf{63.5}} & \cellcolor[HTML]{FFFF00} \color{green} \textit{\textbf{116.3}} & \cellcolor[HTML]{FFFF00} \color{green} \textit{\textbf{194.0}} \\
Full & \cellcolor[HTML]{FFFF00} \color{green} \textit{\textbf{2.2}} & \cellcolor[HTML]{FFFF00} \color{green} \textit{\textbf{4.3}} & \cellcolor[HTML]{FFFF00} \color{green} \textit{\textbf{7.8}} & \cellcolor[HTML]{FFFF00} \color{red} \textit{\textbf{14.8}} & \cellcolor[HTML]{FFFF00} \color{red} \textit{\textbf{27.1}} & \cellcolor[HTML]{FFFF00} \color{red} \textit{\textbf{49.5}} & \cellcolor[HTML]{FFFF00} \color{red} \textit{\textbf{82.1}} & \cellcolor[HTML]{FFFF00} \color{red} \textit{\textbf{166.0}} \\
\bottomrule
\end{tabular}
\end{table}

%% file: tables_raw/v_elastic_raw.tex
\begin{table}[!htbp]
\caption{ViscoElastic so-04 kernel throughput (Gpts/s)}
\label{tab:vel:so-04}
\begin{tabular}{m{0.5cm}m{0.5cm}m{0.5cm}m{0.6cm}m{0.6cm}m{0.6cm}m{0.6cm}m{0.6cm}m{0.6cm}m{0.6cm}}
\toprule
 & 1 & 2 & 4 & 8 & 16 & 32 & 64 & 128 \\
\midrule
Basic & \cellcolor[HTML]{FFFF00} \color{red} \textit{\textbf{1.2}} & 2.3 & 4.4 & 8.1 & 14.5 & 23.9 & 44.1 & \cellcolor[HTML]{FFFF00} \color{green} \textit{\textbf{78.3}} \\
Diag & \cellcolor[HTML]{FFFF00} \color{green} \textit{\textbf{1.3}} & \cellcolor[HTML]{FFFF00} \color{green} \textit{\textbf{2.4}} & \cellcolor[HTML]{FFFF00} \color{green} \textit{\textbf{4.6}} & \cellcolor[HTML]{FFFF00} \color{green} \textit{\textbf{8.3}} & \cellcolor[HTML]{FFFF00} \color{green} \textit{\textbf{15.5}} & \cellcolor[HTML]{FFFF00} \color{green} \textit{\textbf{25.8}} & \cellcolor[HTML]{FFFF00} \color{green} \textit{\textbf{44.2}} & 77.8 \\
Full & \cellcolor[HTML]{FFFF00} \color{red} \textit{\textbf{1.2}} & \cellcolor[HTML]{FFFF00} \color{red} \textit{\textbf{2.2}} & \cellcolor[HTML]{FFFF00} \color{red} \textit{\textbf{4.0}} & \cellcolor[HTML]{FFFF00} \color{red} \textit{\textbf{7.4}} & \cellcolor[HTML]{FFFF00} \color{red} \textit{\textbf{13.5}} & \cellcolor[HTML]{FFFF00} \color{red} \textit{\textbf{20.5}} & \cellcolor[HTML]{FFFF00} \color{red} \textit{\textbf{31.5}} & \cellcolor[HTML]{FFFF00} \color{red} \textit{\textbf{51.0}} \\
\bottomrule
\end{tabular}
\end{table}
\begin{table}[!htbp]
\caption{ViscoElastic so-08 kernel throughput (Gpts/s)}
\label{tab:vel:so-08}
\begin{tabular}{m{0.5cm}m{0.5cm}m{0.5cm}m{0.6cm}m{0.6cm}m{0.6cm}m{0.6cm}m{0.6cm}m{0.6cm}m{0.6cm}}
\toprule
 & 1 & 2 & 4 & 8 & 16 & 32 & 64 & 128 \\
\midrule
Basic & \cellcolor[HTML]{FFFF00} \color{red} \textit{\textbf{1.1}} & 2.0 & 3.8 & 7.0 & 11.6 & 22.2 & 36.2 & 67.1 \\
Diag & \cellcolor[HTML]{FFFF00} \color{green} \textit{\textbf{1.2}} & \cellcolor[HTML]{FFFF00} \color{green} \textit{\textbf{2.2}} & \cellcolor[HTML]{FFFF00} \color{green} \textit{\textbf{4.4}} & \cellcolor[HTML]{FFFF00} \color{green} \textit{\textbf{7.6}} & \cellcolor[HTML]{FFFF00} \color{green} \textit{\textbf{12.8}} & \cellcolor[HTML]{FFFF00} \color{green} \textit{\textbf{23.8}} & \cellcolor[HTML]{FFFF00} \color{green} \textit{\textbf{41.3}} & \cellcolor[HTML]{FFFF00} \color{green} \textit{\textbf{72.2}} \\
Full & 1.1 & \cellcolor[HTML]{FFFF00} \color{red} \textit{\textbf{1.9}} & \cellcolor[HTML]{FFFF00} \color{red} \textit{\textbf{3.5}} & \cellcolor[HTML]{FFFF00} \color{red} \textit{\textbf{6.5}} & \cellcolor[HTML]{FFFF00} \color{red} \textit{\textbf{10.6}} & \cellcolor[HTML]{FFFF00} \color{red} \textit{\textbf{17.5}} & \cellcolor[HTML]{FFFF00} \color{red} \textit{\textbf{30.3}} & \cellcolor[HTML]{FFFF00} \color{red} \textit{\textbf{44.0}} \\
\bottomrule
\end{tabular}
\end{table}
\begin{table}[!htbp]
\caption{ViscoElastic so-12 kernel throughput (Gpts/s)}
\label{tab:vel:so-12}
\begin{tabular}{m{0.5cm}m{0.5cm}m{0.5cm}m{0.6cm}m{0.6cm}m{0.6cm}m{0.6cm}m{0.6cm}m{0.6cm}m{0.6cm}}
\toprule
 & 1 & 2 & 4 & 8 & 16 & 32 & 64 & 128 \\
\midrule
Basic & \cellcolor[HTML]{FFFF00} \color{red} \textit{\textbf{1.0}} & 1.9 & 3.3 & 6.2 & 11.0 & 18.3 & 33.3 & 54.3 \\
Diag & \cellcolor[HTML]{FFFF00} \color{green} \textit{\textbf{1.1}} & \cellcolor[HTML]{FFFF00} \color{green} \textit{\textbf{2.0}} & \cellcolor[HTML]{FFFF00} \color{green} \textit{\textbf{3.7}} & \cellcolor[HTML]{FFFF00} \color{green} \textit{\textbf{6.8}} & \cellcolor[HTML]{FFFF00} \color{green} \textit{\textbf{12.4}} & \cellcolor[HTML]{FFFF00} \color{green} \textit{\textbf{22.1}} & \cellcolor[HTML]{FFFF00} \color{green} \textit{\textbf{37.4}} & \cellcolor[HTML]{FFFF00} \color{green} \textit{\textbf{62.1}} \\
Full & \cellcolor[HTML]{FFFF00} \color{red} \textit{\textbf{1.0}} & \cellcolor[HTML]{FFFF00} \color{red} \textit{\textbf{1.8}} & \cellcolor[HTML]{FFFF00} \color{red} \textit{\textbf{3.2}} & \cellcolor[HTML]{FFFF00} \color{red} \textit{\textbf{5.5}} & \cellcolor[HTML]{FFFF00} \color{red} \textit{\textbf{8.7}} & \cellcolor[HTML]{FFFF00} \color{red} \textit{\textbf{14.6}} & \cellcolor[HTML]{FFFF00} \color{red} \textit{\textbf{23.7}} & \cellcolor[HTML]{FFFF00} \color{red} \textit{\textbf{35.6}} \\
\bottomrule
\end{tabular}
\end{table}
\begin{table}[!htbp]
\caption{ViscoElastic so-16 kernel throughput (Gpts/s)}
\label{tab:vel:so-16}
\begin{tabular}{m{0.5cm}m{0.5cm}m{0.5cm}m{0.6cm}m{0.6cm}m{0.6cm}m{0.6cm}m{0.6cm}m{0.6cm}m{0.6cm}}
\toprule
 & 1 & 2 & 4 & 8 & 16 & 32 & 64 & 128 \\
\midrule
Basic & \cellcolor[HTML]{FFFF00} \color{red} \textit{\textbf{0.7}} & \cellcolor[HTML]{FFFF00} \color{red} \textit{\textbf{1.3}} & \cellcolor[HTML]{FFFF00} \color{red} \textit{\textbf{2.7}} & 4.9 & 8.6 & 14.8 & 27.0 & 42.0 \\
Diag & \cellcolor[HTML]{FFFF00} \color{green} \textit{\textbf{0.9}} & \cellcolor[HTML]{FFFF00} \color{green} \textit{\textbf{1.8}} & \cellcolor[HTML]{FFFF00} \color{green} \textit{\textbf{3.4}} & \cellcolor[HTML]{FFFF00} \color{green} \textit{\textbf{5.9}} & \cellcolor[HTML]{FFFF00} \color{green} \textit{\textbf{10.5}} & \cellcolor[HTML]{FFFF00} \color{green} \textit{\textbf{19.1}} & \cellcolor[HTML]{FFFF00} \color{green} \textit{\textbf{32.0}} & \cellcolor[HTML]{FFFF00} \color{green} \textit{\textbf{49.5}} \\
Full & 0.8 & 1.5 & 2.8 & \cellcolor[HTML]{FFFF00} \color{red} \textit{\textbf{4.6}} & \cellcolor[HTML]{FFFF00} \color{red} \textit{\textbf{7.9}} & \cellcolor[HTML]{FFFF00} \color{red} \textit{\textbf{13.6}} & \cellcolor[HTML]{FFFF00} \color{red} \textit{\textbf{22.8}} & \cellcolor[HTML]{FFFF00} \color{red} \textit{\textbf{33.5}} \\
\bottomrule
\end{tabular}
\end{table}

%% file: tables_raw/acoustic_gpu_raw.tex
\begin{table}[htbp]
\caption{Acoustic so-04 kernel GPU throughput (Gpts/s)}
\label{tab:ac_gpu:so-04}
\begin{tabular}{m{0.5cm}m{0.5cm}m{0.5cm}m{0.6cm}m{0.6cm}m{0.6cm}m{0.6cm}m{0.6cm}m{0.6cm}m{0.6cm}}
\toprule
 & 1 & 2 & 4 & 8 & 16 & 32 & 64 & 128 \\
\midrule
Basic & 34.3 & 65.6 & 123.3 & 200.2 & 348.6 & 583.0 & 985.2 & 1535.0 \\
\bottomrule
\end{tabular}
\end{table}
\begin{table}[htbp]
\caption{Acoustic so-08 kernel GPU throughput (Gpts/s)}
\label{tab:ac_gpu:so-08}
\begin{tabular}{m{0.5cm}m{0.5cm}m{0.5cm}m{0.6cm}m{0.6cm}m{0.6cm}m{0.6cm}m{0.6cm}m{0.6cm}m{0.6cm}}
\toprule
 & 1 & 2 & 4 & 8 & 16 & 32 & 64 & 128 \\
\midrule
Basic & 31.2 & 59.4 & 121.7 & 199.2 & 333.1 & 565.5 & 970.1 & 1474.5 \\
\bottomrule
\end{tabular}
\end{table}
\begin{table}[htbp]
\caption{Acoustic so-12 kernel GPU throughput (Gpts/s)}
\label{tab:ac_gpu:so-12}
\begin{tabular}{m{0.5cm}m{0.5cm}m{0.5cm}m{0.6cm}m{0.6cm}m{0.6cm}m{0.6cm}m{0.6cm}m{0.6cm}m{0.6cm}}
\toprule
 & 1 & 2 & 4 & 8 & 16 & 32 & 64 & 128 \\
\midrule
Basic & 28.8 & 61.0 & 104.7 & 160.2 & 271.2 & 434.6 & 742.2 & 1140.7 \\
\bottomrule
\end{tabular}
\end{table}
\begin{table}[htbp]
\caption{Acoustic so-16 kernel GPU throughput (Gpts/s)}
\label{tab:ac_gpu:so-16}
\begin{tabular}{m{0.5cm}m{0.5cm}m{0.5cm}m{0.6cm}m{0.6cm}m{0.6cm}m{0.6cm}m{0.6cm}m{0.6cm}m{0.6cm}}
\toprule
 & 1 & 2 & 4 & 8 & 16 & 32 & 64 & 128 \\
\midrule
Basic & 25.8 & 47.9 & 90.7 & 143.7 & 242.4 & 387.8 & 666.2 & 1017.3 \\
\bottomrule
\end{tabular}
\end{table}

%% file: tables_raw/elastic_gpu_raw.tex
\begin{table}[htbp]
\caption{Elastic so-04 kernel GPU throughput (Gpts/s)}
\label{tab:el_gpu:so-04}
\begin{tabular}{m{0.5cm}m{0.5cm}m{0.5cm}m{0.6cm}m{0.6cm}m{0.6cm}m{0.6cm}m{0.6cm}m{0.6cm}m{0.6cm}}
\toprule
 & 1 & 2 & 4 & 8 & 16 & 32 & 64 & 128 \\
\midrule
Basic & 6.5 & 11.7 & 22.0 & 34.2 & 58.0 & 95.4 & 143.9 & 198.9 \\
\bottomrule
\end{tabular}
\end{table}
\begin{table}[htbp]
\caption{Elastic so-08 kernel GPU throughput (Gpts/s)}
\label{tab:el_gpu:so-08}
\begin{tabular}{m{0.5cm}m{0.5cm}m{0.5cm}m{0.6cm}m{0.6cm}m{0.6cm}m{0.6cm}m{0.6cm}m{0.6cm}m{0.6cm}}
\toprule
 & 1 & 2 & 4 & 8 & 16 & 32 & 64 & 128 \\
\midrule
Basic & 5.2 & 9.4 & 16.8 & 27.2 & 45.5 & 72.7 & 114.1 & 164.2 \\
\bottomrule
\end{tabular}
\end{table}
\begin{table}[htbp]
\caption{Elastic so-12 kernel GPU throughput (Gpts/s)}
\label{tab:el_gpu:so-12}
\begin{tabular}{m{0.5cm}m{0.5cm}m{0.5cm}m{0.6cm}m{0.6cm}m{0.6cm}m{0.6cm}m{0.6cm}m{0.6cm}m{0.6cm}}
\toprule
 & 1 & 2 & 4 & 8 & 16 & 32 & 64 & 128 \\
\midrule
Basic & 4.0 & 7.2 & 13.3 & 21.7 & 35.8 & 57.2 & 92.7 & 131.9 \\
\bottomrule
\end{tabular}
\end{table}
\begin{table}[htbp]
\caption{Elastic so-16 kernel GPU throughput (Gpts/s)}
\label{tab:el_gpu:so-16}
\begin{tabular}{m{0.5cm}m{0.5cm}m{0.5cm}m{0.6cm}m{0.6cm}m{0.6cm}m{0.6cm}m{0.6cm}m{0.6cm}m{0.6cm}}
\toprule
 & 1 & 2 & 4 & 8 & 16 & 32 & 64 & 128 \\
\midrule
Basic & 2.5 & 4.6 & 8.6 & 15.4 & 26.0 & 42.4 & 68.9 & 100.7 \\
\bottomrule
\end{tabular}
\end{table}

%% file: tables_raw/tti_gpu_raw.tex
\begin{table}[htbp]
\caption{TTI so-04 kernel throughput (Gpts/s)}
\label{tab:tti_gpu:so-04}
\begin{tabular}{m{0.5cm}m{0.5cm}m{0.5cm}m{0.6cm}m{0.6cm}m{0.6cm}m{0.6cm}m{0.6cm}m{0.6cm}m{0.6cm}}
\toprule
 & 1 & 2 & 4 & 8 & 16 & 32 & 64 & 128 \\
\midrule
Basic & 10.5 & 20.3 & 37.8 & 63.8 & 109.6 & 200.1 & 354.9 & 541.8 \\
\bottomrule
\end{tabular}
\end{table}
\begin{table}[htbp]
\caption{TTI so-08 kernel throughput (Gpts/s)}
\label{tab:tti_gpu:so-08}
\begin{tabular}{m{0.5cm}m{0.5cm}m{0.5cm}m{0.6cm}m{0.6cm}m{0.6cm}m{0.6cm}m{0.6cm}m{0.6cm}m{0.6cm}}
\toprule
 & 1 & 2 & 4 & 8 & 16 & 32 & 64 & 128 \\
\midrule
Basic & 8.5 & 16.2 & 31.0 & 53.1 & 90.6 & 163.8 & 289.1 & 460.7 \\
\bottomrule
\end{tabular}
\end{table}
\begin{table}[htbp]
\caption{TTI so-12 kernel throughput (Gpts/s)}
\label{tab:tti_gpu:so-12}
\begin{tabular}{m{0.5cm}m{0.5cm}m{0.5cm}m{0.6cm}m{0.6cm}m{0.6cm}m{0.6cm}m{0.6cm}m{0.6cm}m{0.6cm}}
\toprule
 & 1 & 2 & 4 & 8 & 16 & 32 & 64 & 128 \\
\midrule
Basic & 7.5 & 14.4 & 27.4 & 46.0 & 78.0 & 138.9 & 250.3 & 405.1 \\
\bottomrule
\end{tabular}
\end{table}
\begin{table}[htbp]
\caption{TTI so-16 kernel throughput (Gpts/s)}
\label{tab:tti_gpu:so-16}
\begin{tabular}{m{0.5cm}m{0.5cm}m{0.5cm}m{0.6cm}m{0.6cm}m{0.6cm}m{0.6cm}m{0.6cm}m{0.6cm}m{0.6cm}}
\toprule
 & 1 & 2 & 4 & 8 & 16 & 32 & 64 & 128 \\
\midrule
Basic & 5.8 & 11.2 & 21.3 & 38.2 & 65.7 & 115.8 & 205.2 & 322.4 \\
\bottomrule
\end{tabular}
\end{table}

%% file: tables_raw/v_elastic_gpu_raw.tex
\begin{table}[!htbp]
\caption{ViscoElastic so-04 kernel throughput (Gpts/s)}
\label{tab:vel_gpu:so-04}
\begin{tabular}{m{0.5cm}m{0.5cm}m{0.5cm}m{0.6cm}m{0.6cm}m{0.6cm}m{0.6cm}m{0.6cm}m{0.6cm}m{0.6cm}}
\toprule
 & 1 & 2 & 4 & 8 & 16 & 32 & 64 & 128 \\
\midrule
Basic & 3.4 & 6.3 & 11.9 & 19.2 & 33.6 & 57.4 & 90.8 & 128.1 \\
\bottomrule
\end{tabular}
\end{table}
\begin{table}[!htbp]
\caption{ViscoElastic so-08 kernel throughput (Gpts/s)}
\label{tab:vel_gpu:so-08}
\begin{tabular}{m{0.5cm}m{0.5cm}m{0.5cm}m{0.6cm}m{0.6cm}m{0.6cm}m{0.6cm}m{0.6cm}m{0.6cm}m{0.6cm}}
\toprule
 & 1 & 2 & 4 & 8 & 16 & 32 & 64 & 128 \\
\midrule
Basic & 2.8 & 5.3 & 9.4 & 16.0 & 27.9 & 46.0 & 73.7 & 107.8 \\
\bottomrule
\end{tabular}
\end{table}
\begin{table}[!htbp]
\caption{ViscoElastic so-12 kernel throughput (Gpts/s)}
\label{tab:vel_gpu:so-12}
\begin{tabular}{m{0.5cm}m{0.5cm}m{0.5cm}m{0.6cm}m{0.6cm}m{0.6cm}m{0.6cm}m{0.6cm}m{0.6cm}m{0.6cm}}
\toprule
 & 1 & 2 & 4 & 8 & 16 & 32 & 64 & 128 \\
\midrule
Basic & 2.5 & 4.7 & 8.5 & 13.1 & 23.0 & 37.4 & 60.4 & 88.4 \\
\bottomrule
\end{tabular}
\end{table}
\begin{table}[!htbp]
\caption{ViscoElastic so-16 kernel throughput (Gpts/s)}
\label{tab:vel_gpu:so-16}
\begin{tabular}{m{0.5cm}m{0.5cm}m{0.5cm}m{0.6cm}m{0.6cm}m{0.6cm}m{0.6cm}m{0.6cm}m{0.6cm}m{0.6cm}}
\toprule
 & 1 & 2 & 4 & 8 & 16 & 32 & 64 & 128 \\
\midrule
Basic & 1.6 & 3.1 & 6.2 & 10.7 & 18.6 & 31.0 & 48.9 & 71.6 \\
\bottomrule
\end{tabular}
\end{table}